\documentclass[aps,prd,twocolumn,groupedaddress,nofootinbib,showpacs]{revtex4-1}

\usepackage{graphicx,amssymb,subfigure}
\usepackage{amsfonts}
\usepackage{amsmath}
\usepackage{amssymb}
\usepackage{graphicx}
\usepackage{color}
\usepackage{subfigure}
\usepackage{epstopdf}

\begin{document}

\addtolength{\hoffset}{-0.175cm}
\addtolength{\textwidth}{0.35cm}
\title{Increasing the Lensing Figure of Merit through Higher Order Convergence Moments}

\author{Martina Vicinanza$^{1,2,3,4}$}
\email{mvicinanza@fc.ul.pt}
\author{Vincenzo F. Cardone$^3$}
\email{winnyenodrac@gmail.com}
\author{Roberto Maoli$^{2,5}$}
\email{roberto.maoli@roma1.infn.it}
\author{Roberto Scaramella$^{3,5}$}
\email{roberto.scaramella@oa-roma.inaf.it}
\author{Xinzhong Er$^3$}
\email{er.xinzhong@oa-roma.inaf.it}

\affiliation{$^1$Dipartimento di Fisica, Universit\`{a} di Roma "Tor Vergata", via della Ricerca Scientifica 1, 00133 - Roma, Italy}
\affiliation{$^2$Dipartimento di Fisica, Universit\`{a} di Roma "La Sapienza", Piazzale Aldo Moro, 00185 - Roma, Italy}
\affiliation{$^3$I.N.A.F.\,-\,Osservatorio Astronomico di Roma, via Frascati 33, 00040 - Monte Porzio Catone (Roma), Italy}
\affiliation{$^4$Instituto de Astrof\'{i}sica e Ci\^{e}ncias do Espa\c{c}o, Universidade de Lisboa, Faculdade de Ci\^{e}ncias, Campo Grande, PT1749-016 Lisbon, Portugal}
\affiliation{$^5$I.N.F.N.\,-\,Sezione di Roma 1, Piazzale Aldo Moro, 00185 - Roma, Italy}
\date{Received \today; published -- 00, 0000}

\begin{abstract}

The unprecedented quality, the increased dataset, and the wide area of ongoing and near future weak lensing surveys allows to move beyond the standard two points statistics thus making worthwhile to investigate higher order probes. As an interesting step towards this direction, we expolore the use of higher order moments (HOM) of the convergence field as a way to increase the lensing Figure of Merit (FoM). To this end, we rely on simulated convergence to first show that HOM can be measured and calibrated so that it is indeed possible to predict them for a given cosmological model provided suitable nuisance parameters are introduced and then marginalized over. We then forecast the accuracy on cosmological parameters from the use of HOM alone and in combination with standard shear power spectra tomography. It turns out that HOM allow to break some common degeneracies thus significantly boosting the overall FoM. We also qualitatively discuss possible systematics and how they can be dealt with. 

\end{abstract}

\pacs{98.80.-k, 98.80.Es, 95.36.+x}

\maketitle

\section{Introduction}

That the universe is spatially flat and undergoing a phase of accelerated expansion is nowadays a well known fact confirmed by an overwhelming flood of hard to confute evidences (see, e.g., \cite{W13} and refs. therein for a comprehensive altough not updated review). What is responsible for this speed up is, on the contrary, still frustrating unknown notwithstanding the considerable amount of papers proposing theoretical solutions to this problem. On the one hand, the concordance $\Lambda$CDM model \cite{P03} composed by a Cold Dark Matter (CDM) component causing clustering and a cosmological constant $\Lambda$ causing acceleration, makes an excellent work at fitting the data on cosmological scales \cite{planck,6dFGS,DR7,BW,SNeIa} but it is plagued by both theoretical shortcomings and unsatisfactory agreement with data on galactic scales \cite{burgess2013}. On the other hand, dynamical models, collectively referred to as dark energy \cite{CST06,LucaBook,TWG}, including, e.g., an evolving scalar field rolling down its self\,-\,interaction potential, performs well at reproducing the observed data, but lack a solid foundation somewhat appearing as a way to shift the problem from what is driving speed up to what is sourcing the field and its evolution. A radically different approach is possible if one takes cosmic speed up as a first signature of a failure in our understanding of the laws of gravity. Rather than being evidence for something missing in the cosmic pie, acceleration then becomes a way Nature is pointing us at a more general formulation of the theory of gravity with General Relativity. This opens up the way to a full set of modified gravity theories \cite{Clif13,Joy15} able to closely follows the same background evolution of the $\Lambda$CDM model hence being in agreement with the same data. 

Discriminating among these two rival approaches is the aim of modern cosmology with ongoing and planned surveys trying to achieve this goal by both reducing statistical and systematic errors (hence constraining the dark energy equation of state) and looking for observational probes able to test the growth of perturbations which evolves in different ways in the two competing scenarios. Thanks to the possibility offered to probe both the kinematics and the dynamics of the universe, gravitational lensing has long been considered as an ideal tool to investigate the nature and the nurture of dark energy and to look for signatures of modified gravity. It is therefore not surprising that cosmic shear tomography has been recommended as the most promising technique to solve the problem of cosmic acceleration. After the pioneering small area surveys such as the CFHTLenS \cite{Hey12} and DeepLens \cite{DeepLens} programs which have shown the feasibility of the proposal, large area surveys have started to fully exploit the lensing potentialities with first interesting results already coming from surveys in progress, such as DES \cite{DES}, KIDS \cite{KIDS} and HSC \cite{HSC}. Near future surveys, both ground (LSST \cite{LSST}) and space based (Euclid \cite{RB}, WFIRST \cite{WFIRST}), will make it possible to make the error budget systematics dominated so that, provided these latter are taken under control, the hunt to the responsibile of cosmic speed up could likely come to a successful conclusion.

From a practical point of view, all the above quoted surveys aim at mapping cosmic shear from the ellipticity of large galaxy samples. Shear catalogs are then used to determine the two points correlation function splitting sources in redshift bins to perform tomography. This is then contrasted against theoretical expectations thus constraining the cosmological parameters (see, e.g., \cite{shear}). Alternatively, one can rely on a 3D analysis based on spherical Fourier\,-\,Bessel decompisition to constrain both dark energy models \cite{Tom14} and modified gravity theories \cite{Pr16}. It is worth noting that both these techniques mainly probe the growth of structures in the linear and quasi\,-\,linear regime so that the constraints on cosmological parameters are affected by severe degeneracies such as, e.g., the well known $\Omega_M$\,-\,$\sigma_8$ one. In order to break this degeneracy, a possible way out consists in moving to a different tracer. Indeed, cosmic shear measurements can also be used to reconstruct the convergence field which quantifies the projection along the line of sight of the density contrast $\delta({\bf x}) = [\rho({\bf x}) - \bar{\rho}]/\bar{\rho}$. What makes convergence so attractive is the different scales which is sensible to. Indeed, the convergence field mainly probes the nonlinear scales \cite{JSW2000,MJ2000,V2000,TJ2004,TW2004,VMB2005} thus offering complementary information to the shear field. Moreover, on these scales, the collapse of structures introduce deviations from the Gaussianity which are strongly related to the underlying cosmological model and theory of gravity. The non Gaussianity of the field can be quantified through higher than second order moments which can then be used to break the $\Omega_M$\,-\,$\sigma_8$ degeneracy \citep{Berny1997,JaSel1997}. Actually, higher order moments depend on the full set of cosmological parameters so that it is worth investigating whether they can represent a valuable help to narrow down the uncertainties on the dark energy equation of state too. In other words, one can wonder whether convergence moments can help to increase the Figure of Merit (FoM) if used alone and/or in combination with the standard second order cosmic shear tomography.

The plan of the paper is as follows. In Sect.\,II, we derive a phenomenological yet well motivated relation connecting the theoretical moments with the observed ones taking care of issues related to noise and map reconstruction. How to predict moments from a given cosmological model is detailed in Sect.\,III, while Sect.\,IV presents the simulated maps we use to show that moments can be measured from wide enough surveyes. Armed with these mock data, we can then validate the relation between theoretical and observed moments which is carried out in Sect.\,V thus making us confident that moments can indeed be used as cosmological probes. What can be gained by such a probe is investigated in Sect.\,VI through a Fisher matrix based forecast analysis. A discussion of possible systematics and how they could impact the results and taken into account for is presented in Sect.\,VII, while Sect.\,VIII is devoted to conclusions. Some further material is given in Appendix A and B for the interested reader.

\section{Convergence moments}

The lensing effect of the large scale structure is responsible for the generation of the convergence field $\kappa$. However, we do not actually measure the convergence $\kappa$, but rather the ellipticity of the galaxies which are then used as input to reconstruct the shear field $\gamma$. A suitable algorithm is then used to get the convergence map. Provided the code is correctly working, one can relate the input field $\kappa$ to the observed one $\kappa_{obs}$ through the following linear relation

\begin{equation}
\kappa_{obs} = (1 + m) \kappa + c + {\cal{N}}
\label{eq: kappaobs}
\end{equation}
where $(m, c)$ are the multiplicative and additive bias and ${\cal{N}}$ is a noise term. Since both the signal and the noise vanish when averaged over a sufficiently large area, i.e., $\langle \kappa \rangle = \langle {\cal{N}} \rangle = 0$, one usually deals with $\tilde{\kappa}_{obs} = \kappa_{obs} - \langle \kappa_{obs} \rangle$ so that $\langle \tilde{\kappa}_{obs} \rangle = 0$. From Eq.(\ref{eq: kappaobs}), one naively get

\begin{equation}
\tilde{\kappa}_{obs} = (1 + m) \kappa + {\cal{N}}
\label{eq: kappatilde}
\end{equation}
which is the starting point for the derivation of the moments. Hereafter, to shorten the notation, we will drop the tilde sign from $\tilde{\kappa}_{obs}$ and denote with $\kappa_{obs}$ the observed convergence after subtracting the mean value. 

Let us now compute powers of $\kappa_{obs}$ using Eq.(\ref{eq: kappatilde}) to get

\begin{displaymath}
\kappa_{obs}^2 = (1 + m)^2 \kappa^2 + 2(1 + m) \kappa {\cal{N}} + {\cal{N}}^2 \ ,
\end{displaymath}

\begin{displaymath}
\kappa_{obs}^3 = (1 + m)^3 \kappa^3 + 3(1 + m)^2 \kappa^2 {\cal{N}} + 3 (1 + m) \kappa {\cal{N}}^2 + {\cal{N}}^3 \ ,
\end{displaymath}

\begin{eqnarray}
\kappa_{obs}^4 & = & (1 + m)^4 \kappa^4 + 4(1 + m)^3 \kappa^3 {\cal{N}} + 6 (1 + m)^2 \kappa^2 {\cal{N}}^2 \nonumber \\
& + & 4(1 + m) \kappa {\cal{N}}^3 + {\cal{N}}^4 \nonumber \ .
\end{eqnarray}
Since $\kappa$ and ${\cal{N}}$ are uncorrelated, the expectation values of products as $\kappa^n {\cal{N}}^m$ will simply be the product of the corresponding expectation values. Since both $\kappa$ and ${\cal{N}}$ have zero mean, it is then easy to get the following expressions for the moments of the observed convergence field\,:

\begin{displaymath}
\langle \kappa_{obs}^{2} \rangle = (1 + m)^2 \langle \kappa^2 \rangle + \langle {\cal{N}}^2 \rangle \ , 
\end{displaymath}

\begin{displaymath}
\langle \kappa_{obs}^{3} \rangle = (1 + m)^3 \langle \kappa^3 \rangle + \langle {\cal{N}}^3 \rangle \ , 
\end{displaymath}

\begin{displaymath}
\langle \kappa_{obs}^{4} \rangle = (1 + m)^4 \langle \kappa^4 \rangle + 6 (1 + m)^2 \langle \kappa^2 \rangle \langle {\cal{N}}^2 \rangle \  + \langle {\cal{N}}^4 \rangle  \ . 
\end{displaymath}
The moments of the input convergence field $\kappa$ should be theoretically predicted for a given cosmological model. However, since theory is imperfect (as we will see later), one can postulate the following ansatz

\begin{equation}
\langle \kappa^n \rangle = (1 + \mu_n) \langle \kappa^n \rangle_{th} + \gamma_n
\label{eq: kappathlin}
\end{equation}
where $\langle \kappa^n \rangle_{th}$ is the moment of order $n$ as predicted from the theory, and $(\mu_n, \gamma_n)$ are calibration parameters to match theory with numerical simulations. Inserting these relations into the above ones, we finally end up with the following calibration relations

\begin{equation}
\langle \kappa_{obs}^2 \rangle = (1 + m_2) \langle \kappa^2 \rangle + c_2 + \langle {\cal{N}}^2 \rangle \ ,
\label{eq: cal2nd}
\end{equation}

\begin{equation}
\langle \kappa_{obs}^3 \rangle = (1 + m_3) \langle \kappa^3 \rangle + c_3 + \langle {\cal{N}}^3 \rangle \ ,
\label{eq: cal3rd}
\end{equation}

\begin{eqnarray}
\langle \kappa_{obs}^4 \rangle & = & (1 + m_4) \langle \kappa^4 \rangle + 
c_4 + \langle {\cal{N}}^4 \rangle  \nonumber \\
 & + & 6 [(1 + m_2) \langle \kappa^2 \rangle + c_2] \langle {\cal{N}}^2 \rangle \ , 
\label{eq: cal4th}
\end{eqnarray}
where we have dropped the label {\it th} to shorten the notation, and defined

\begin{displaymath}
1 + m_n = (1 + m)^n (1 + \mu_n) \ \ , \ \ c_n = (1 + m)^n \gamma_n \ \ .
\end{displaymath}
It is worth noting how, although the calibration parameters are in total seven, namely $(\mu_2, \gamma_2, \mu_3, \gamma_3, \mu_4, \gamma_4, m)$, they are actually degenerate since they enter Eqs.(\ref{eq: cal2nd})\,-\,(\ref{eq: cal4th}) only through their combinations defining the six nuisance parameters $(m_2, c_2, m_3, c_3, m_4, c_4)$. One could naively think that the noise moments $(\langle {\cal{N}}^2 \rangle, \langle {\cal{N}}^3 \rangle, \langle {\cal{N}}^4 \rangle)$ are fully degenerate with $(c_2, c_3, c_4)$ since both enter as additive terms. However, that this is not the case can be understood noting that what moments of the observed convergence field are not directly measured on the map as it is, but only after smoothing it with a filter of aperture $\theta$. As such, in the above relations, while $(m_n, c_n)$ are constants, both the observed and theoretical moments $(\langle \kappa_{obs}^n \rangle, \langle \kappa \rangle)$ and the noise moments $\langle {\cal{N}}^n \rangle$ must be considered as function of the smoothing angle $\theta$. While the dependence of $\langle \kappa^n \rangle$ on $\theta$ will be derived later from a theoretical motivated approach, we can postulate here the following scaling for the noise moments

\begin{equation}
\langle {\cal{N}}^n \rangle = \nu_n (\theta/\theta_0)^{-n}
\label{eq: noisescale}
\end{equation}
with $\theta_0$ an arbitrary reference scale (which we will set to $1 \ {\rm arcsec}$) and $\nu_n$ the value of the noise moment of order $n$ on the smoothed map. Note that Eq.(\ref{eq: noisescale}) is correct for a white noise, but can be taken as an intuitive general result no matter the noise power spectrum. In the particular case of a Gaussian noise, it is 

\begin{displaymath}
\nu_3 = 0 \ \ , \ \ \nu_4 = 3 \nu_2^2
\end{displaymath}
so that the number of unknown noise quantities reduce to one. However, as a conservative choice, we will not rely on this assumption in the following. We therefore end up with a total of nine nuisance parameter to match the theoretically expected moments to the observed one as measured on a convergence map reconstructed from noisy shear data and smoothed with a filter of aperture $\theta$. 

\section{Moments from theory}

Two main actors enter the scene of Eqs.(\ref{eq: cal2nd})\,-\,(\ref{eq: cal4th}). We investigate here how one of them can be estimated for a given cosmological model. To this end, we first remind that the weak lensing convergence field is the result of inhomogenities in the large scale matter distribution along the line of sight to a distant source. The 2D convergence field $\kappa$ is then related to the 3D density contrast $\delta$ as

\begin{equation}
\kappa = \int_{0}^{\chi_s}{d\chi W(\chi) \delta(\chi)} = \frac{c}{H_0} \int_{0}^{z_h}{\frac{W(z) \delta(z)}{E(z)} dz}
\label{eq: kappavsdelta}
\end{equation}
where

\begin{equation}
\chi(z) = \frac{c}{H_0} \int_{0}^{z}{\frac{dz^{\prime}}{E(z^{\prime})}}
\label{eq: defchi}
\end{equation}
is the comoving distance to a source at redshift $z$ (having assumed a spatially flat universe), and $E(z) = H(z)/H_0$ with $H(z)$ the cosmology dependent Hubble parameter and a subscript $0$ denotes present day values. In Eq.(\ref{eq: kappavsdelta}), the integral extends up to the redshift of last scattering surface $z_h$ to take into account the cumulative effect of the full matter distribution. However, the sum is actually weighthed by the lensing kernel $W(z)$ which takes into account the normalized source redshift distribution $n(z)$ through

\begin{equation}
W(z) = \frac{3 \Omega_M H_0^2}{2 c^2} (1 + z) \chi(z) \int_{z}^{z_h}{\frac{\chi(z^{\prime}) - \chi(z)}{\chi(z^{\prime})} n(z^{\prime}) dz^{\prime}} \ .
\label{eq: defkern}
\end{equation}
Using a Fourier decomposition and the Limber flat sky approximation is then possible to compute the two points projected correlation function (see, e.g., \cite{MJ01} and refs. therein). Its average smoothed over an angle $\theta$ with a filter ${\cal{W}}(\theta)$ then gives the second order moment of the convergence field. A similar procedure can then be used to infer higher order moments starting from multi points correlation functions and their corresponding multi\,-\,spectra. Under the assumption of validity of the hierarchical ansatz (which is the case in highly non linear regime and in the quasi linear one in the limit of vanishing variance), all higher order multi\,-\,spectra can be written as weighted sums of the products of the matter power spectrum. it is then possible to write the moments up to the fourth order of the convergence field as \cite{MJ01}

\begin{equation}
\langle \kappa^2 \rangle(\theta) = {\cal{C}}_2(\kappa_{\theta}) \ ,
\label{eq: kappa2nd}
\end{equation}

\begin{equation}
\langle \kappa^3 \rangle(\theta) = 3 {\cal{Q}}_3 {\cal{C}}_3(\kappa_{\theta}^2) \ ,
\label{eq: kappa3rd}
\end{equation}

\begin{equation}
\langle \kappa^4 \rangle(\theta) = (12 {\cal{R}}_a + 4 {\cal{R}}_b) {\cal{C}}_4(\kappa_{\theta}^3) \ ,
\label{eq: kappa4th}
\end{equation}
where we have defined

\begin{equation}
\kappa_{\theta} = 2\pi \int{P[\ell/\chi(z), z] {\cal{W}}(\ell \theta_0) \ell d\ell} \ ,
\label{eq: defkappatheta0}
\end{equation}

\begin{equation}
{\cal{C}}_t(\kappa^n_{\theta}) = \frac{c}{H_0} \int_{0}^{z_h}{\frac{W^t(z) \kappa_{\theta}^{n}(z)}{\chi^{2(t - 1)}(z) E(z)} dz} \ .
\label{eq: defcfun}
\end{equation}
In Eq.(\ref{eq: defkappatheta0}), $P(k, z)$ is the matter power spectrum evaluated in $k = \ell/\chi(z)$ because of the flat sky approximation, while the coefficients $({\cal{Q}}_3, {\cal{R}}_a, {\cal{R}}_b)$ depends on the amplitude of the different topologies contributing to the multi point correlation functions. Their values should be tailored against suitably designed N\,-\,body simulations which are not always available. We will therefore set their values to the following fiducial values

\begin{displaymath}
{\cal{Q}}_3^{fid} = 1.00 \ \ , \ \ {\cal{R}}_a^{fid} = 7.29 \ \ , \ \ {\cal{R}}_b^{fid} = 16.23 \ \ .
\end{displaymath}
Eqs.(\ref{eq: kappa2nd})\,-\,(\ref{eq: kappa4th}) can be used as input in the calibration relations (\ref{eq: cal2nd})\,-\,(\ref{eq: cal4th}). However, these have been derived postulating that the actual theoretical quantities are linearly related to the ones we have derived here as in Eq.(\ref{eq: kappathlin}). Ideally, the theory summarized here should lead to a perfect match with the (unknown) true moments so that $\mu_n = \gamma_n = 0$ for all $n$. However, there are different arguments advicing to be conservative and allow for $(\mu_n, \gamma_n)$ deviate from null values. We sketch them qualitatively below.

\begin{itemize}

\item[i.]{Eqs.(\ref{eq: kappa3rd}) and (\ref{eq: kappa4th}) define the third and fourth order moments up to a multiplicative term given by ${\cal{Q}}_3$ and ${\cal{Q}}_4 = 12 {\cal{R}}_a + 4 {\cal{R}}_b$. We have set these quantities to fiducial values, but they should be actually set matching multi point correlation functions to those measured from high resolution N\,-\,body simulations. We can easily parameterize the systematic uncertainty on them through the multiplicative bias terms

\begin{displaymath}
1 + \mu_3 = {\cal{Q}}_3/{\cal{Q}}_3^{fid} \ \ , \ \ 1 + \mu_4 = {\cal{Q}}_4/{\cal{Q}}_4^{fid} \ \ .
\end{displaymath}
Note that, according to this argument, we should set $\mu_2 = 0$. Although we do not force this constraint for reasons explained below, we nevertheless note that we will indeed find that $\mu_2$ is quite small. On the contrary, nothing prevents $(\mu_3, \mu_4)$ to significantly deviate from the null value.}

\item[ii.]{The convergence moments are estimated from functionals of the smoothed mean convergence $\kappa_{\theta}$ defined in Eq.(\ref{eq: defkappatheta0}). Here, one should integrate over all the scales $k$, but what one actually does is to truncate the integration range to avoid extrapolating the nonlinear model for the matter power spectrum to scales where it is not tested. Although one expects that the scales outside the integration range give a little contribution, this term can be boosted by the functional defining $\langle \kappa^n \rangle$. It is therefore worth allowing for this possibility introducing the additive bias term $\gamma_n$.} 

\item[iii.]{Notwithstanding the great efforts in modeling nonlinerities in the matter power spectrum, the matching between the theoretical one $P(k, z)$ and that the estimated from N\,-\,body simulations is still not perfect on all scales and redshift of interest. If we denote with $P_{N}(k, z)$ this latter and take that this as the best representation of the unknown actual one, we can assume that the following relation hold true

\begin{displaymath}
P_{N}(k, z) = (1 + \pi_1) P(k, z) + \pi_0 P(k_0, z_0)
\end{displaymath}
where $(\pi_0, \pi_1)$ are constant quantities, $P(k_0, z_0)$ the value of the power spectrum in an arbitrary set reference point $(k_0, z_0)$, and the linear approximation is motivated by the assumption that the deviations are small enough. It is then only a matter of algebra to show that the moments evaluated using $P_N(k, z)$ and those from $P(k. z)$ are related by a linear relation as far as one neglects higher order terms in $(\pi_0, \pi_1)$.}

\end{itemize}
These considerations provide a valid support to Eq.(\ref{eq: kappathlin}) and hence to the calibration relations derived from it. We will nevertheless validate the linear relation (\ref{eq: kappathlin}) later through a comparison with moments evaluated on a convergence map inferred from numerical simulations.

\section{Moments from simulated maps}

In order to validate the theoretical approach worked out above and find out the calibration parameters $(m_n, c_n)$ , we need to contrast them against moments measured on a convergence map reconstructed by shear noisy data. Although such a kind of dataset are starting to be available (see, e.g., \cite{Ludo2013} from CFHTLenS, and \cite{DesMap} from DES Science Verification data), the survey area is still too small to allow for reducing the statistical uncertainty to an acceptable level and prevent from cosmic variance effects. Moreover, noiseless convergence maps are also needed in order to validate Eq.(\ref{eq: kappathlin}) which can obviously not be inferred from actual data. We are therefore forced to rely on a simulated dataset.

\subsection{The MICE lensing catalog}

N\,-\,body light cone simulations are an ideal tool to build all\,-\,sky lensing maps. We rely, in particular, on the the MICE Grand Challenge (MICE\,-\,GC) containing about 70 billion dark matter particles in a $(3 h^{-1} \ {\rm Gpc})^3$ volume \cite{Fos2015a,Crocce2015b}. The parent simulation is run in a flat $\Lambda$CDM with cosmological parameters set

\begin{displaymath}
(\Omega_M, \Omega_b, h, n_s, \sigma_8) = (0.25, 0.044, 0.70, 0.95, 0.80) \ ,
\end{displaymath}
being $(\Omega_M, \Omega_b, h, n_s, \sigma_8) $ the matter and baryon density parameters, the Hubble constant in units of $100 \ {\rm km/s/Mpc}$, the spectral index and the variance of perturbations on the scale $R = 8 h^{-1} \ {\rm Mpc}$, respectively. Galaxies are associated to dark matter haloes using a Halo Occupation Distribution and a Halo Abundance Matching technique \cite{Carr2015} whose parameters are set to match local observational constraints, such as the local luminosity function \citep{Blan2003,Blan2005a}, the galaxy clustering as a function of luminosity and colour \citep{Zeha2011} and colour	\,-\,magnitude diagrams \citep{Blan2005b}.

Given its large volume and fine mass resolution, the MICE\,-\,GC simulation also allows an accurate modelling of the lensing observables from upcoming wide and deep galaxy surveys. Following the {\it Onion Universe} approach \cite{Fos2015b}, all\,-\,sky lensing maps are constructed with sub\,-\,armiminute scale resolution.  These lensing maps allow to model galaxy lensing properties, such as the convergence, shear, and lensed magnitudes and positions. Tests have been performed to show that the galaxy lensing mocks can be used to accurately model lensing observables down to arcminute scales. We use the MICECAT v2.0 (kindly avaialblet to us from P. Fosalba) which updates the public release MICECAT v1.0 catalog to include less massive and hence lower luminosities galaxies. The galaxy catalog is complete up to $i \sim 24$ for $0 \le z < 1.4$ so that we will hereafter focus our attention to this redshift range only.

\begin{figure*}
\centering
\includegraphics[width=5.5cm]{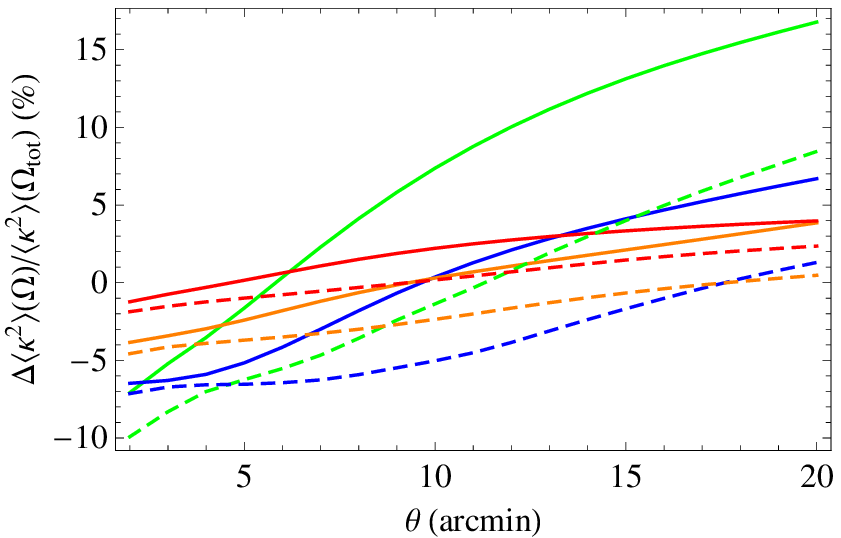}
\includegraphics[width=5.5cm]{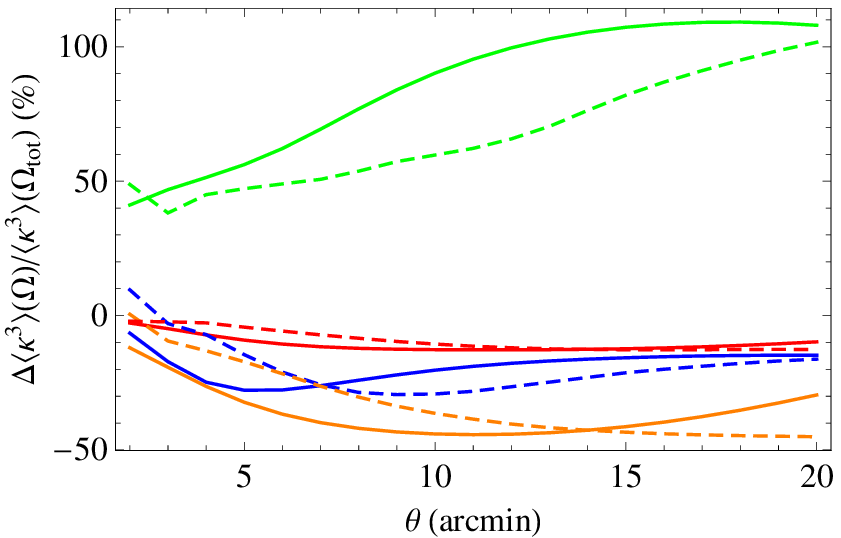}
\includegraphics[width=5.5cm]{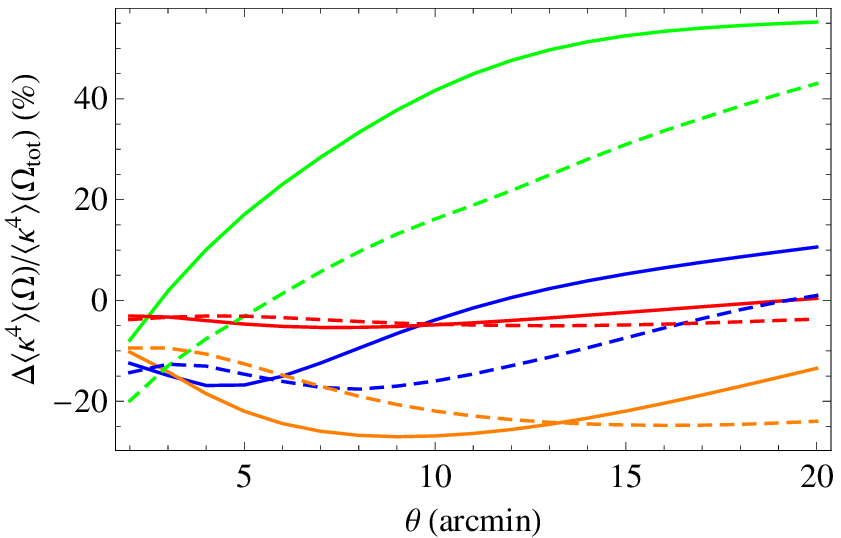} \\
\includegraphics[width=5.5cm]{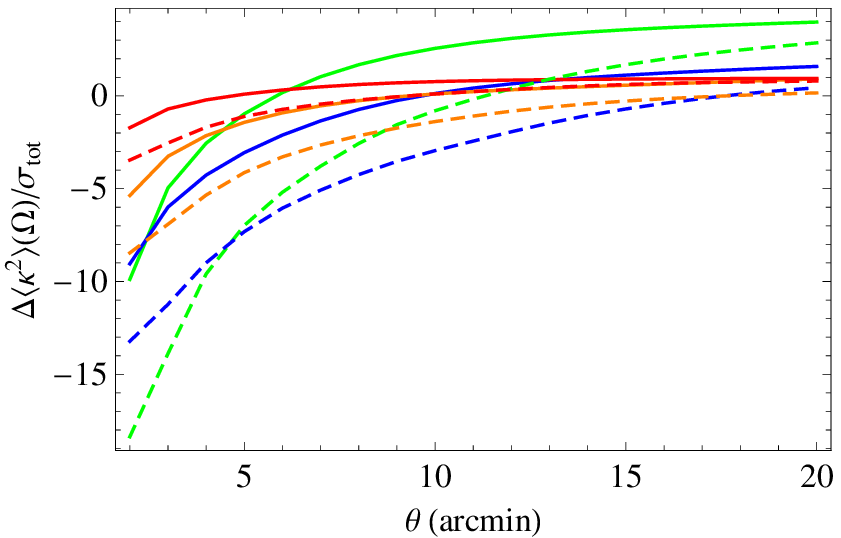}
\includegraphics[width=5.5cm]{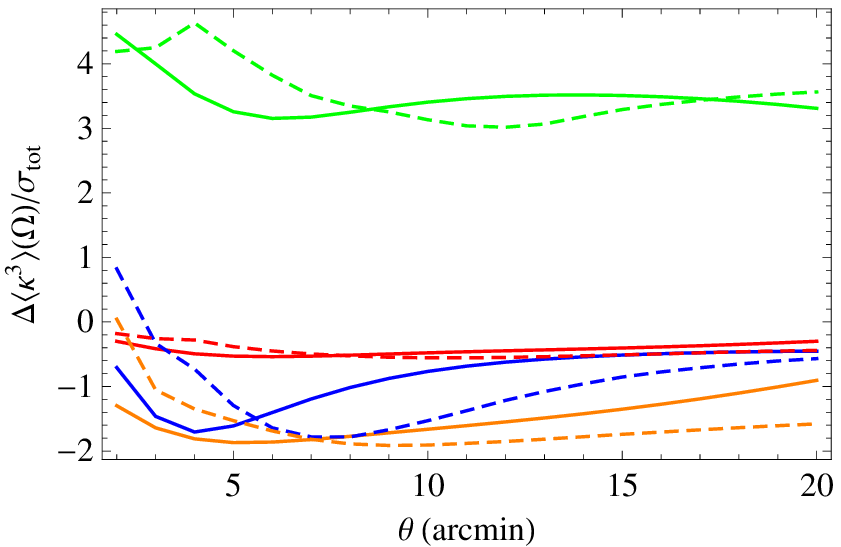}
\includegraphics[width=5.5cm]{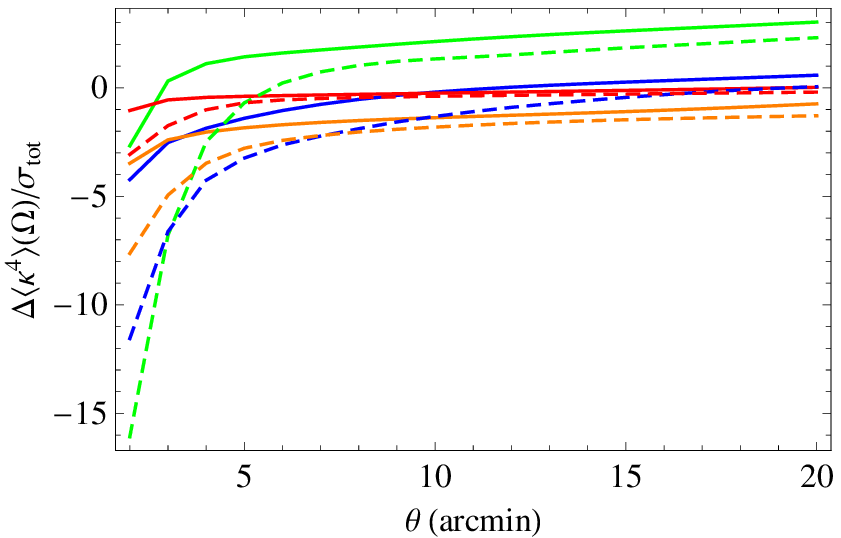} \\
\caption{Convergence of the HOM for different values of the survey area $\Omega$. {\it Top.} Percentage deviation $\Delta \langle \kappa^n \rangle(\Omega)/\langle \kappa^n \rangle(\Omega_{tot})$ as function of the smoothing angle $\theta$ for $\Omega = \{100, 500, 1500, 2500\} \ {\rm sq \ deg}$ (green, blue, orange, red lines) and $\Omega_{tot} = 3500 \ {\rm sq \ deg}$. {\it Bottom.} Same as before but with the difference normalized with respect to the statistical error. Solid (dashed) lines refer to the results for the Gaussian (top hat) smoothing filter.}
\label{fig: homvsarea}
\end{figure*}

\subsection{Convergence maps and moments estimate}

In order to build the convergence and shear maps we need for validation and calibration, we first cut $5 \times 5 \ {\rm sq deg}$ approximately square patches taking care that they are well separated so that they can be considered indepdendent. This reduces the usable area to $\Omega = 3500 \ {\rm sq \ deg}$ divided in 140 subfields. Each subfield has an area large enough to ensure good statistics, but small enough to make deviations from flat sky approximation (used in the theoretical derivation) negligible. The MICECATv2.0 catalog reports the redshift $z$ and the (RA, Dec) coordinates on the sky of each galaxy. We use the redshift $z$ to only select galaxies in the redshift range $(0.1, 1.4)$, and use a sinusoidal projection to convert the (RA, Dec) coordinates defined on the curved sky into the Cartesian ones $(x, y)$ defined on the tangential plane. Following \cite{Ludo2013}, we then arrange galaxies in approximately square pixels with side length $0.85 \ {\rm arcmin}$. Given that the number density of galaxies is about $n_g \simeq 27 \ {\rm gal/arcmin}^2$, this size ensures that there is large number of objects in each pixel without compromising the map resolution.

We construct two kind of maps. First, we use the convergence values reported in the MICECATv2.0 catalog without any further modification. This can be considered the actual convergence map which can be used to infer the actual moments the theory should match. We will therefore use it to validate the linear relation (\ref{eq: kappathlin}), and estimate the calibration parameters $(\mu_n, \gamma_n)$. Note that, strictly speaking, we do not need $(\mu_n, \gamma_n)$ since they are fully degenerate with $(m, c)$ so that only the combined parameters $(m_n, c_n)$ finally enter the calibration relations (\ref{eq: cal2nd})\,-\,(\ref{eq: cal4th}). We therefore use this ideal convergence map only as a way to show that the theory reproduces the high order convergence moments up to a linear transformation.

Actually, one does not observe the convergence itself, but rather the shear components\footnote{Actually, what is observed are the two components $(e_1, e_2)$ of the ellipticity, but with no noise and intrisic alignment there is no difference with the shear components.} $(\gamma_1, \gamma_2)$. These quantities are available in the MICECAT v2.0 catalog, but we perturb them adding random values of the intrinsic ellipticity on both the shear components to mimic the noise due to intrinsic ellipiticity. We then smooth the map and apply the KS93 method \cite{KS93} to get what we can consider the observed convergence map $\kappa_{obs}$. Moments estimated from this map are those which enter the left hand side of Eqs.(\ref{eq: cal2nd})\,-\,(\ref{eq: cal4th}). Note that the smoothing procedure, which is typically used to reduce noise, critically depends on which filter is adopted. To investigate the impact of this choice on the results, we consider two popular cases, namely a Gaussian and a top hat filter. In the Fourier space, their window functions respectively read

\begin{equation}
\tilde{W}(\ell, \theta) = \left \{
\begin{array}{l}
\displaystyle{\exp{(- \ell^2 \theta^2/2)}} \\
 \\ 
\displaystyle{2 J_1(\ell \theta)/(\ell \theta)} \\
\end{array}
\right . 
\label{eq: filters}
\end{equation}
where $J_{\nu}$ is the Bessel function of order $\nu$ and $\theta$ the filter aperture which we vary over the range (2, 20) arcmin.

The catalogs thus obtained are used as input for the estimate of high order convergence moments which are simply given by 

\begin{displaymath}
\langle \kappa^n \rangle = {\cal{N}}_{pix}^{-1} \sum{\tilde{\kappa}_i^n} \ \ (n = 2, 3, 4)
\end{displaymath}
where the sum is over the ${\cal{N}}_{pix}$ pixels in the map and $\tilde{\kappa}_i$ is the value of the convergence in the $i$\,-\,th pixel after smoothing the map with a given filter. Note that we first subtract a constant offset in order to have $\langle \kappa \rangle = 0$ to be consistent with our derivation of the calibration relations. 

This procedure is repeated for each of the 140 subfields thus getting a list of $\langle \kappa^n \rangle$ values. We finally estimate the moments and the covariance matrix as

\begin{equation}
\langle \kappa^n \rangle_{obs} =\frac{\sum{\langle \kappa^n \rangle(k)}}{{\cal{N}}_{f}} \ \ (n = 2, 3, 4) 
\label{eq: kappaobsend}
\end{equation}
where $\langle \kappa^n \rangle(k)$ is the $n$\,-\,th order moment estimated from the $k$\,-\,th subfield, and the sum is over the ${\cal{N}}_f$ subfields. For later applications, we remind the reader that it is ${\cal{N}}_f = \Omega/25$ with $\Omega$ the used survey area ($\Omega = 3500 \ {\rm sq \ deg}$ giving ${\cal{N}}_f = 140$ for our reference MICECAT based case.

In practical applications, one will fit moments as function of the smoothing scale $\theta$ so that the data vector is
\begin{eqnarray}
{\bf D}_{obs} & = & \left \{ \langle \kappa^2 \rangle_{obs}(\theta_1), \langle \kappa^2 \rangle_{obs}(\theta_2), \ldots, 
\langle \kappa^2 \rangle_{obs}(\theta_N), \right . \nonumber \\
 &  & \ \ 
\langle \kappa^3 \rangle_{obs}(\theta_1), \langle \kappa^3 \rangle_{obs}(\theta_2), \ldots, 
\langle \kappa^3 \rangle_{obs}(\theta_N), \nonumber \\
 &  & \ \ \left .
\langle \kappa^4 \rangle_{obs}(\theta_1), \langle \kappa^4 \rangle_{obs}(\theta_2), \ldots, 
\langle \kappa^4 \rangle_{obs}(\theta_N) \right \}
\label{eq: dobsvec}
\end{eqnarray}
where we will explore different choices for both the smoothing angle range $(\theta_{min}, \theta_{max})$ and the sampling $d\theta$ taking equispaced values giving $N = (\theta_{max} - \theta_{min})/d\theta$. 

In order to estimate the data covariance matrix, one should run different simulations of the same survey which is actually not possible. However, mimicking what is done with actual surveys (see, e.g., \cite{Ludo2013}) and in previous works based on simulated maps \cite{Petri}, we can use the ${\cal{N}}_f$ subfields and estimate the covariance matrix elements as

\begin{equation}
Cov_{ij}^{obs} = 
\frac{1}{{\cal{N}}_f} \frac{\sum{\left [ D_{obs}^{i, k} - D_{obs}^{i} \right ] \left [ D_{obs}^{j, k} - D_{obs}^{j} \right ]}}
{{\cal{N}}_f - 1}
\label{eq: dobscov}
\end{equation}
where $D_{obs}^{i, k}$ is the $i$\,-\,th component of the data vector ${\bf D}_{obs}$ estimated on the convergence map of the $k$\,-\,th subfield, $D_{obs}^{i}$ the value of the same component from the final vector (\ref{eq: dobsvec}), and the sum is over the ${\cal{N}}_f$ subfields. it is worth spending some more words on Eq.(\ref{eq: dobscov}). This is based on the implicit underlying assumption that all the subfields are statistically faithful realization of the same underlying properties. As such, the covariance matrix for each subfield can be evaluated as given by the second multiplicative term in Eq.(\ref{eq: dobscov}). This will give us the covariance for a field of area equal to the subfield one $\Omega_f$ so that a further $\Omega_f/\Omega = 1/{\cal{N}}_f$ factor is needed to scale the covariance to the full survey area $\Omega$.

\begin{figure*}
\centering
\includegraphics[width=5.5cm]{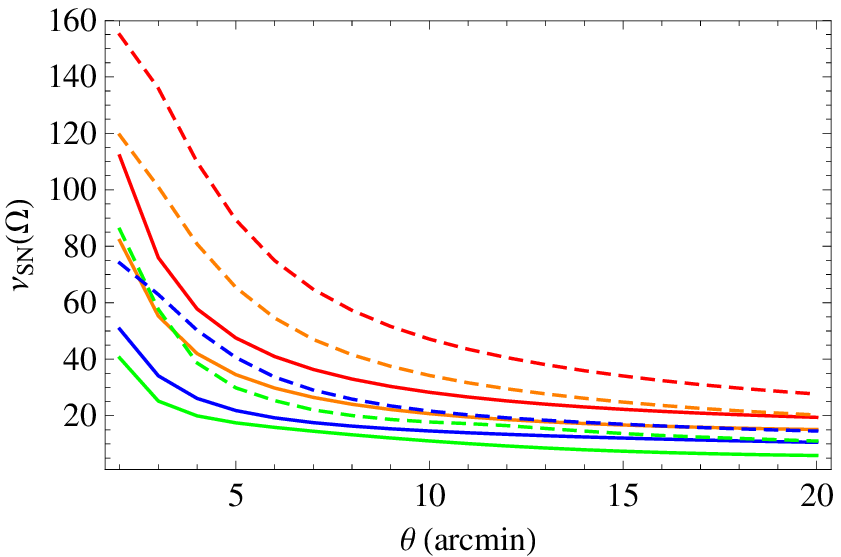}
\includegraphics[width=5.5cm]{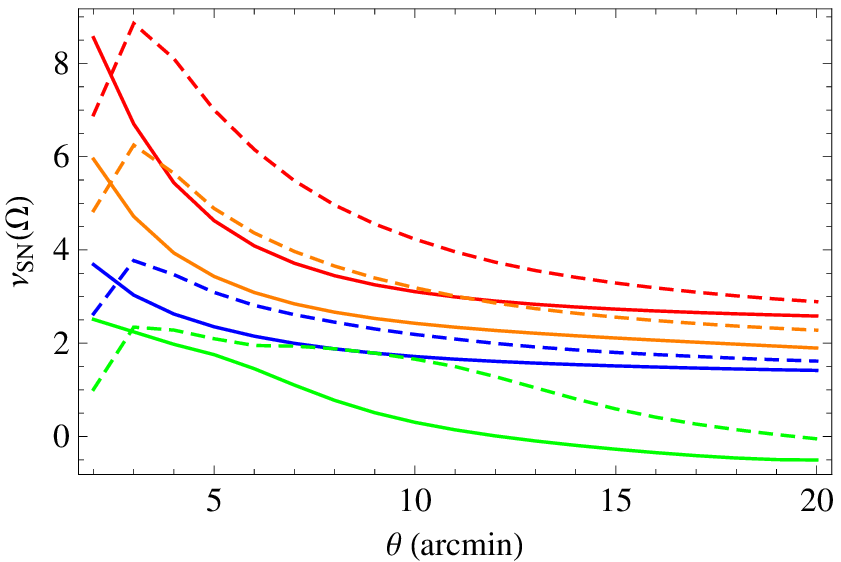}
\includegraphics[width=5.5cm]{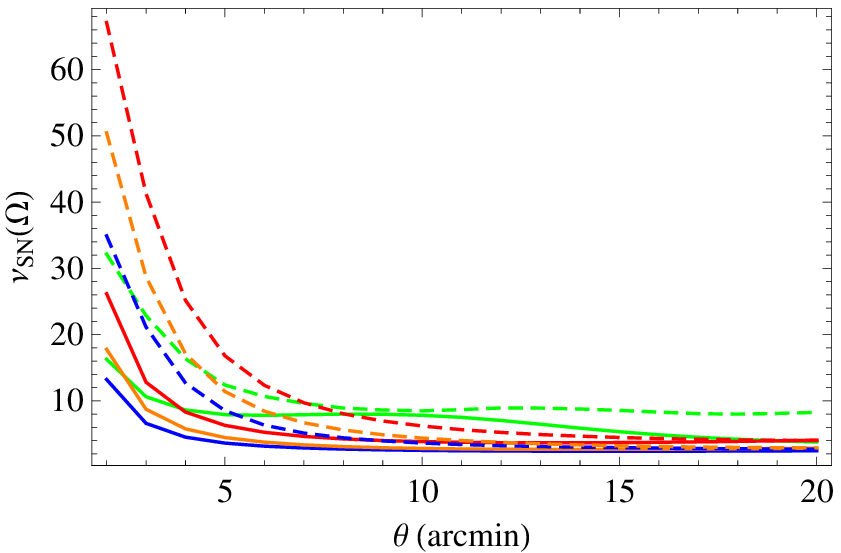} \\
\includegraphics[width=5.5cm]{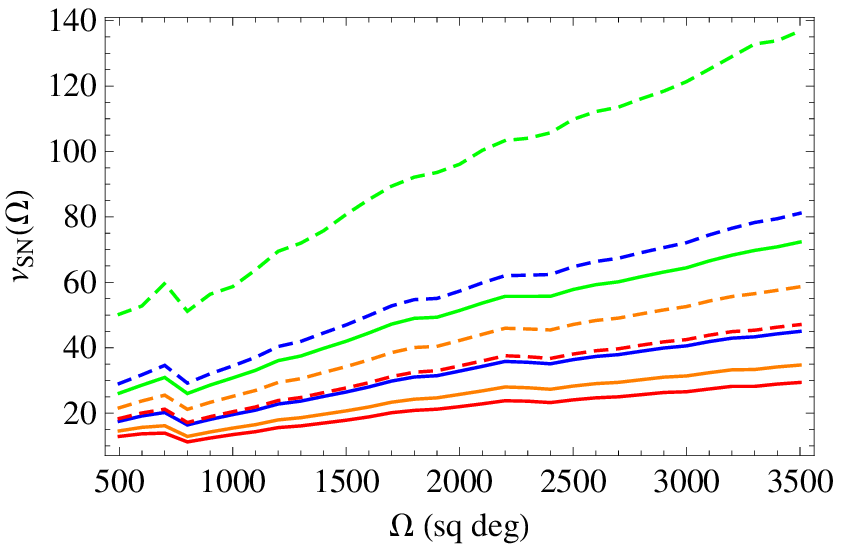}
\includegraphics[width=5.5cm]{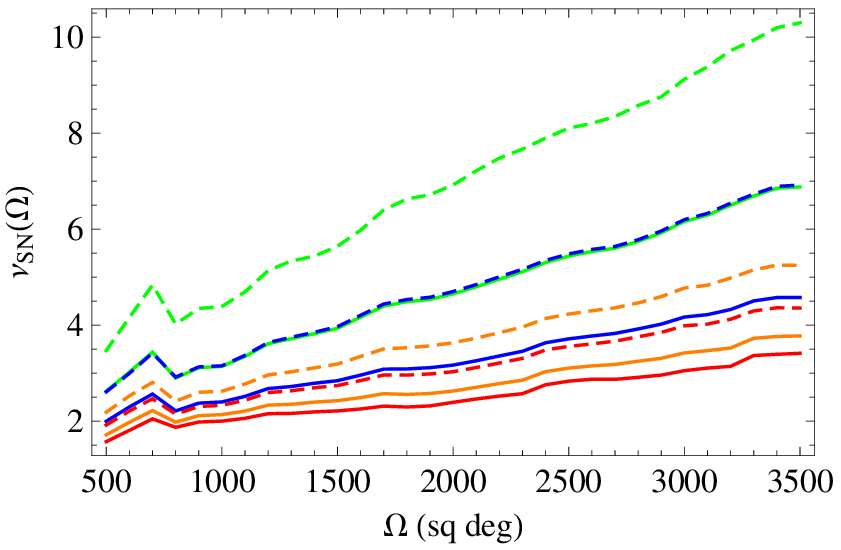}
\includegraphics[width=5.5cm]{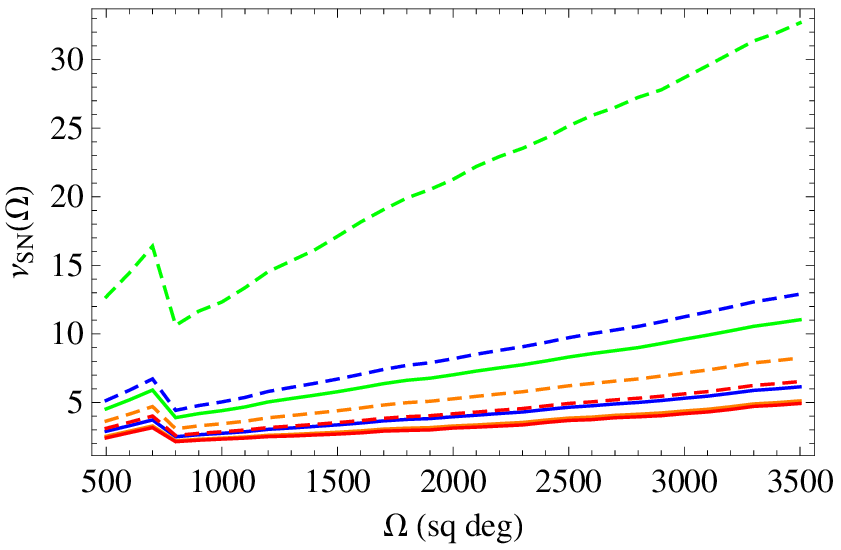} \\
\includegraphics[width=5.5cm]{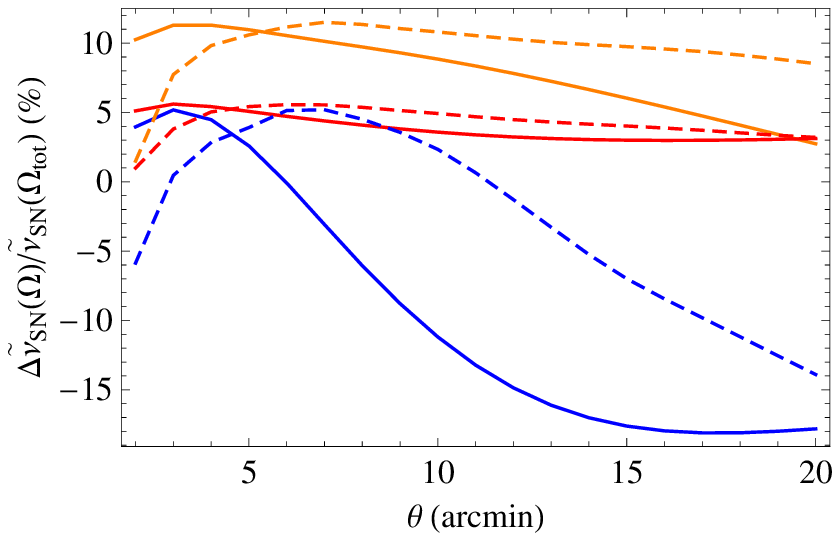}
\includegraphics[width=5.5cm]{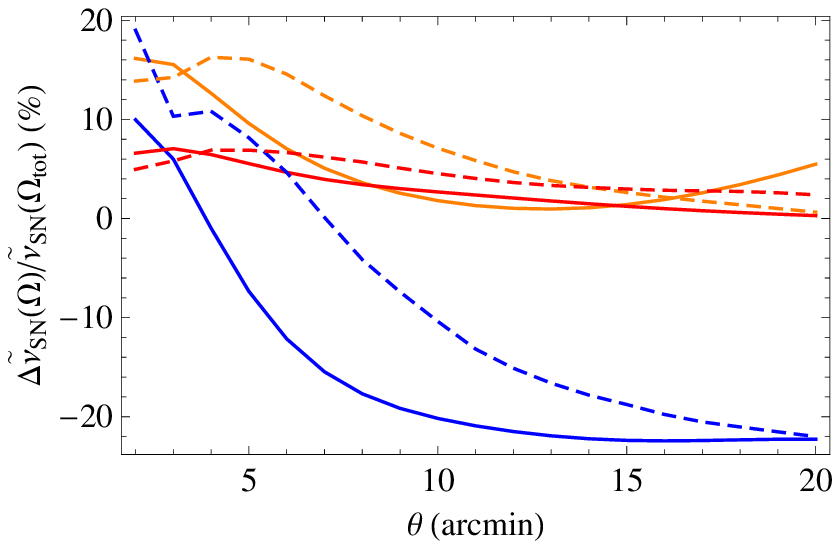}
\includegraphics[width=5.5cm]{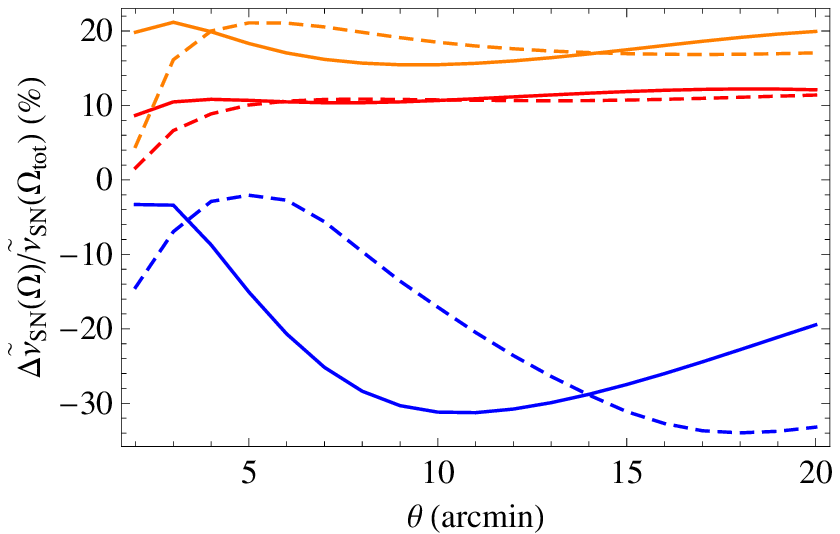} \\
\caption{Scaling of the HOM $\langle \kappa^2 \rangle$ (left), $\langle \kappa^3 \rangle$ (centre), $\langle \kappa^4 \rangle$ (right) signal\,-\,to\,-\,noise ratio $\nu_{SN}$ with smoothing angle $\theta$ and survey area $\Omega$. {\it Top.} S/N ratio as function of $\theta$ for $\Omega = \{100, 500, 1500, 2500\} \ {\rm sq \ deg}$ (green, blue, orange, red lines). {\it Centre.}  S/N ratio as function of $\Omega$ for $\theta = \{4, 7, 10, 13\} \ {\rm arcmin}$ (green, blue, orange, red lines). {\it Bottom.}  Percentage deviation of the rescaled S/N ratio (see text) from the value estimated from the full survey area. In all panels, solid (dashed) lines refer to result for Gaussian (top hat) smoothing filter.}
\label{fig: stonvsarea}
\end{figure*}

\begin{figure*}
\centering
\includegraphics[width=7.5cm]{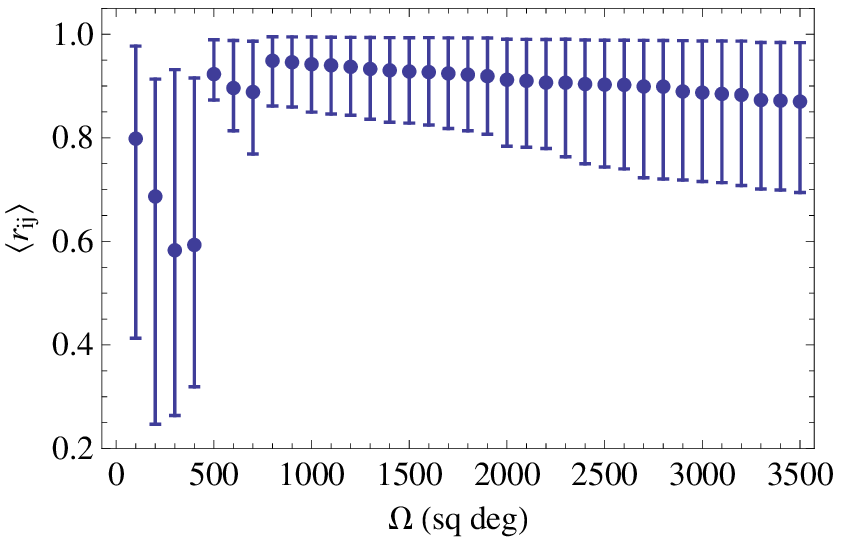}
\includegraphics[width=7.5cm]{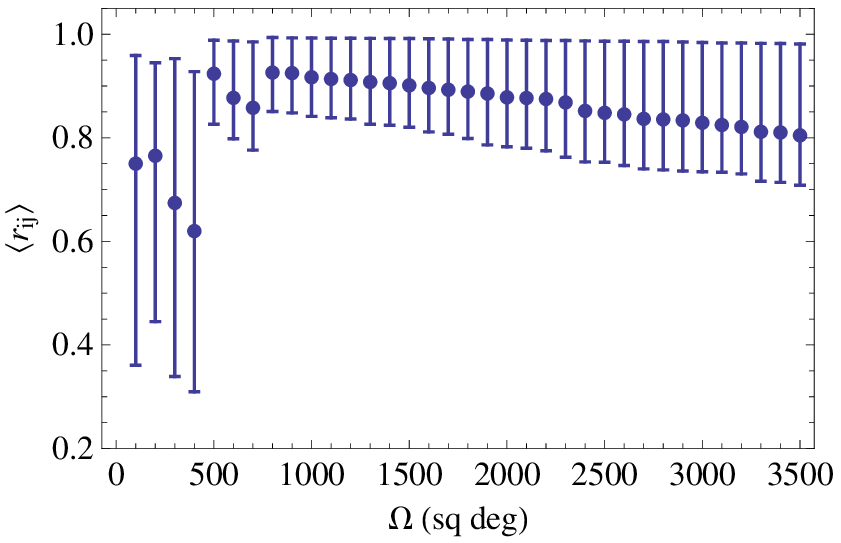} \\
\includegraphics[width=7.5cm]{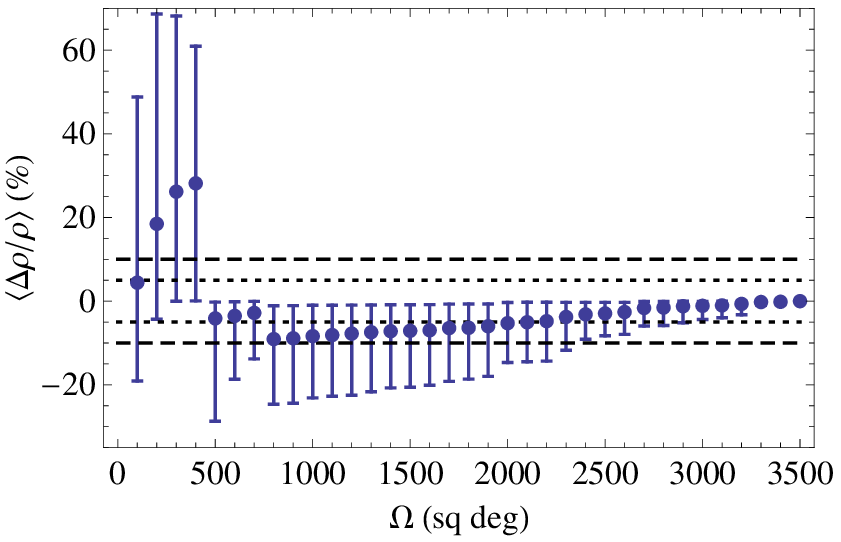}
\includegraphics[width=7.5cm]{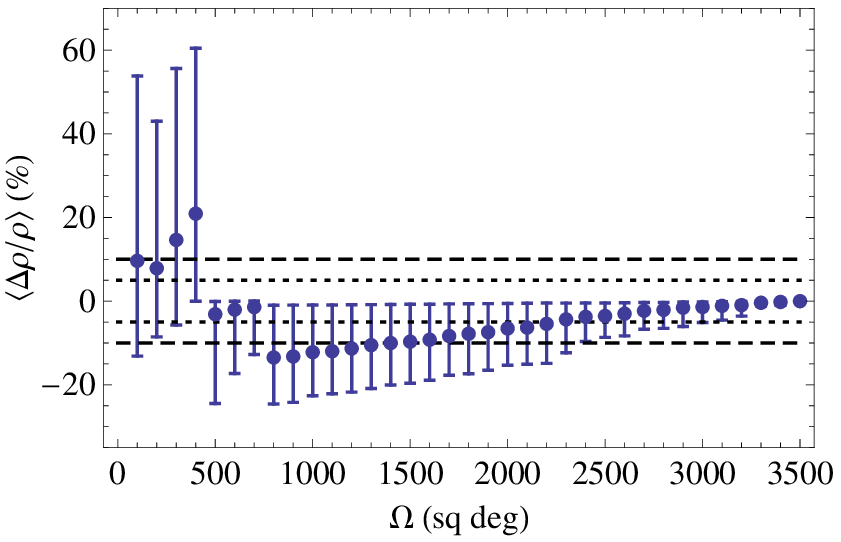} \\
\caption{Testing the convergence of the correlation matrix ${\bf R}$. {\it Top.} Average value and $68\%$ confidence range of the independent element of ${\bf R}$ for the Gaussian (left) and top hat (right) filters. {\it Bottom.} Average value and $68\%$ confidence range of the percentage deviation of the ${\bf R}$ elements from the value they get for $\Omega = 3500 \ {\rm sq \ deg}$. Dashed (dotted) lines mark the $\pm 10 (\pm 5)\%$ limits.}
\label{fig: rijarea}
\end{figure*}

\subsection{Convergence of moments}

Higher order moments (hereafter HOM) probe the shape of the convergence probability distribution function (pdf). One can qualitatively expect that the higher the order, the better the pdf is characterized, but the larger must the statistics be to catch the details of its shape. This translates in a requirement on the survey area, i.e., one could wonder how large must the total area be so that HOM are reliably recovered from the convergence maps. To this end, we estimate HOM as a function of the survey area $\Omega$ which, in our case, actually means estimating moments using only a number of subfields ${\cal{N}}_f \le 140$. Should the moments be reliably estimated, the functions\footnote{Hereafter, unless confusion is possible, we drop the label $obs$ to shorten the notation. Moreover, we will set $(\theta_{min}, \theta_{max}, d\theta) = (2, 20, 1) \ {\rm arcmin}$. Unless otherwise stated, all the results are qualitatively valid also for other choices.} $\langle \kappa^n \rangle(\theta)$  must be independent of the survey area $\Omega$ as far as $\Omega \ge \Omega_{min}$ with $\Omega_{min}$ the final requirement on the survey area. How to set $\Omega_{min}$ actually depends on what one is interested in. To understand this point, let us look at Fig.\,\ref{fig: homvsarea}. Top panels show that the deviation of HOM from their fiducial value (i.e., the one estimated from the full area) can be quite significant with the difference being particularly dramatic for the third order moment (which probes the asymmetry of the convergence pdf). If we demand that $\Delta \langle \kappa^n \rangle/\langle \kappa^n \rangle$ is less than $10\%$ for all orders, we get $\Omega_{min} \sim 2500 \ {\rm sq \ deg}$. However, this is a quite conservative limit which does not take into account of the statistical errors. Indeed, one can relax the constraint asking that the systematic error on the HOM estimate are smaller than the statistical uncertainties $\sigma$. Bottom panels show that such a requirement is easier to fullfil depending on both the survey area (with smaller areas leading to larger $\sigma$ hence less demanding constraints), and the smoothing scale $\theta$. For instance, if a Gaussian filter is used to smooth the convergence maps, $\Omega_{min} = 500 \ {\rm sq \ deg}$ are enough to measure moments with a systematic bias smaller than the $2 \sigma$ error as soon as $\theta > 10 \ {\rm arcmin}$. This area increases to $\Omega_{min} = 1500 \ {\rm sq \ deg}$ if one asks for all moments to be measured with a bias smaller than $2 \sigma$ for smoothing scales $\theta > 5 \ {\rm arcmin}$. As a general result, we also find that the requirements on the minimum area are stronger for a top hat filter which turns out to be less efficient in cancelling features introduced by the reconstruction from noisy data.

When fitting a model to the data, one should be confident that not only the HOM central values, but also the errors have been estimated in order to avoid unrealistically tight constraints on model parameters because of underestimate of the uncertainties. We therefore investigate how the signal\,-\,to\,-\,noise ratio $\nu_{SN}$ scales with the survey area using the square root of the diagonal elements\footnote{Actually, one should take into account also the non diagonal elements and define the noise taking care of the correlations between moments of the same order evaluated at different smoothing scales or different orders at the same smoothing. We prefer to rely on diagonal elements only to mimic the case when one is using a single order only. For instance, one could be interested in using only $\langle \kappa^2 \rangle(\theta_0)$ with $\theta_0$ a particular smoothing scale. In this case, the S/N ratio would be defined using only the error on this quantity so that one should check how this scale with the survey area which is what we are investigating here.} of the data covariance matrix to define $\nu_{SN}$. Top and central panels in Fig.\,\ref{fig: stonvsarea} show the expected behaviour of $\nu_{SN}(\theta)$ with both the smoothing scale and the survey area. On the one hand, the smaller the smoothing scales, the larger are the moments as one can naively understand considering the case of a Gaussian distribution. On the other hand, Eq.(\ref{eq: dobscov}) suggests a scaling of the noise with $\sqrt{1/\Omega}$ so that we expect $\nu_{SN}(\theta) \propto \Omega^{\alpha}$ with $\alpha > 0$ and deviating from the naive $\alpha = 1/2$ value because also the signal depends on the area as inferred from Fig.\,\ref{fig: homvsarea}. It is worth wondering which is the minimal area to make the S/N ratio estimate reliable. Note that for this test we have first to rescale the S/N for the different area that is why we consider $\tilde{\nu}_{SN}(\theta) = \sqrt{\Omega} \nu_{SN}(\theta)$, and look for the minimal area needed to make this quantity deviate from the final value less than $20\%$. Bottom panels in Fig.\,\ref{fig: stonvsarea} shows a conservative cut $\Omega > 1500 \ {\rm sq \ deg}$ ensures reliable estimates of both HOM and their uncertainties.

It is worth noting that our definition of S/N, although quite intuitive, is somewhat optimistic since it neglects the strong correlation among the components of the data vector. We should therefore be confident that the correlation matrix ${\bf R}$ too is stable. Being its entries defined as 

\begin{displaymath}
r_{ij} = \frac{Cov_{ij}^{obs}}{\sqrt{Cov_{ii}^{obs} Cov_{jj}^{obs}}} \ ,
\end{displaymath}
there are $3N (3N + 1)/2$ independent elements so that it is not immediate to understand whether the area is large enough to consider reliable the estimate of ${\bf R}$. As a qualitative test, we therefore look at $\langle r_{ij} \rangle$, i.e., the average value of its off diagonal elements which is plotted as a function of the survey area $\Omega$ in top panels of Fig.\,\ref{fig: rijarea} where the error bars denote the $68\%$ confidence range. Taken at face value, these results show that there is no convergence at all with $\langle r_{ij} \rangle$ slowly but constantly decreasing with $\Omega$. However, the $68\%$ confidence ranges well overlap for $\Omega \ge 1000 \ {\rm sq \ deg}$ suggesting that the variation can indeed be neglected. On the other hand, bottom panels show that $\langle \Delta \rho/\rho \rangle = \langle 100 \times [1 - r_{ij}(\Omega)/r_{ij}(\Omega_{tot})] \rangle$ becomes smaller than $10\%$ as far as $\Omega \ge 1500 \ {\rm sq \ deg}$. Considering that this is only a qualitative test, we can conclude that an area as large as $\Omega \sim 1500 \ {\rm sq \ deg}$ allows a reliable estimate of the HOM, the on diagonal elements of the covariance matrix, and the full data correlation matrix.

As a final remark, we remind the reader that all the results in this paragraph have been obtained relying on the MICECAT catalog which has a definite redshift range and a galaxy number density $n_g \simeq 27 \ {\rm gal/arcmin^2}$. Should one or both of these characteristics be different, one might redo the analysis which is, however, outside our aim here.

\begin{figure*}
\centering
\includegraphics[width=5.5cm]{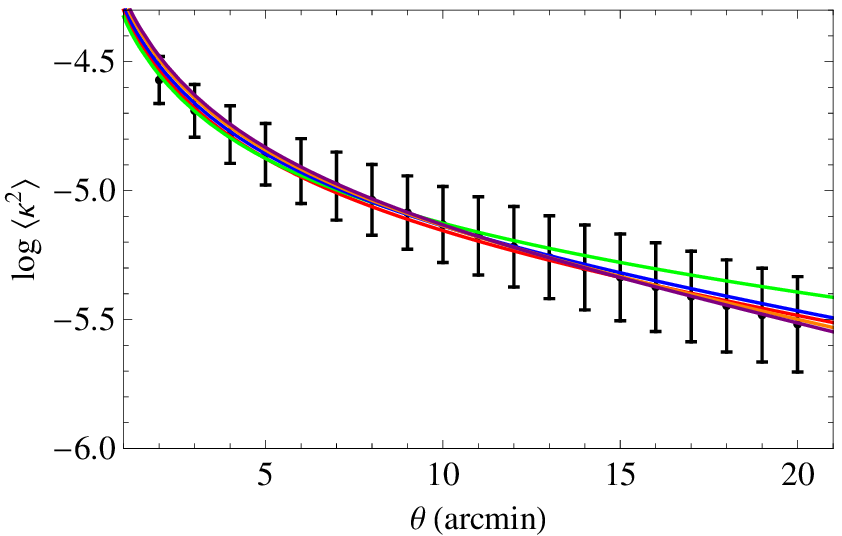}
\includegraphics[width=5.5cm]{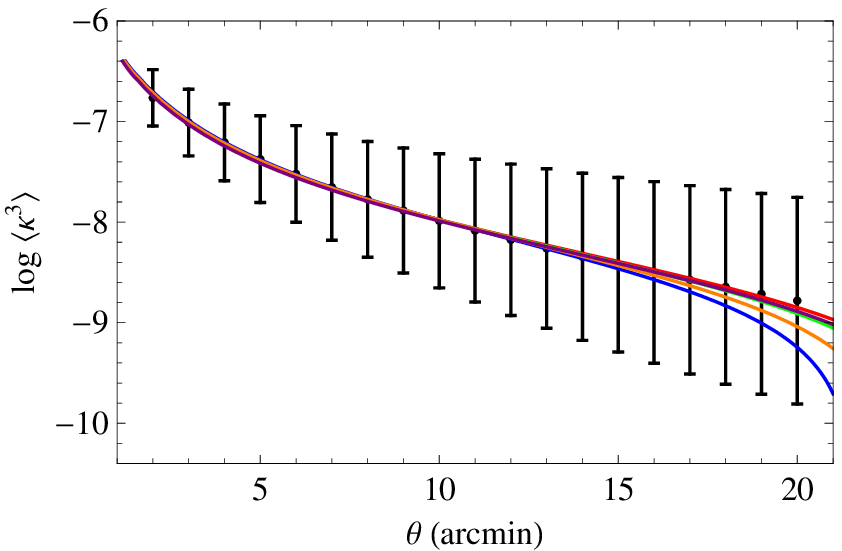}
\includegraphics[width=5.5cm]{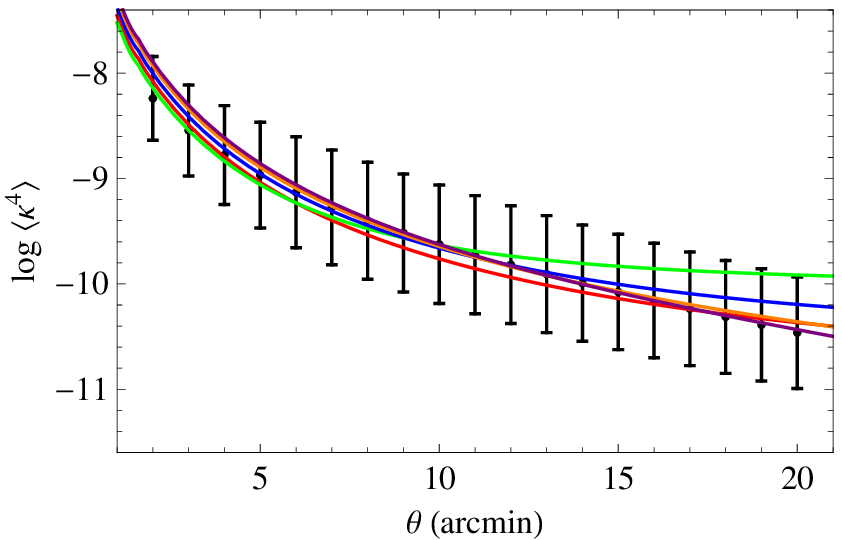} \\
\includegraphics[width=5.5cm]{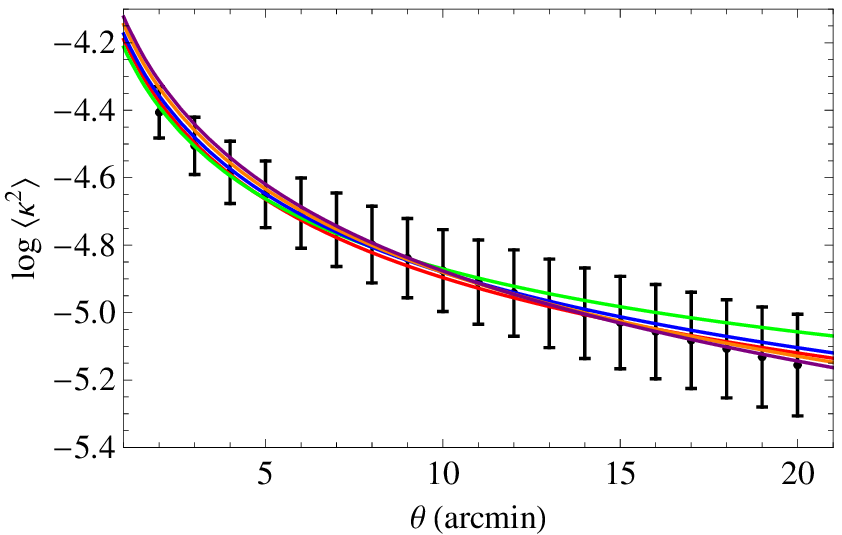}
\includegraphics[width=5.5cm]{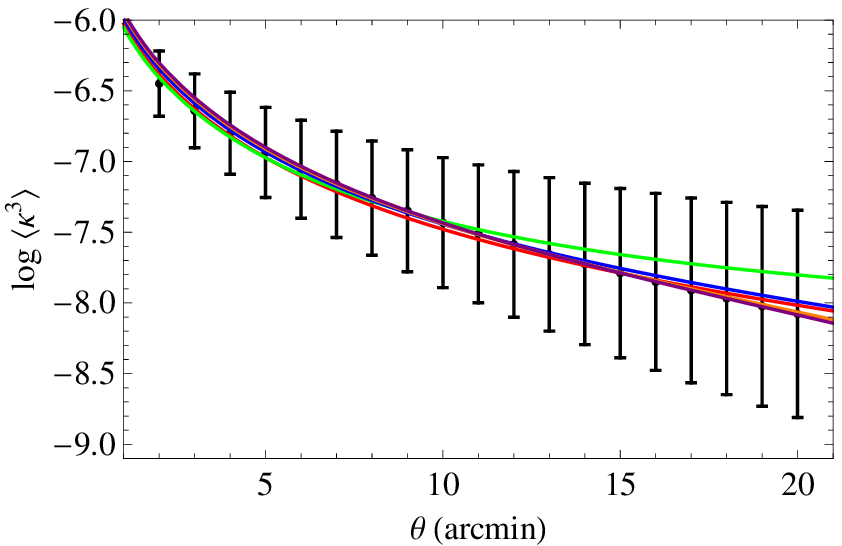}
\includegraphics[width=5.5cm]{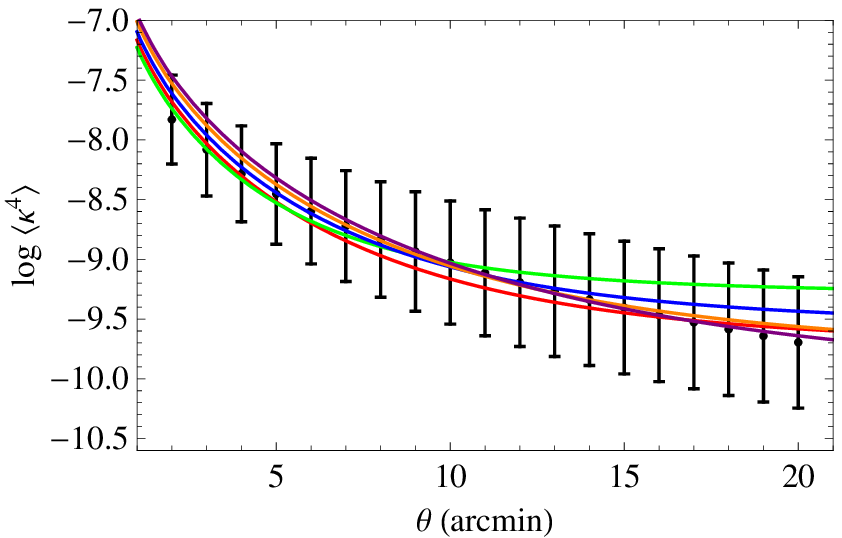} \\
\caption{HOM as estimated on the noiseless convergence map for Gaussian (top) and top hat (bottom) filters. Best fit curves are superimposed to the data with red, green, blue, orange, purple lines referring to the best fit paramters to the smoothing angle range $(2, 20)$, $(2, 12)$, $(4, 14)$, $(6, 16)$, $(8, 18) \ {\rm arcmin}$, respectively.}
\label{fig: calth}
\end{figure*}

\section{Validating the HOM calibration}

In order to make moments a viable tool to constrain cosmological parameters, two preliminary steps are mandatory. First, one has to show that HOM can be reliably estimated from convergence maps reconstructed from noisy shear data. Second, one has to find out a way to match theoretical predictions with observations through a phenomenological recipe taking care of what can not be described within the theory. While the first step has been successfully completed in the above section, the second one is encapsulated in the calibration relations (\ref{eq: cal2nd})\,-\,(\ref{eq: cal4th}). Our aim here is to both validate their derivation and demonstrate that they can indeed fit the measured moments.

\subsection{Idealized convergence moments}

To this end, let us first consider the HOM as estimated from the convergence map provided by the MICECAT catalog itself. Since there is no noise and no reconstruction from shear is performed, HOM measured on this map may be considered the true ones which theory has to match. As explained in Sect.\,III, our theoretical predictions are given by Eq.(\ref{eq: kappathlin}) with $\langle \kappa^n \rangle_{th}$ computed as in Eqs.(\ref{eq: kappa2nd})\,-\,(\ref{eq: kappa4th}). Validating these relations therefore means finding the parameters $(\mu_n, \gamma_n)$ from a fit of the $\langle \kappa^n \rangle(\theta)$ data. The results are shown in Fig.\,\ref{fig: calth} where we plot the best fit curves\footnote{Best fit values and $68\%$ confidence ranges from all the fits described in this section are given in Appendix A.} superimposed to the estimated HOM\footnote{Note that these plots and the following ones in Figs.\,5 and 6 refer to the data as estimated from a single subfield so that the error bars are larger by a factor ${\cal{N}}_{f}^{1/2} \simeq 12$. We make this choice in order to show that the fit to the full dataset reproduces the data of each subfield too so that one can rely on the found calibration parameters even if only a smaller area is available.}. We As it is apparent, the fit is quite good with the linear model passing through the points well within the statistical uncertainty. Moreover, the rms of the percentage residuals, defined as $\rho_{rms}(n) = 100  \times (1 - \langle \kappa^n \rangle_{fit}/\langle \kappa^n \rangle)$, turns out to be 

\begin{displaymath}
\rho_{rms}(n) = \left \{
\begin{array}{l}
(4.88, 9.34, 21.5) \% \\
 \\
(4.55, 9.22, 21.2) \% \\
\end{array}
\right . 
\end{displaymath}
with the three values referring to moments of order $n = 2, 3, 4$, and the first (second) row for a Gaussian (top hat) filter. These small numbers definitely validate our assumption that the theoretical approach outlined in Sect.\,III is indeed able to reproduce the N\,-\,body derived HOM thus providing a first validation of Eqs.(\ref{eq: cal2nd})\,-\,(\ref{eq: cal4th}).

\begin{figure*}
\centering
\includegraphics[width=5.5cm]{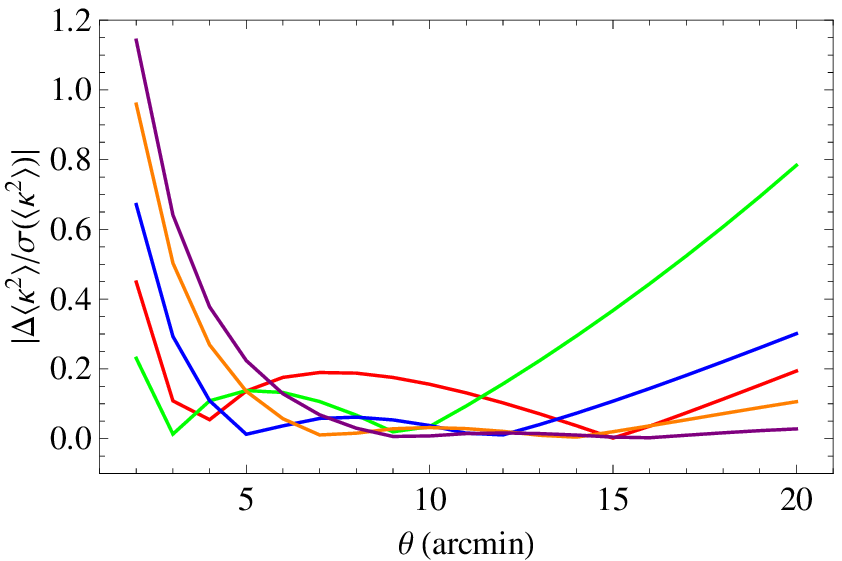}
\includegraphics[width=5.5cm]{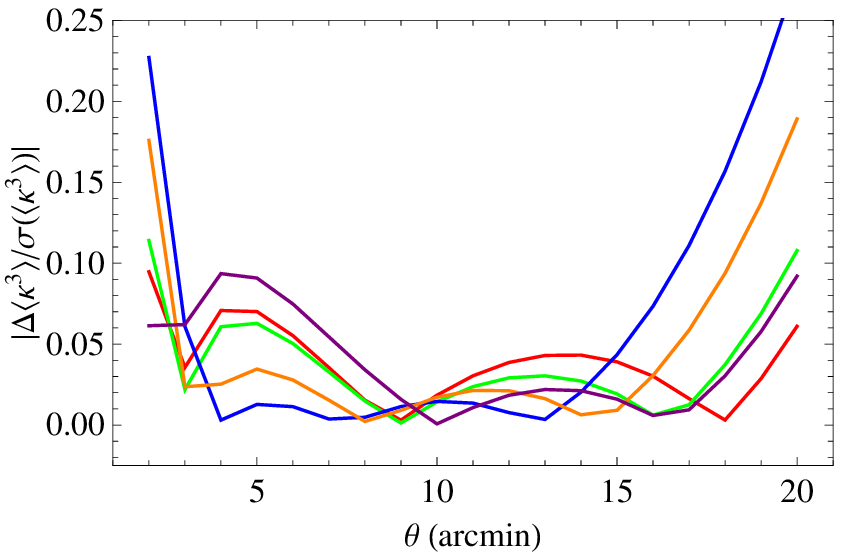}
\includegraphics[width=5.5cm]{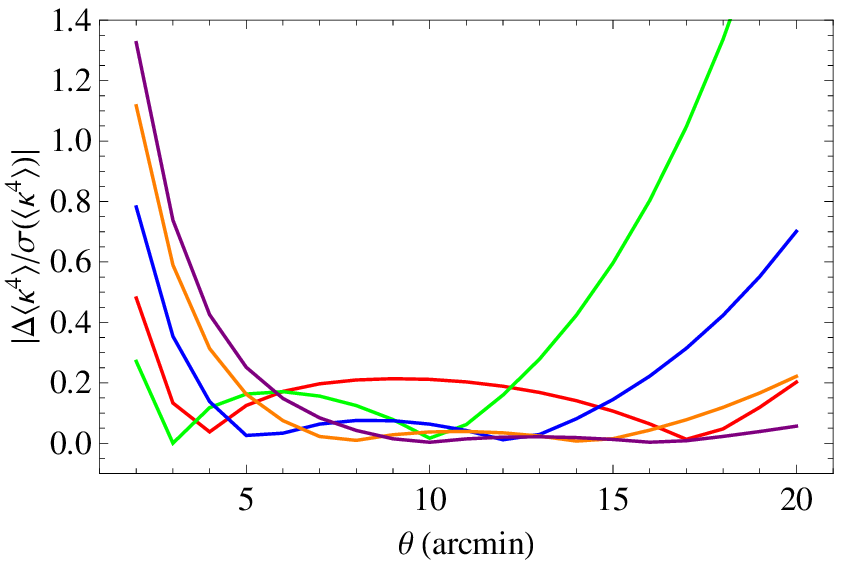} \\
\includegraphics[width=5.5cm]{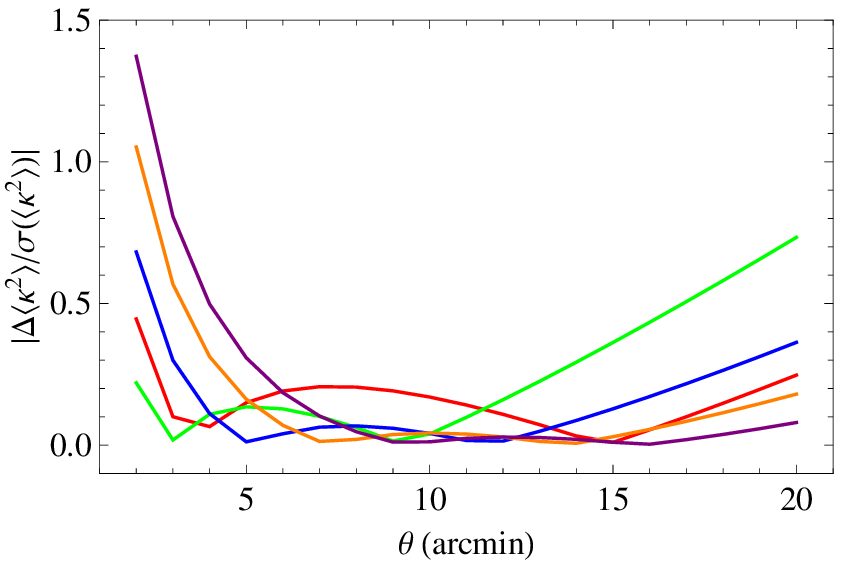}
\includegraphics[width=5.5cm]{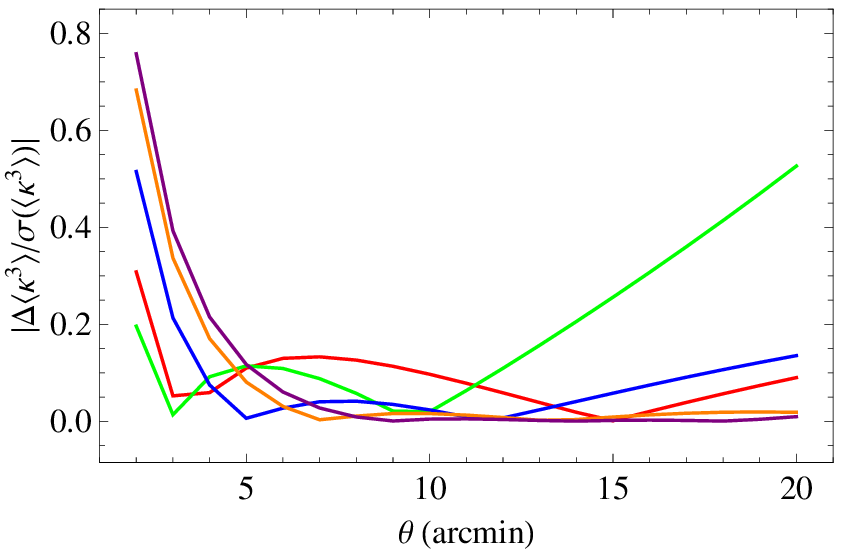}
\includegraphics[width=5.5cm]{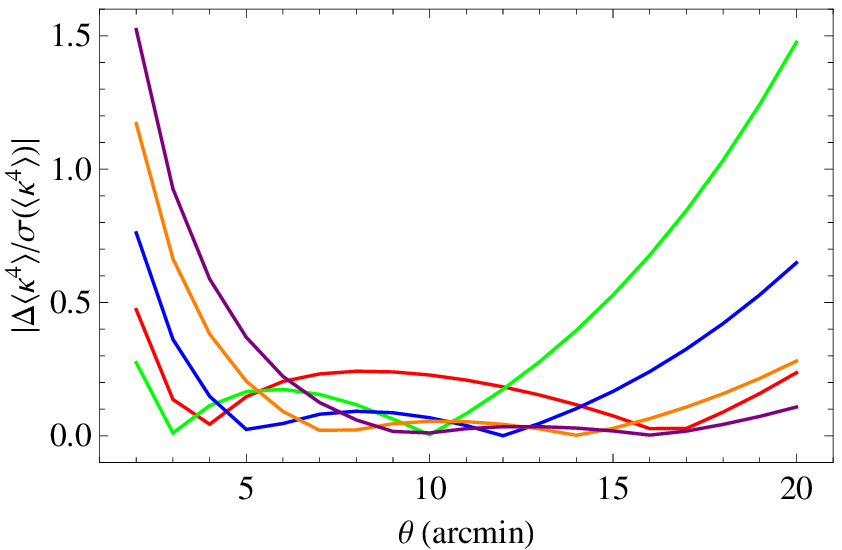} \\
\caption{Same as Fig.\,\ref{fig: calth} but now we show the error normalized fit residuals (in absolute value). }
\label{fig: calthres}
\end{figure*}

Although the exact values of $(\mu_n, \gamma_n)$ are not of any use in the following, it is nevertheless instructive looking at the best fit values. For the Gaussian filter, we get 

\begin{displaymath}
\mu_2 = -0.002 \ \ , \ \ \gamma_2 = -1.45 \times 10^{-6}  \ \ , 
\end{displaymath}
\begin{displaymath}
\mu_3 = 0.460 \ \ , \ \ \gamma_3 = -2.82 \times 10^{-9}  \ \ ,
\end{displaymath}
\begin{displaymath}
\mu_4 = -0.191 \ \ , \ \ \gamma_4 = 1.50 \times 10^{-11}  \ \ .
\end{displaymath}
As expected, the multiplicative bias for the 2nd order moment is negligibly small, while the larger values of $(\mu_3, \mu_4)$ corrects for deviations of the $({\cal{Q}}_3, {\cal{Q}}_4)$ constant factors from the assumed fiducial values. On the other hand, the additive bias $(\gamma_2, \gamma_3, \gamma_4)$ are smaller than the lowest data values confirming our expectation that cutting the integration range has not a dramatic impact on the estimate of theoretical moments. Similar considerations also apply to the results for the top hat filter which read

\begin{displaymath}
\mu_2 = -0.161 \ \ , \ \ \gamma_2 = 0.74 \times 10^{-6}  \ \ , 
\end{displaymath}
\begin{displaymath}
\mu_3 = 0.298 \ \ , \ \ \gamma_3 = -1.45 \times 10^{-9}  \ \ ,
\end{displaymath}
\begin{displaymath}
\mu_4 = -0.483 \ \ , \ \ \gamma_4 = 1.72 \times 10^{-10}  \ \ .
\end{displaymath}
The only remarkable difference is the multiplicative bias for the 2nd order moment which turns out to be definetely larger than for the Gaussian case suggesting that a top hat filter is less efficient in smoothing nonlinearities not correctly taken into account in the theoretical derivation. 

Although the percentage residuals are already acceptably small, one can nevertheless wonder whether it is possible to further reduce them cutting the smoothing angle range. Such an approach would be motivated by the consideration that the theory is inevitably based on assumptions which can break down for small $\theta$, i.e., when small scales features in the convergence map are not smoothed out. We therefore repeat the fitting procedure limiting the smoothing angle range setting

\begin{displaymath}
(\theta_{min}, \theta_{max}) = (2, 12), (4, 14), (6, 16), (8, 18) \ {\rm arcmin} \ ,
\end{displaymath}
thus getting the green, blue, orange, purple curves in Fig.\,\ref{fig: calth}. To better illustrate the results, in Fig.\,\ref{fig: calthres}, we plot the error normalized best fit residuals (in absolute value) showing that the fit definetely improves, i.e., the green, blue, orange, purple curves stay below the red one over the corresponding fitting range. Moreover, the rms of the percentage residuals decreases up to an order of magnitude with the increase of $\theta_{min}$. This result can be easily explained noting that, the larger is the smoothing angle, the less nonlinearities contribute to the moments. Since these are the main source of uncertainties in the theoretical modeling, it is not surprising that decreasing their impact makes the linear approximation more and more better thus reducing $\rho_{rms}(n)$. 

\begin{figure*}
\centering
\includegraphics[width=5.5cm]{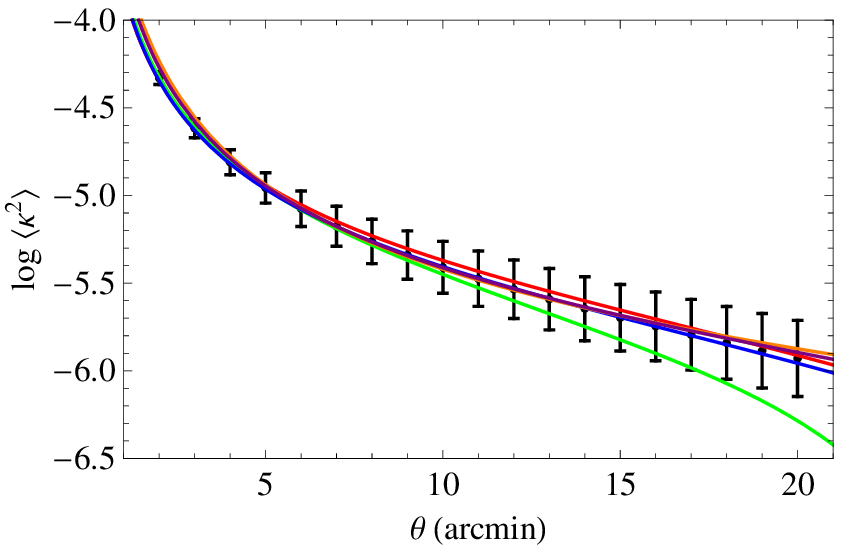}
\includegraphics[width=5.5cm]{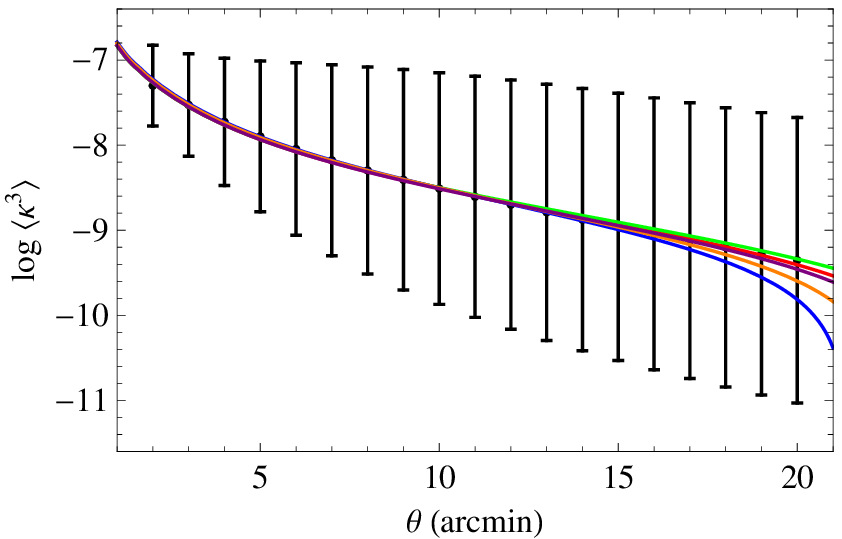}
\includegraphics[width=5.5cm]{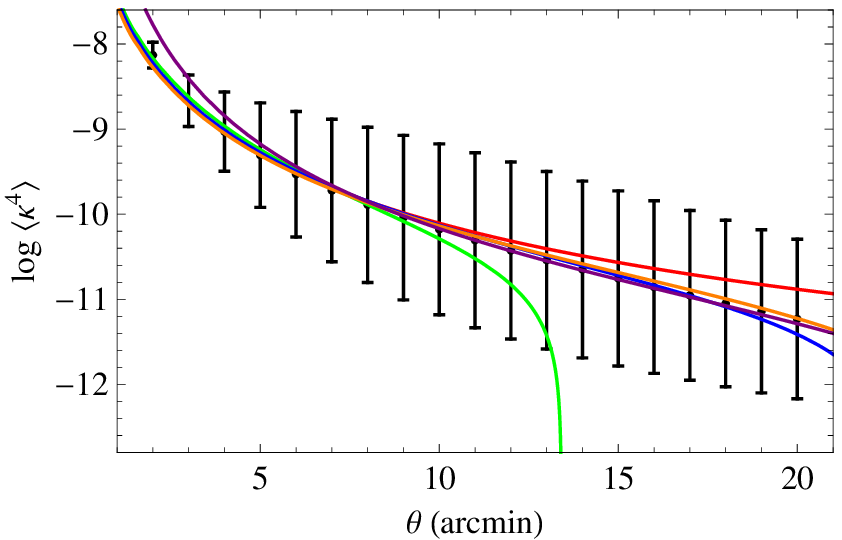} \\
\includegraphics[width=5.5cm]{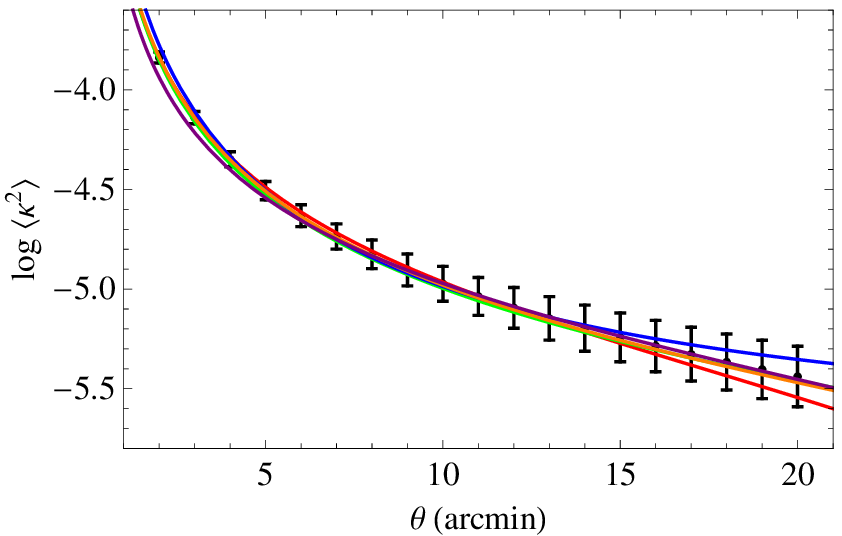}
\includegraphics[width=5.5cm]{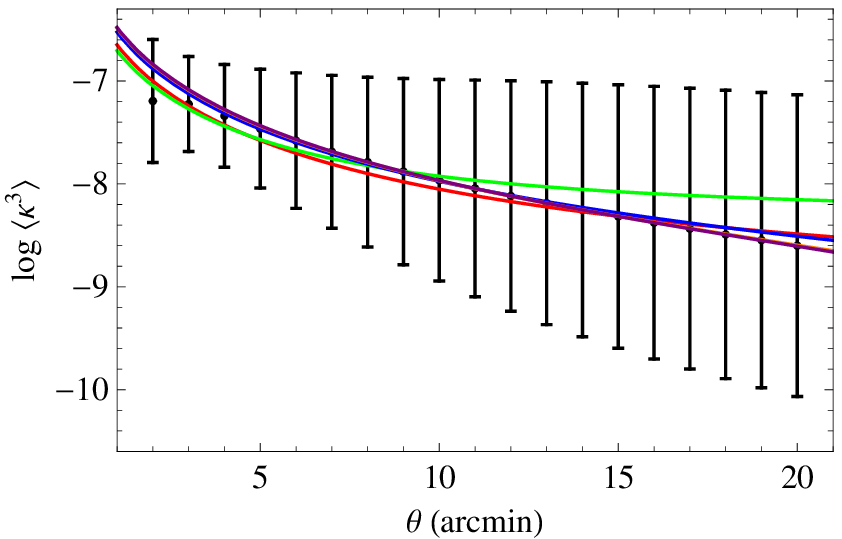}
\includegraphics[width=5.5cm]{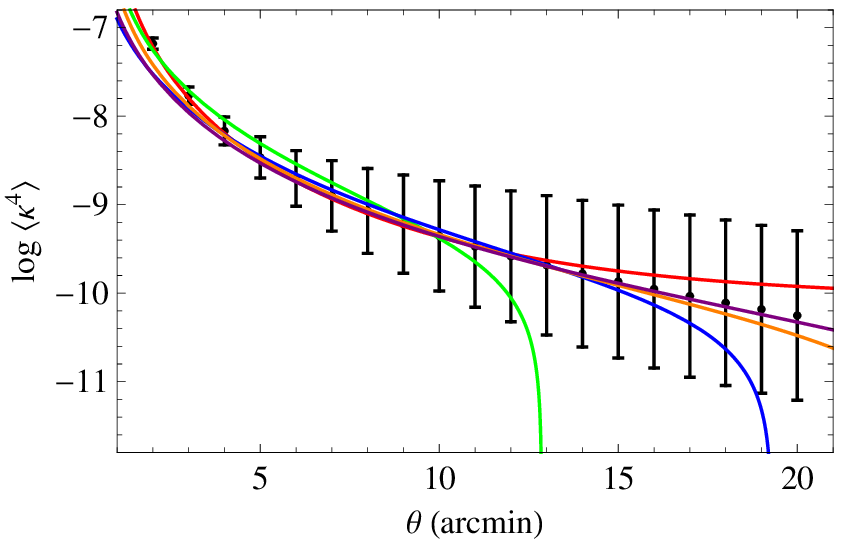} \\
\caption{Same as Fig.\,\ref{fig: calth} but for moments from the map reconstructed from noisy shear data.}
\label{fig: calobs}
\end{figure*}

\subsection{Observed moments}

The above results validate Eq.(\ref{eq: kappathlin}) showing that the theoretical modeling may be reconciled with actual measurements on idealized convergence map through a linear transformation. However, this is only half of the story. We have indeed still to demonstrate that the full calibration relations work at reproducing the observed HOM. These are estimated from the convergence map as reconstructed applying the KS93 method \cite{KS93} to noisy shear data. We therefore fit Eqs.(\ref{eq: cal2nd})\,-\,(\ref{eq: cal4th}) to the data\footnote{Note that we actually perform fits separately the 2nd and 4th order moments and the 3rd order one since the calibration parameters $(m_3, c_3, \nu_3)$ only enters in the $\langle \kappa^3 \rangle_{obs}$ relation. Moreover, we do not assume the noise is Gaussian so that the three parameters $(\nu_2, \nu_3, \nu_4)$ are all free to vary.}  measured on our simulated reconstructed maps getting the results in Fig.\,\ref{fig: calobs}.

As a reassuring result, we find that the fit is again quite good even if the rms of percentage residuals is definitely larger than before. Indeed, for the Gaussian and top hat filters (first and second row below), we find

\begin{displaymath}
\rho_{rms}(n) = \left \{
\begin{array}{l}
(7.84, 6.10, 53.3) \% \\
 \\
(8.57, 21.5, 42.2) \% \\
\end{array}
\right . 
\end{displaymath}
for $n = (2, 3, 4)$. Although the rms for the 4th order moment is quite large, it is nevertheless worth stressing that the best fit curve stays well within the error bars so that the fit is still acceptable. A look at the $\langle \kappa^4 \rangle(\theta)$ plot, however, shows that most of the deviations takes place at large $\theta$ so that one could reduce the rms by cutting this range. We therefore repeat the fit for the four smaller ranges used before and actually find a somewhat contradictory result. Although $\rho_{rms}(n = 4)$ indeed reduces when fitting to the $(2, 12)$ and $(4, 14)$ ranges only, the best results are obtained for $\theta_{min} \ge 6 \ {\rm arcmin}$ suggesting that the theory has problems on small angular scales. This is again consistent with what we get in the previous paragraph. However, when fitting to the full range, the algorithm tries to improve tha agreemeent on small $\theta$ since this is the range where the S/N is larger. We should therefore advocate against the use of small smoothing scales to get 4th order moment data, but this comes at the cost of giving away half the dataset. We will investigate later which strategy (i.e., improving the rms of percentage residuals or retaining as much data as possible) is more convenient.

It is interesting to look at the best fit values of the calibration parameters. Considering in the rest of this paragraph only the Gaussian filter (since the results are qualitatively the same for the top hat case) and the fit to the full range\footnote{See Appendix A for the full list of calibration parameters.}, we get

\begin{displaymath}
m_2 = -0.19_{-0.25}^{+0.28} \ \ , \ \ c_2 = -0.69_{-1.27}^{+1.35} \times 10^{-6}  \ \ , 
\end{displaymath}
\begin{displaymath}
m_3 = -0.53_{0.10}^{+0.38} \ \ , \ \ c_3 = -0.61_{-0.52}^{+0.71} \times 10^{-9}  \ \ ,
\end{displaymath}
\begin{displaymath}
m_4 = -0.11_{-0.33}^{+0.34} \ \ , \ \ c_4 = -0.10_{-3.15}^{+3.06} \times 10^{-11}  \ \ ,
\end{displaymath}
while the noise reference moments are 

\begin{displaymath}
\nu_2 = 0.21_{-0.21}^{+1.13} \times 10^{-4} \ , 
\end{displaymath}
\begin{displaymath}
\nu_3 = -0.56_{-0.07}^{+0.27} \times 10^{-10} \ ,
\end{displaymath}
\begin{displaymath}
\nu_4 = 2.24_{-2.24}^{+1.28} \times 10^{-10} \  .
\end{displaymath}
One could wonder whether these numbers are consistent with expectations. Using the expressions derived before, we can estimate the multiplicative bias $m$ as 

\begin{displaymath}
m = \left ( \frac{1 + m_n}{1 + \mu_n} \right )^{1/n} - 1 \ .
\end{displaymath}
Ideally, we should find the same result no matter which order we choose for setting the calibration parameters. The error on $m$ can then be qualitatively estimated propagating the errors on $(m_n, \mu_n)$. Since the confidence ranges are asymmetric, we first follow \cite{dago}, and correct for asymmetric errors with the following replacement rules

\begin{displaymath}
m_n \rightarrow m_n + (\sigma_{n}^{+} - \sigma_{n}^{-}) \ \ ,  \ \ 
\sigma(m_n) \rightarrow (\sigma_{n}^{+} + \sigma_{n}^{-})/2 
\end{displaymath}
with $\sigma_{n}^{\pm}$ the positive and negative error on $m_n$. We thus get

\begin{displaymath}
m(m_n, \mu_n) = \left \{
\begin{array}{ll}
-0.08 \pm 0.15 & n = 2 \\
 & \\
-0.20 \pm 0.09 & n = 3 \\ 
 & \\
0.03 \pm 0.10 & n = 4 \\
\end{array}
\right .  \  .
\end{displaymath}
These numbers are consistent with each other within less than $2 \sigma$ (taken as the sum in quadrature of the errors on the terms compared) which is a quite reassuring result if one also remind that the fitting algorithm does not enforce the consistency of $m$ amining only at finding the values of the calibration parameters $(m_n, c_n)$ which maximize the likelihood.

Somewhat more difficult is to judge whether the values of $(\nu_2, \nu_3, \nu_4)$ are reasonable or not being this question actually ill defined. On the one hand, we note that our simulated maps contains a Gaussian noise originating from the dispersion of galaxies intrinsic ellipiticity. According to \cite{Hama2004}, we should expect a noise second order moment on the convergence map given by $\sigma_{\kappa}^2 = \sigma_{\epsilon}^2/2 \theta_{pix}^2 n_g$. Setting $\sigma_{\epsilon} = 0.28$, $\theta_{pix} = 0.85 \ {\rm arcmin}$, $n_g = 27 \ {\rm gal/arcmin^2}$, we get $\sigma_{\kappa}^2 = 2.0 \times 10^{-3}$ which is two orders of magnitude larger than our estimated $\nu_2$. However, $\sigma_{\kappa}^2$ does not take into account either the smoothing or the reconstruction so that a disagreement is far from being worrisome. On the other hand, we should find $(\nu_3, \nu_4) = (0, 3 \nu_2^2)$ since we know that the noise is Gaussian. While $\nu_3$ may indeed be considered vanishing since its value is two orders of magnitude smaller than the lowest $\langle \kappa^3 \rangle(\theta)$, $\nu_4$ is definitely different from its expected Gaussian counterpart $(\nu_4 = 1.32 \times 10^{-9}$). The difference may come from both the reconstruction procedure, and the fitting algorithm which does not enforce any relation between $\nu_2$ and $\nu_4$ thus leaving a degeneracy between these latter and $c_4$. One could have changed the fitting code by forcing $(\nu_3, \nu_4)$ to their Gaussian values (which also reduces the number of parameters). However, in practical applications, one can not assume a priori that the noise is Gaussian so that we prefer to be conservative, and not introduce any prior on the noise parameters. 

Motivated by these considerations, we deem as successfully passed the validation test of  Eqs.(\ref{eq: cal2nd})\,-\,(\ref{eq: cal4th}). 

\section{HOM as cosmological tools}

Having shown that it is possible to correctly match theoretical and observed moments, we can now investigate whether HOM are actually of any help in constraining cosmological parameters ${\bf p}_c$. As a preliminary remark, it is worth stressing that ${\bf p}_c$ are not the only quantities which enter theory predictions. Indeed, one can not assume a priori that the calibration parameters $(m_n, c_n)$ are the same for every cosmological model so that having determined them from a fit to simulated data mimicking the actual ones is far from being sufficient to deem them as known quantities. Moreover, the noise properties of the field are fully unknown so that one has to add the noise reference moments $(\nu_2, \nu_3, \nu_4)$ to the list of quantities to be fitted for. Summarizing, the full parameters vector ${\bf p}$ will be the union of the cosmological one 

\begin{displaymath}
{\bf p}_c = \{\Omega_M, \Omega_b, w_0, w_a, h, n_s, \sigma_8 \} \ ,
\end{displaymath}
and the nuisance one

\begin{displaymath}
{\bf p}_n = \{m_2, c_2, m_3, c_3, m_4, c_4, \nu_2, \nu_3, \nu_4\} \ .
\end{displaymath}
Given the large number of parameters, we do not expect HOM alone to be able to put severe constraints on all of them. We will therefore set some parameters when considering HOM only as dataset. On the other hand, it has been suggested from long time that HOM can help breaking degeneracies affecting second order statistisc such as the cosmic shear tomography power spectra. We will therefore also investigate how constraints on the full set of parameters change when HOM are added to shear tomography. 

\subsection{Fisher matrix formalism}

In order to answer the above questions, we rely on the Fisher matrix formalism \cite{TTH97}. We therefore compute the Fisher matrix elements given by

\begin{equation}
F_{ij} = \left \langle - \frac{\partial^2 {\ln{\cal{L}}}({\bf p})}{\partial p_i \partial p_j} \right \rangle 
\label{eq: fijdef}
\end{equation} 
where the average is approximated by the likelihood value estimated in the fiducial model and nuisance parameters ${\bf p}_{fid}$, and, assuming a multivariate Gaussian, the likelihood function reads

\begin{equation}
-2 \ln{{\cal{L}}({\bf p})} \propto
({\bf D}_{obs} - {\bf D}_{th})^T {\bf Cov}^{-1} ({\bf D}_{obs} - {\bf D}_{th}) \ .
\label{eq: deflike}
\end{equation}
where ${\bf D}_{obs}$ has been defined in Eq.(\ref{eq: dobsvec}), and ${\bf D}_{th}$ is its theoretical counterpart evaluated through Eqs.(\ref{eq: cal2nd})\,-\,(\ref{eq: cal4th}).  The Fisher matrix elements therefore read

\begin{equation}
F_{ij} = \frac{\partial {\bf D}_{th}}{\partial p_i} {\bf Cov}^{-1} \frac{\partial {\bf D}_{th}}{\partial p_j}
\label{eq: fijhom}
\end{equation}
showing that a key role is played by the data covariance matrix which we split up as 

\begin{equation}
{\bf Cov} = {\bf Cov}^{obs} + {\bf Cov}^{sys}
\label{eq: defcovtot}
\end{equation}
with the first term giving the statistical errors as computed from Eq.(\ref{eq: dobscov}). The second term is introduced to take into account the inaccuracies in the calibration procedures, i.e., the fact that we can predict HOM from theory using Eqs.(\ref{eq: cal2nd})\,-\,(\ref{eq: cal4th}) only up to a scatter quantified by the rms of percentage residuals. This systematics covariance matrix is diagonal\footnote{The systematics covariance matrix could actually be non diagonal. Indeed, the expected values of the calibration parameters $(m_n, c_n)$ are correlated being all related to the same multiplicative bias $m$. However, we have determined the calibration parameters from separate fits without forcing $(m_2, m_3, m_4)$ to obey theoretical relations among them. This is expected to cancel any correlation among them and among the residuals of the calibration relations.} so that its elements read

\begin{equation}
Cov_{ij}^{sys} = \rho_{rms}(i) \rho_{rms}(j) D_{obs}(i) D_{obs}(j) \delta^{K}_{ij}
\label{eq: covijsys}
\end{equation}
with $\delta^{K}_{ij}$ the Kronecker $\delta$, and $\rho_{rms}(i)$ the rms of the percentage residuals for the order of the moment corresponding to the $i$\,-\,th element  of the data vector ${\bf D}_{obs}$. What actually enters the likelihood is the precision matrix, i.e., the inverse of the covariance matrix. However, it is known that this quantity may be biased if estimated from a not large enough number of independent realizations. Following \cite{H07}, we therefore compute is as 

\begin{equation}
{\bf Cov}^{-1} = \frac{{\cal{N}}_f - {\cal{N}}_d - 2}{{\cal{N}}_f - 1} ({\bf Cov}^{obs} + {\bf Cov}^{sys})^{-1}
\label{eq: covinv}
\end{equation}
where the multiplicative term corrects for the finite number of realizations with ${\cal{N}}_d$ the length of the data vector. An important caveat is in order here. Since we have divided the full survey area $\Omega$ in subfields of $25 \ {\rm sq \ deg}$ each, it is ${\cal{N}}_f = \Omega/25$. On the other hand, the data vector length is ${\cal{N}}_d = \eta (\theta_{max} - \theta_{min})/d\theta = \eta \Delta \theta/d\theta $, with $\eta$ the number of moments used (e.g., $\eta =1$ if only moments of order 2 are used, or $\eta = 3$ if one uses all orders). The choice of $(\Omega, \Delta \theta, d\theta)$ must therefore be done taking care that the condition ${\cal{N}}_f > {\cal{N}}_d + 2$ is fulfilled in order to avoid having a physically meaningless negative definite inverse covariance matrix. For our fiducial choice, it is 

\begin{displaymath}
(\Omega, \theta_{min}, \theta_{max}, d\theta, \eta) = (3500, 2, 20, 1, 3)
\end{displaymath}
giving ${\cal{N}}_f = 140$ and ${\cal{N}}_d = 57$ so that the multiplicative factor takes the quite small value $({\cal{N}}_f - {\cal{N}}_d - 2)/({\cal{N}}_f - 1) \simeq 0.58$ which reduces the Fisher matrix elements, and hence the forecasted constraints on the cosmological parameters. Since the total area is fixed, a possible way out could be to reduce the subfields area so that ${\cal{N}}_f$ is larger, but we should then check whether the estimated HOM are still reliable. As an alternative, one can reduce the length of the data vector ${\cal{N}}_d$ using a larger sampling $d\theta$ or only a subset of the data (so that $\eta$ is smaller). While both these choices are possible, they both ask for reducing the number of constraints which can have the opposite effect of weakening overall the HOM constraining power.

Another effect also enters the game driving the choice of $(\theta_{min}, \theta_{max}, d\theta)$. As we have seen, $\rho_{rms}(n)$ is a function of $(\theta_{min}, \theta_{max})$ so that the systematics covariance matrix will be different depending on which smoothing angle range is used. The decrease of $\rho_{rms}(n)$ with $\theta_{min}$ would advocate in favour of larger values, but it is possible to show that the derivatives terms into Eq.(\ref{eq: fijhom}) are larger at small $\theta$ consistent with the naive expectation that large smoothing angle cancels both the noise and the signal. Which is the best range to use is, therefore, a compromise between these two opposite behaviours. A similar discussion also applies to the choice of the sampling $d\theta$. As shown before, moments of same or different orders evaluated at two different smoothing angles $(\theta_1, \theta_2)$ may be significantly correlated if the $\theta_2 - \theta_1$ is smaller than few times $d\theta$. As a consequence, the effective number of constraints is actually smaller than the length of the data vector. Moreover, should $d\theta$ be too small, the observed covariance matrix ${\rm Cov}^{obs}$ could become close to degenerate thus making unstable its numerical inversion and hence unreliable the Fisher matrix itself. Whether this happens or not also depend on the $\theta$ range. Indeed, the larger is $\theta$, the more the details of the convergence field are smoothed out and the moments reduce to those of a Gaussian distribution. Should we therefore consider only large $\theta$, the moments will be all correlated being simply a description of the smoothed noise field so that which $d\theta$ value does not matter at all. 

Motivated by these considerations, we will therefore explore different combinations of $(\theta_{min}, \theta_{max}, d\theta)$. In particular, we consider five different ranges, namely

\begin{displaymath}
(\theta_{min}, \theta_{max}) = (2, 20), (2, 12), (4, 14), (6, 16), (8, 18) \ {\rm arcmin} 
\end{displaymath}
and sample them with $d\theta = 1, 2, 3, 4 \ {\rm arcmin}$. Moreover, we will also investigate how the results depend on the smoothing filter used, and whether it is more convenient to use moments of all order or only a subset of them.

As a final preliminary remark, a somewhat technical note is in order. As already said at the beginning of this section, one has to add the set of nuisance parameters ${\bf p}_n$ to the list of quantities to be determined by the fit to the data. As a consequence, they also enter the Fisher matrix so that their fiducial values have to be set. It is, however, convenient to introduce a new nuisance parameter set $\tilde{{\bf p}}_n = {\bf p}_n/{\bf p}_{n}^{fid}$ with ${\bf p}_{n}^{fid}$ the values found from the fit to the simulated dataset for a given fiducial cosmology. Should the calibration be independent on cosmology, we should expect $\tilde{p}_{in} = 1$ for all the components $i$ of the scaled vector. We therefore use $\tilde{{\bf p}}_n$ as nuisance parameter set and fix its fiducial value to unity. In order to break the degeneracies with cosmological parameters, we will add a prior matrix to the Fisher matrix one, i.e., we redefine the Fisher matrix elements as

\begin{equation}
F_{ij} = \frac{\partial {\bf D}_{th}}{\partial p_i} {\bf Cov}^{-1} \frac{\partial {\bf D}_{th}}{\partial p_j} + \pi_{ij} \ ,  
\label{eq: fijhomprior}
\end{equation}
where $\pi_{ij}$ are the elements of the prior matrix. This is a diagonal matrix whose first seven elements are set to zero (since we do not want to put any prior on the cosmological parameters), while the remaining elements are set to $1/\sigma_i^2$ with $\sigma_i$ quantifying our belief in the prior knwoldege of the nuisance parameters. Setting, e.g., $\sigma_i = 0.1$ means we are assuming that the true value of the $i$\,-\,th component of the nuisance parameters vector lies within $10\%$ of the corresponding fiducial one. In order to reduce the quantities to change in the analysis, we will set $\sigma_i = \sigma_p$ for all $i$ and explore how constraints change as a function of $\sigma_p$. 

\begin{figure*}
\centering
\includegraphics[width=4.25cm]{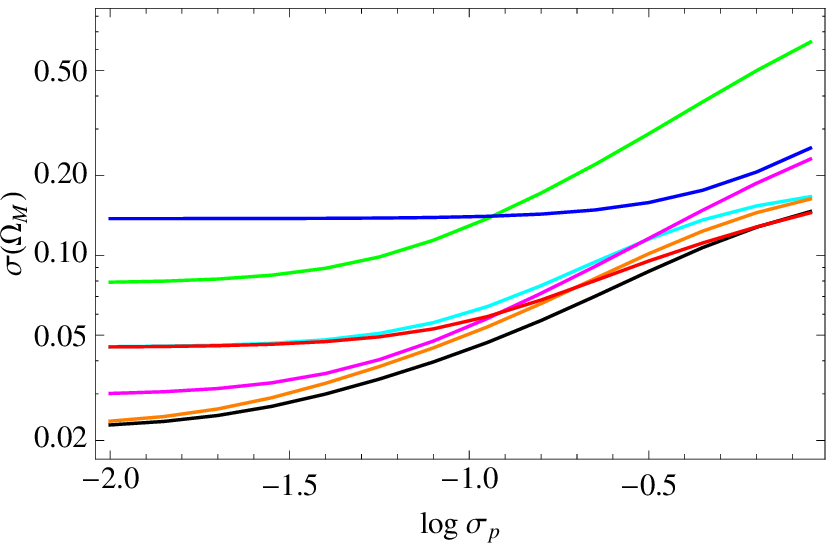}
\includegraphics[width=4.25cm]{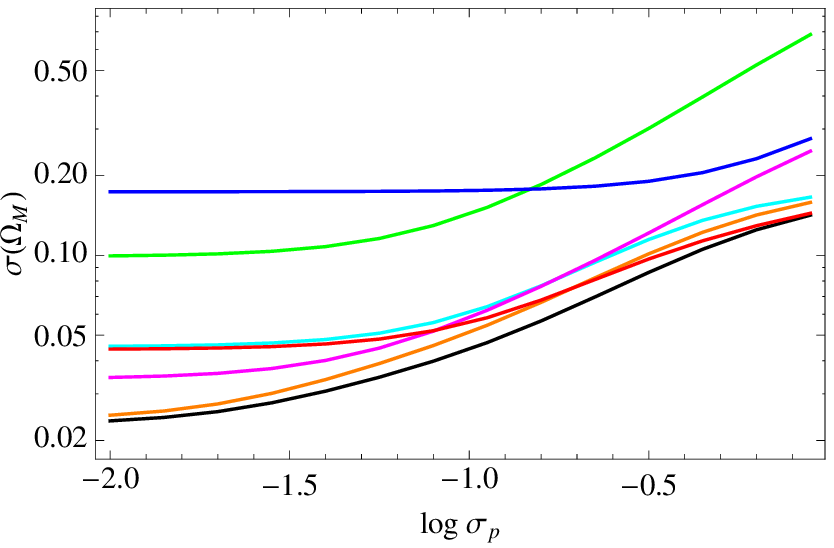}
\includegraphics[width=4.25cm]{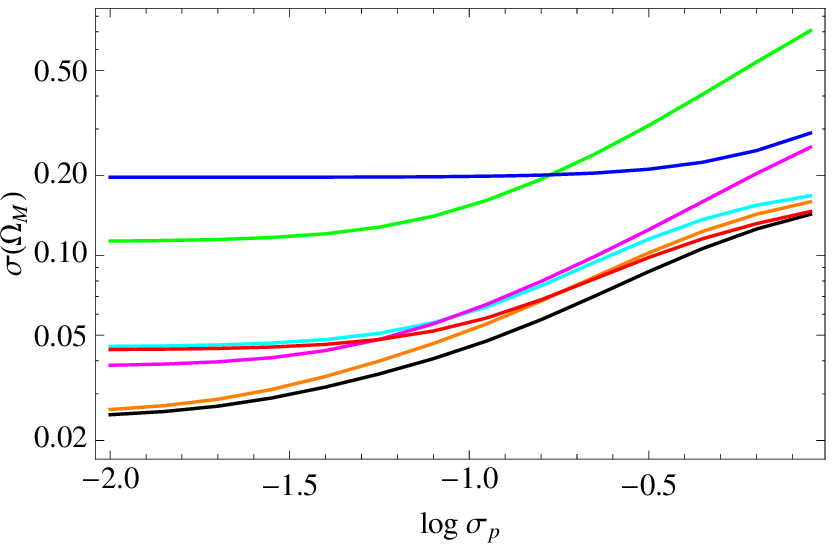}
\includegraphics[width=4.25cm]{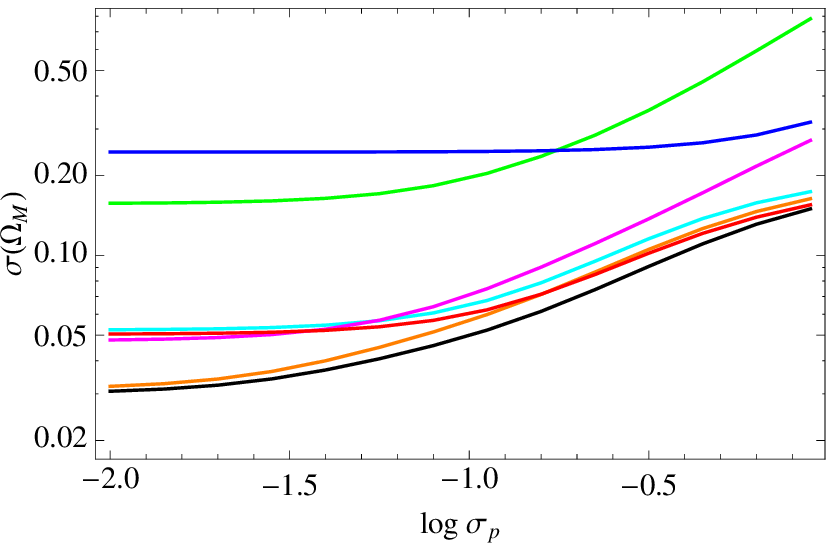} \\
\includegraphics[width=4.25cm]{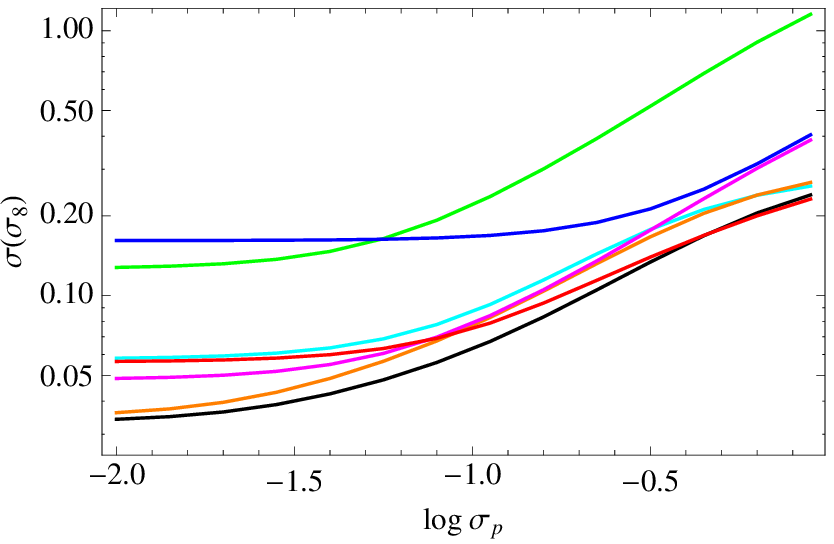}
\includegraphics[width=4.25cm]{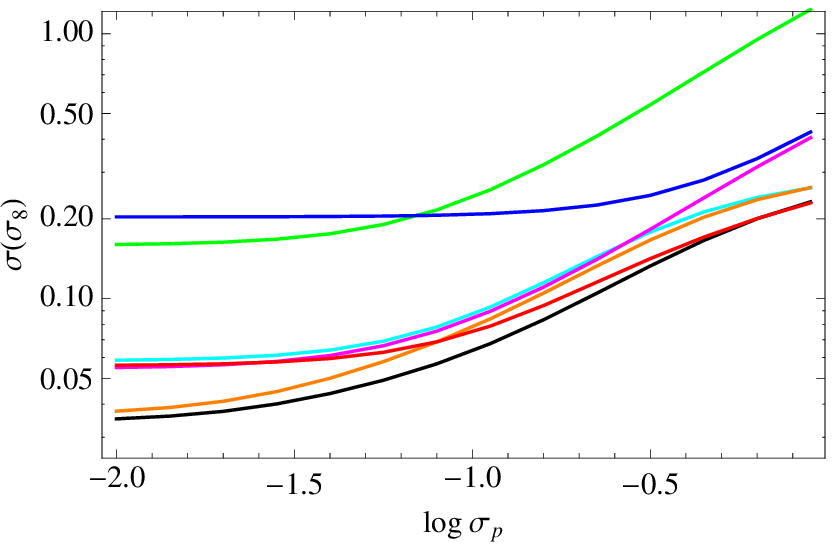}
\includegraphics[width=4.25cm]{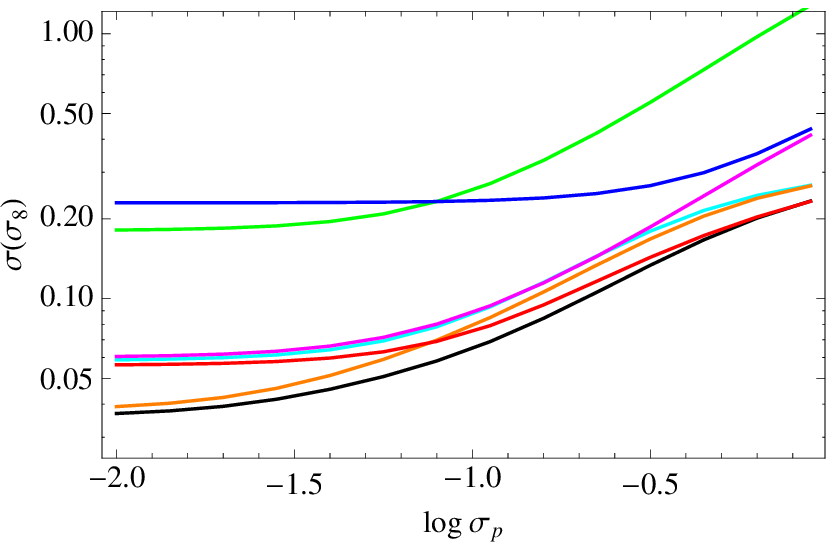}
\includegraphics[width=4.25cm]{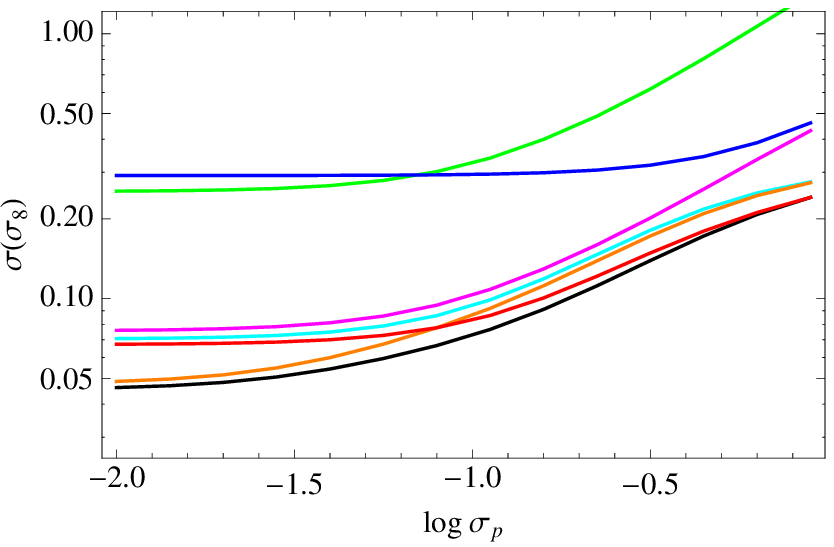} \\
\caption{Fisher matrix forecast of the $1 \sigma$ uncertainty on $\Omega_M$ (top) and $\sigma_8$ (bottom) as a function of the prior parameter $\sigma_p$ from the fit to data covering the smoothing angle range $(\theta_{min}, \theta_{max}) = (2, 20) \ {\rm arcmin}$ sampled with step $d\theta = 1, 2, 3, 4 \ {\rm arcmin}$ (from left to right). A Gaussian filter has been used for smoothing, while black, green, cyan, blue, orange, magenta, red lines refer to the results using all moments, 2nd order only, 3rd order only, 4th order only, 2nd and 3rd, 2nd and 4th, 3rd and 4th, respectively.}
\label{fig: oms8vsmoments}
\end{figure*}

\subsection{Constraints on $(\Omega_M, \sigma_8)$ from HOM only}

In a first application, one will probably be interested in using moments only as constraints. However, because of the large number of parameters (cosmological and nuisance) and the reduced numbers of effective degrees of freedom because of the large correlations among moments at different smoothing scales, one expects too weak constraints if alla parameters are left free. One would therefore focus only on the two parameters lensing is most sensible too, i.e., the matter density $\Omega_M$ and the variance of linear perurbations $\sigma_8$. We therefore compute the Fisher matrix, set all the cosmological parameters but $(\Omega_M, \sigma_8)$, and marginalize over the nuisance parameters for any given choice of the prior $\sigma_p$, the fitting range $(\theta_{min}, \theta_{max})$, and the sampling $d\theta$. Although limited to two parameters only, this analysis will be helpful to elucidate how HOM actually works hence selecting the best strategy. 

\begin{figure*}
\centering
\includegraphics[width=4.25cm]{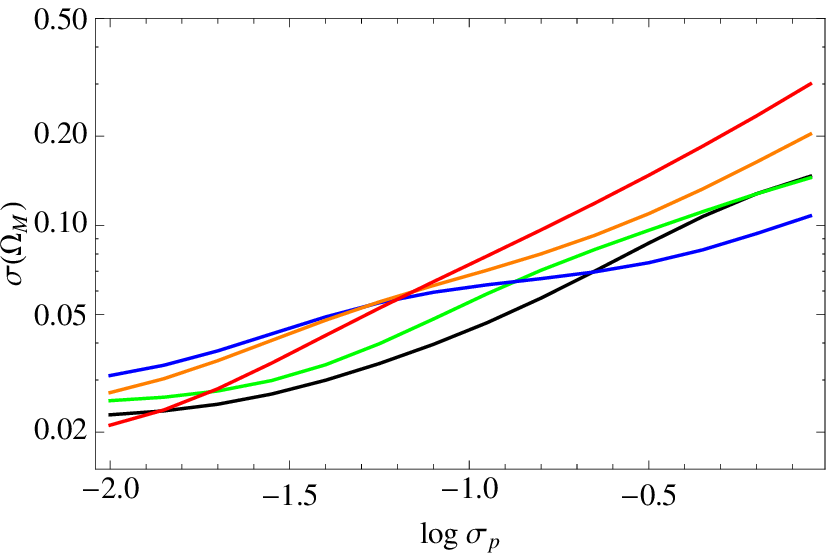}
\includegraphics[width=4.25cm]{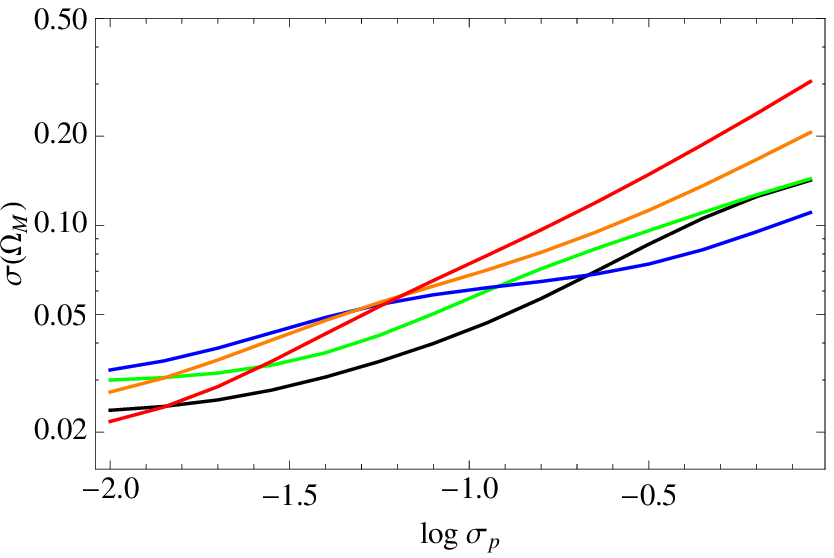}
\includegraphics[width=4.25cm]{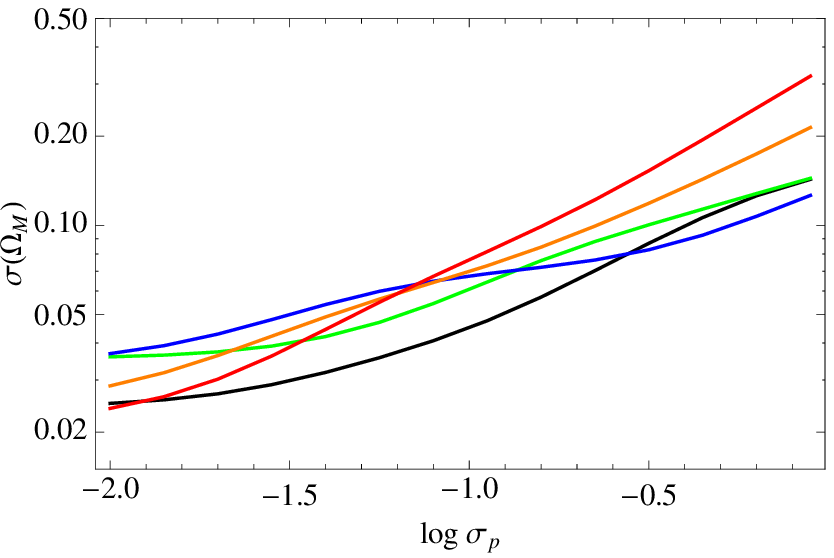}
\includegraphics[width=4.25cm]{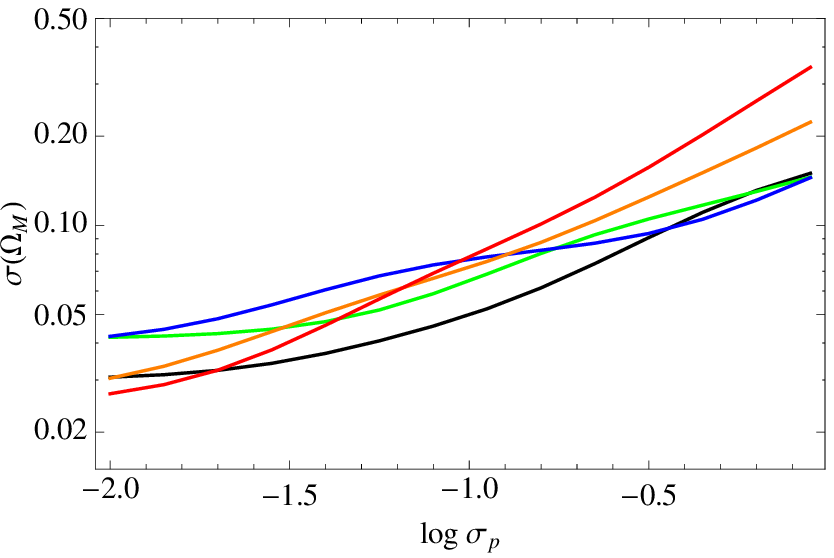} \\
\includegraphics[width=4.25cm]{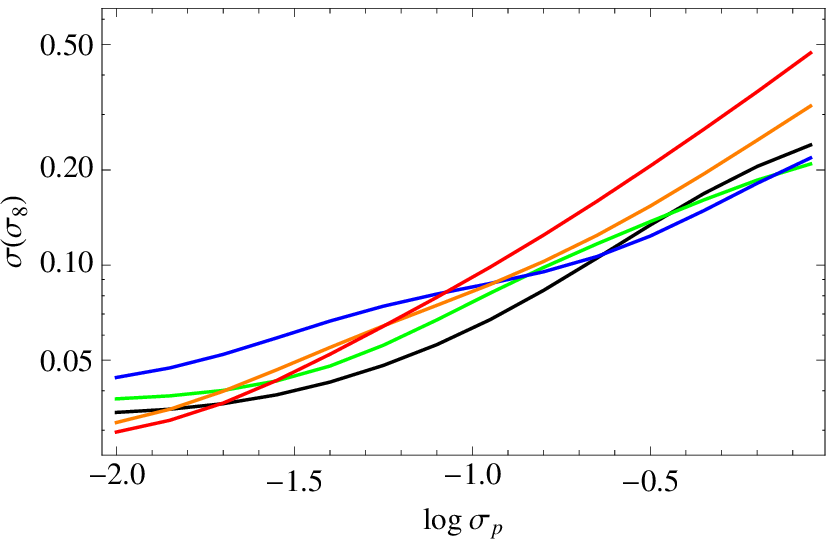}
\includegraphics[width=4.25cm]{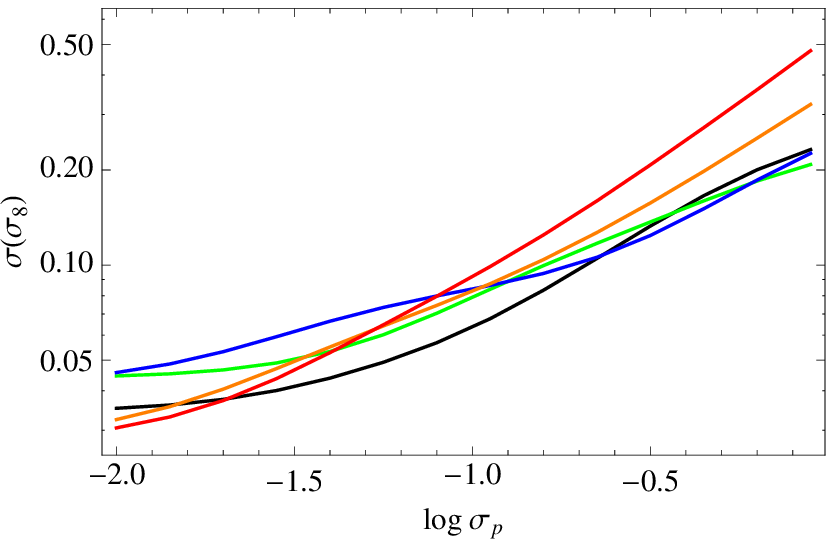}
\includegraphics[width=4.25cm]{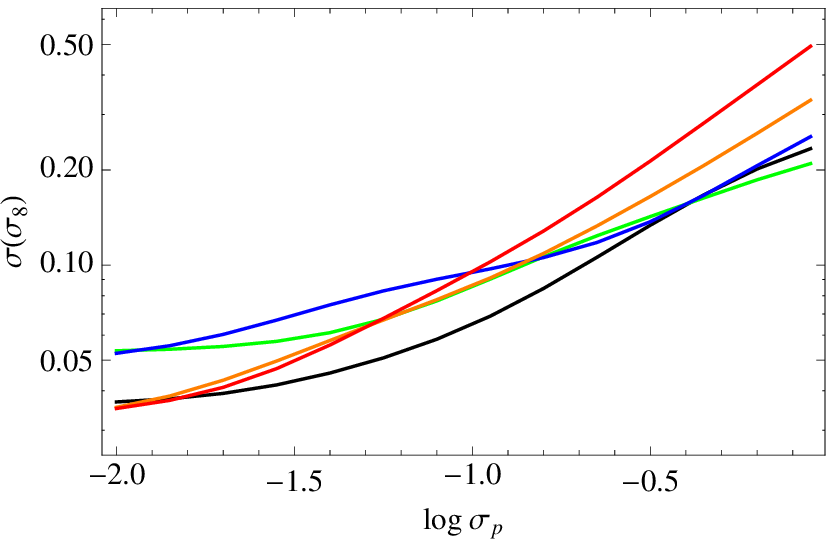}
\includegraphics[width=4.25cm]{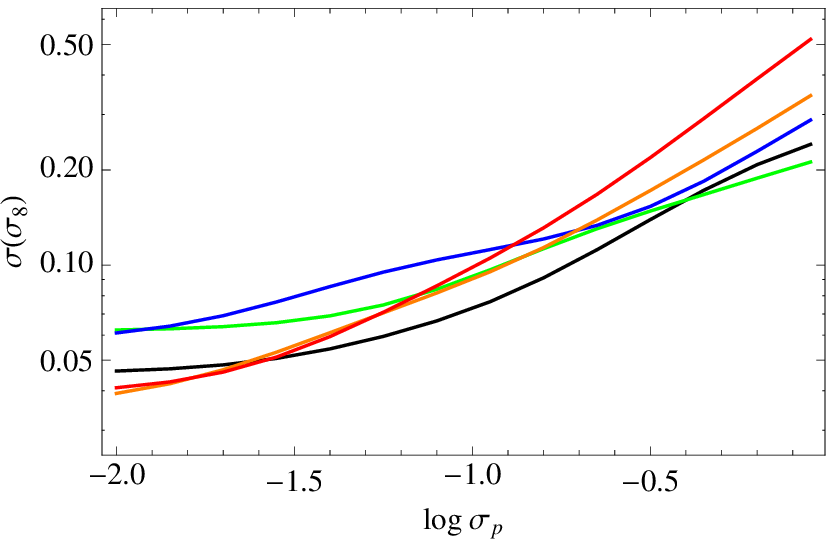} \\
\caption{Same as Fig.\,\ref{fig: oms8vsmoments} but all moments are used, while black, green, blue, orange, red lines refer to using data over the ranges (2, 20), (2, 12), (4, 14), (6, 16), (8, 18) arcmin, respectively, sampled with step $d\theta = 1, 2, 3, 4 \ {\rm arcmin}$ from left to right.}
\label{fig: oms8vsrange}
\end{figure*}

Let us start by investigating whether all moments have to be used or comparable results can be obtained using only a given combination of 2nd, 3rd, and 4th order moments\footnote{Unless otherwise stated, all the results in this section refers to HOM estimated after smoothing the reconstructed convergence map with a Gaussian filter. The resulta are qualitatively unchanged for the top hat case too the only differences being highlighted at the end of this section if necessary.}. Fig.\,\ref{fig: oms8vsmoments} show the forecasted $1 \sigma$ constraints on the parameters $(\Omega_M, \sigma_8)$ setting $(\theta_{min}, \theta_{max}) = (2, 20) \ {\rm arcmin}$ and four values of the sampling step $d\theta$. As expected, the constraints on $(\Omega_M, \sigma_8)$ are quite weak and strongly dependent on the value of the prior $\sigma_p$. A saturation roughly takes place for $\sigma_p < 0.03$ suggesting that, for smaller values, the results are limited by the HOM statistical and systematic errors. For a $10\%$ prior on the nuisance parameters and using all moments, we get $\sigma(p_{\mu})/p_{\mu} \simeq 18 \  (6) \%$ for $p_{\mu} = \Omega_M (\sigma_8)$ which points at HOM as interesting tools to constrain these two cosmological parameters. However, the constraints are strongly dependent on priors with $\sigma(p_{\mu})/p_{\mu} \simeq 37 \  (22) \%$ for $p_{\mu} = \Omega_M (\sigma_8)$ if no prior is set. These are definitely too large to deem HOM as actually useful which, however, does not come as a surprise having HOM being suggested as complementary probes to be used in combination with standard cosmic shear tomography to break degeneracies and not as a single tool. We will nevetheless keep on investigating their use as standalone probes to better highlight which strategy is most suited to strengthen the constraints. 

One can naively expect that the strongest constraints are obtained when all moments are combined. This is actually not so obvious. Indeed, using all moments maximizes ${\cal{N}}_d$ in Eq.(\ref{eq: covinv}) thus reducing the inverse covariance matrix and hence the $F_{ij}$ elements. Moreover, because of the strong correlation among moments of different orders estimated at the same $\theta$ and moments of the same order at close $\theta$ values, the effective number of degrees of freedom is smaller than the length of the data vector ${\bf D}_{obs}$ so that it is not clear whether adding more orders automatically translates in more constraints. On the other hand, the number of nuisance parameters is smaller if only one order or only two out of three are used. That is why we have explored all seven possibilities which are plotted as different coloured lines in Fig.\,\ref{fig: oms8vsmoments}. Notwithstanding the above cautionary considerations, it turns out that the best case scenario is indeed the most obvious one. The best constraints are obtained when the full HOM dataset (black line) is used, but we nevertheless note that almost equivalent results are obtained for the case with only 2nd and 4th order moments used (magenta line). Dropping off the 4th order moments to use only 2nd and 3rd order ones (orange line) leads to constraints on $(\Omega_M, \sigma_8)$ which are comparable but yet definitely weaker, while using only one set of moments (green, cyan, and blue lines) gives significantly weaker constraints. Such results suggest that most of the cosmological information is encapsulated in 2nd and 4th order moments, while 3rd order ones contain less information but can still be used to replace 4th order moments becuase of the smaller systematic uncertainty. 

Let us now move to the question about which smoothing angle range $(\theta_{min}, \theta_{max})$ should be used to minimize the forecasted errors on $(\Omega_M, \sigma_8)$. On one hand, for a fixed sampling $d\theta$, the wider is $\Delta \theta = \theta_{max} - \theta_{min}$, the larger is the length $N$ of the data vector. On the other hand, the larger is $\Delta \theta$, the larger is ${\cal{N}}_d$ thus making the inverse covariance matrix smaller. Moreover, moving to larger $\theta_{min}$ values allows to reduce (up to an order of magnitude) the systematics contribution to the data covariance matrix which should help in strenghtening the constraints. To explore which effects dominate, we use all HOM and plot the constraints as a function of the prior parameter $\sigma_p$ for the five different ranges we have introduced before and four different sampling step values. Results in Fig.\,\ref{fig: oms8vsrange} show thar the strongest constraints are obtained when the full $(2, 20) \ {\rm arcmin}$ (black line) range is used, but comparable results can be reached if the upper limit is cut to $12 \ {\rm arcmin}$ only (green line). It is instructive to note that the worst results are obtained for the intermediate range $(4, 14) \ {\rm arcmin}$ (blue line). Such a result can be qualitatively explained as follows. Most of the cosmological information is encapsulated in the dependence of HOM on cosmological parameters at small smoothing angles. In other words, the derivatives $\partial {\bf D}_{th}/\partial p_i$ entering the Fisher matrix elements $F_{ij}$ are decreasing function of $\theta$ so that $F_{ij}$ is larger when small $\theta$ ranges are used. On the other hand, the overall $\rho_{rms}(n)$ is smaller when the fit is limited to ranges with large $\theta_{min}$ thus increasing ${\bf Cov}^{-1}$ and hence $F_{ij}$. These two different effects dominate in the extreme ranges, while compensate each other in the intermediate one which therefore provides the weakest constraints. 

Fig.\,\ref{fig: oms8vsrange} can also be used to infer how the results change with the sampling step $d\theta$. Again, two competing effects are at work here. On one hand, the larger is $d\theta$, the smaller are the elements of the correlation matrix ${\bf R}$ and the larger is the multiplicative term into Eq.(\ref{eq: covinv}) thus leading to an overall increase of $F_{ij}$. On the contrary, the larger is $d\theta$, the smaller is the length of the data vector ${\bf D}_{obs}$ so that the number of degrees of freedom in the fit is smaller hence weakening the constraints. Fig.\,\ref{fig: oms8vsrange} shows that the two effects approximately compensates each other if the full $(2, 20) \ {\rm arcmin}$ range is used, while it is safer to set $d\theta \le 3 \ {\rm arcmin}$ if one wants to use only  subset of the data.

We finally conclude this analysis with a note on the results when the top hat filter is used. All the results are qualitatively unchanged pointing at a preference for using all HOM data over the full smoothing angle range with the finest sampling. The only remarkable difference is the magnitude of the constraints on $(\Omega_M, \sigma_8)$ which turns out to be dramatically weaker. We indeed find that $\sigma(\Omega_M)$ and $\sigma(\sigma_8)$ can be larger up to a factor 2 for the same value of the prior parameter $\sigma_p$. This is likely related to the top hat filter being too aggressive in performing noise reduction with the smoothing procedure removing also part of the signal where cosmological information would have been usable. We however warn the reader that what is of most interest is not the use of HOM alone, but in combination with shear tomography so that we postopone to the next paragraph the decision whether to dismiss the top hat filter overall.

\begin{figure*}
\centering
\includegraphics[width=4.25cm]{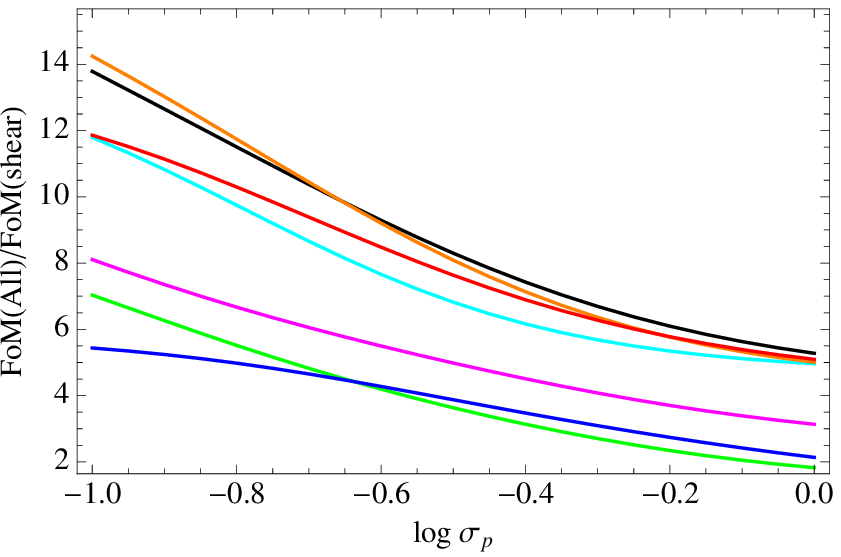}
\includegraphics[width=4.25cm]{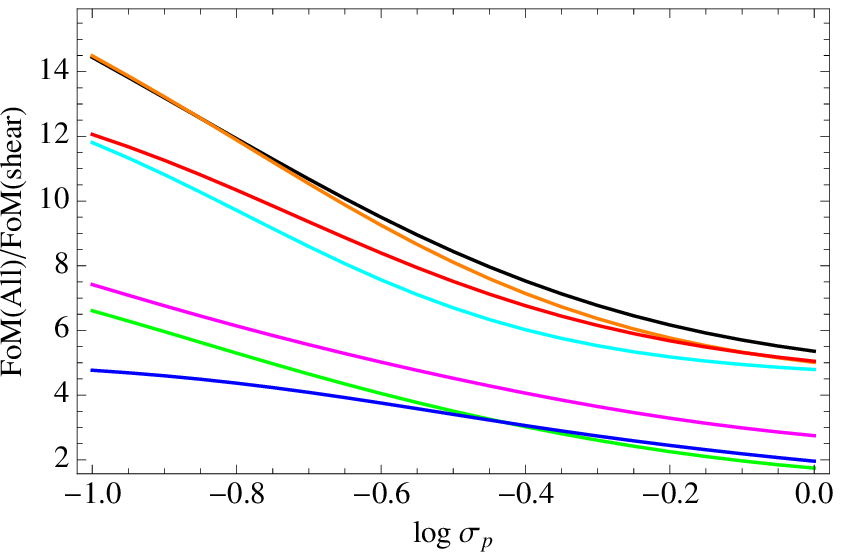}
\includegraphics[width=4.25cm]{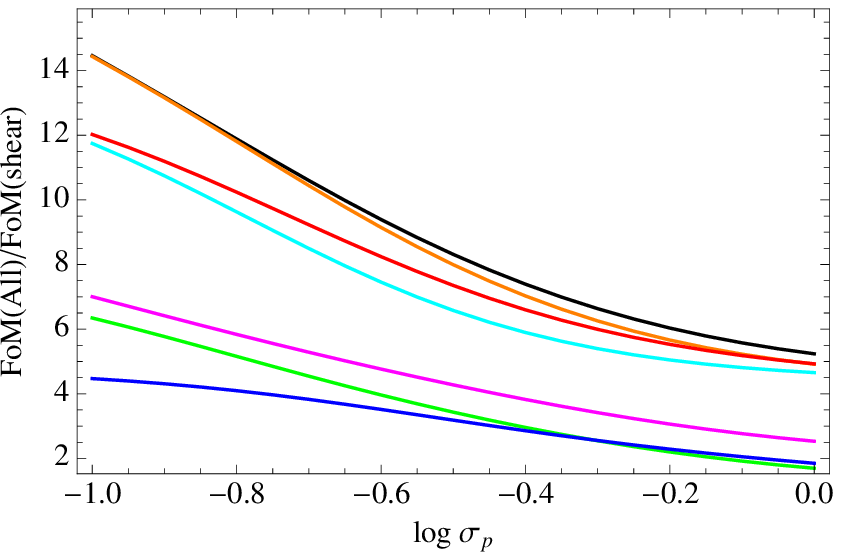}
\includegraphics[width=4.25cm]{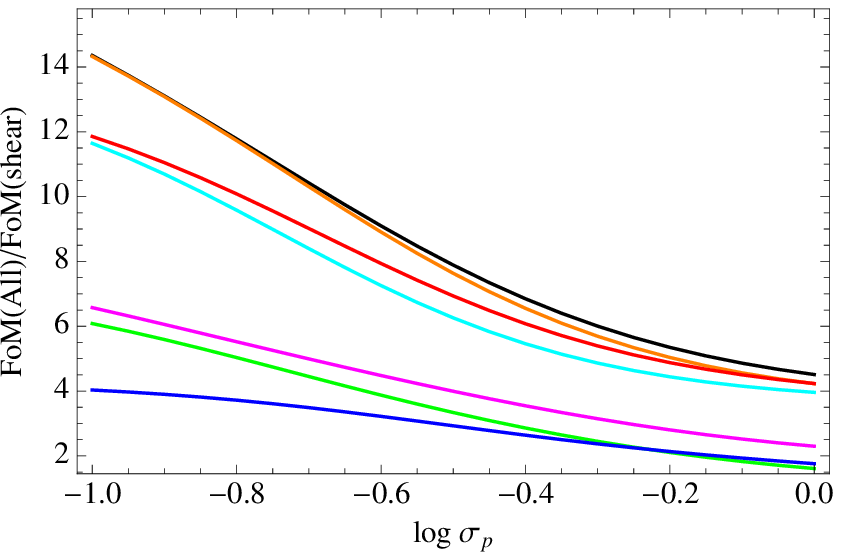} \\
\includegraphics[width=4.25cm]{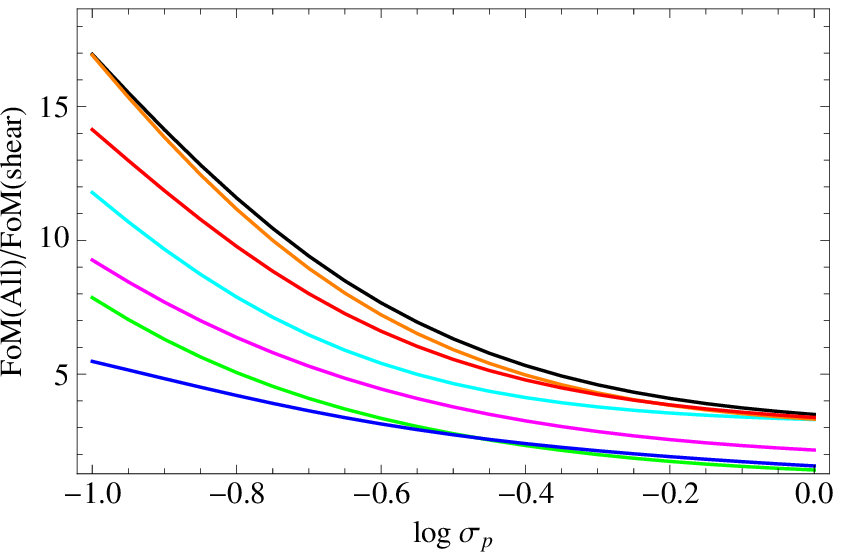}
\includegraphics[width=4.25cm]{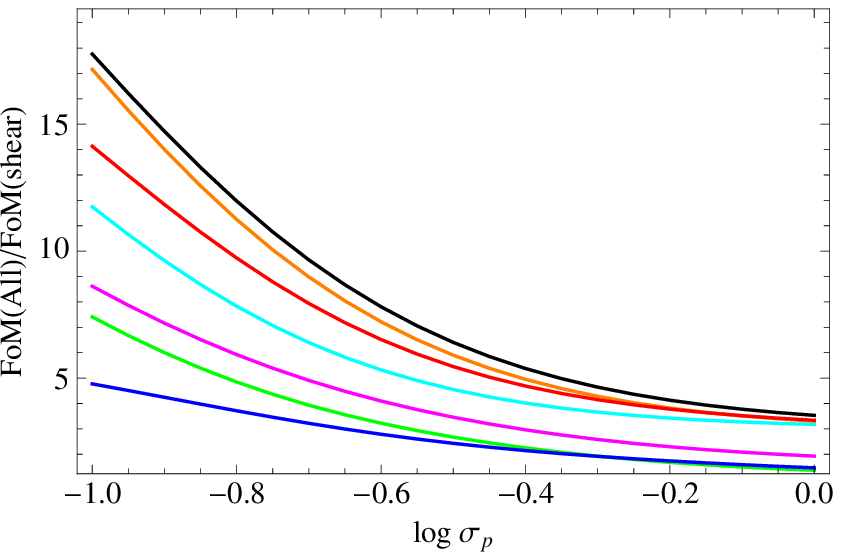}
\includegraphics[width=4.25cm]{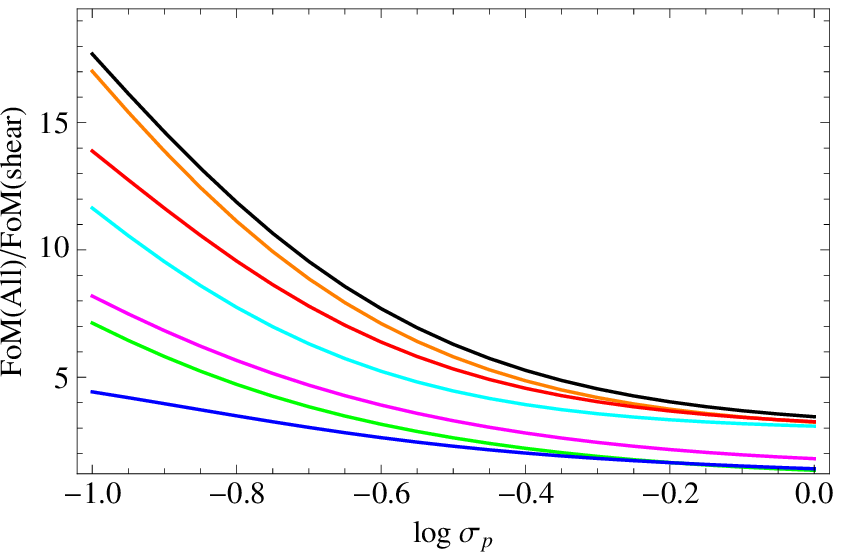}
\includegraphics[width=4.25cm]{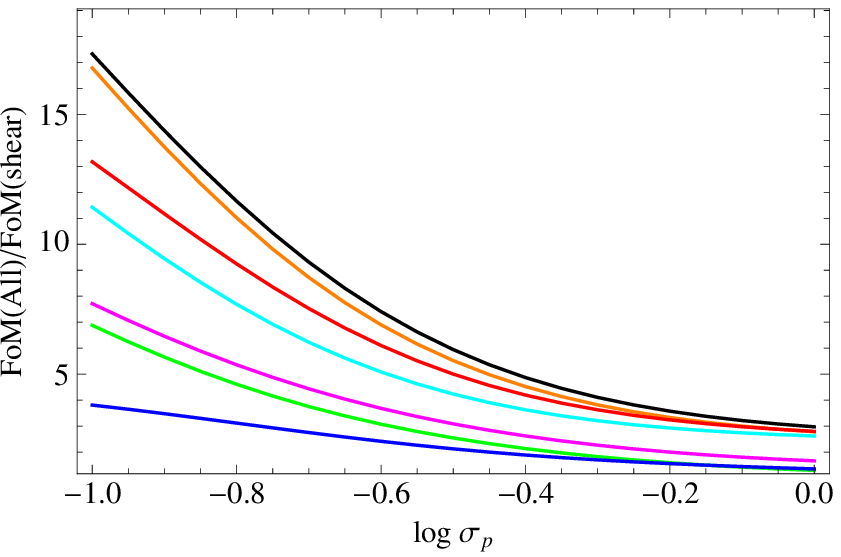} \\
\caption{FoM ratio as function of the prior parameter $\sigma_p$. Black, green, cyan, blue, orange, magenta, red lines refer to using all HOM, 2nd order only, 3rd only, 4th only, 2nd and 3rd, 2nd and 4th, 3rd and 4th in combination with cosmic shear tomography based on three (top) and six (bottom) redshift bins. A Gaussian filter is used to smooth the convergence map with smoothing angles spanning the range $(2, 20) \ {\rm arcmin}$ with sampling step $d\theta = 1, 2, 3, 4 \ {\rm arcmin}$ from left to right.}
\label{fig: fomratiovsmoments}
\end{figure*}

\subsection{Joint use of shear tomography and HOM}

It is long been suggested that higher than 2nd order statistics can help breaking degeneracies among cosmological parameters which limit the efficiency of cosmic shear tomography. It is therefore our aim here to investigate whether this is indeed the case and to which extent the combinaton of shear tomography and HOM improves the constraints on underlying cosmology. To this end, we assume that shear tomography and HOM are independent probes so that the total Fisher matrix simply reads

\begin{displaymath}
{\bf F} = {\bf F}_{HOM} + {\bf F}_{WL} + {\bf \Pi}
\end{displaymath}
with ${\bf F}_{HOM}$ and ${\bf F}_{WL}$ the HOM and shear tomography Fisher matrices, and ${\bf \Pi}$ the priors matrix defined as explained before. We refer the reader to Appendix B for details on the estimate of ${\bf F}_{WL}$, while we only remark here that it is based on an idealized survey having the same area, number density, and redshift distribution of the MICECAT catalogue we have used as input for the HOM Fisher matrix. We will consider two different options for the tomography splitting sources in 3 or 6 redshift bins. 

\begin{figure*}
\centering
\includegraphics[width=4.25cm]{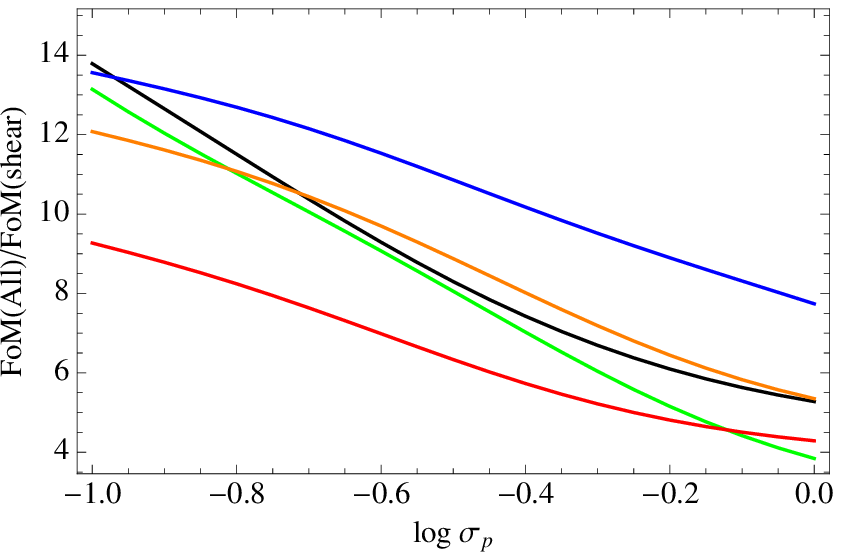}
\includegraphics[width=4.25cm]{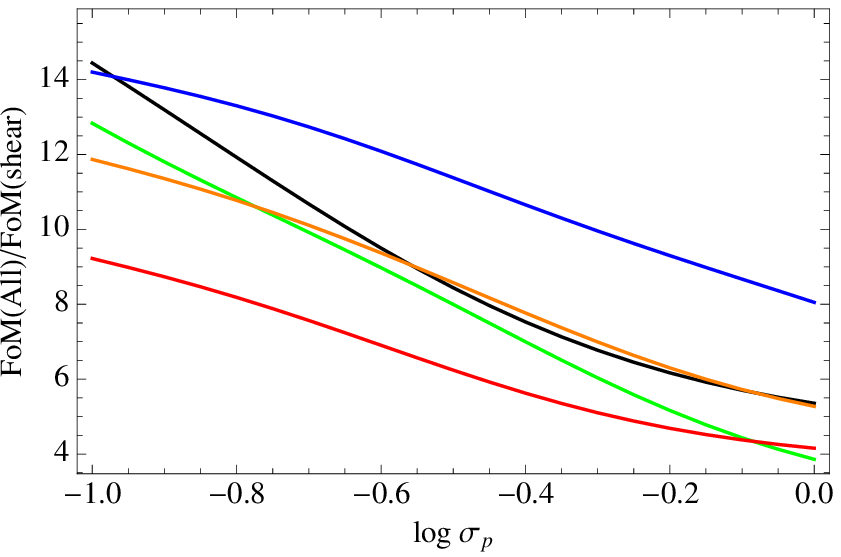}
\includegraphics[width=4.25cm]{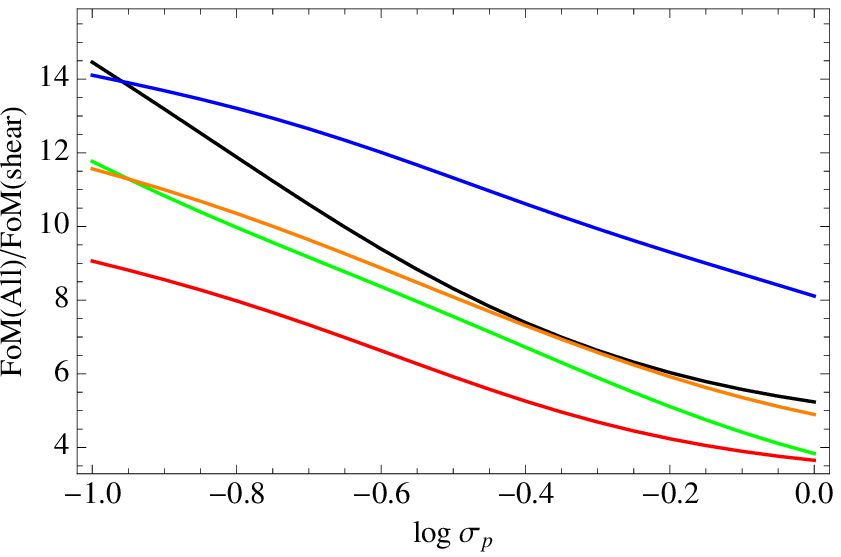}
\includegraphics[width=4.25cm]{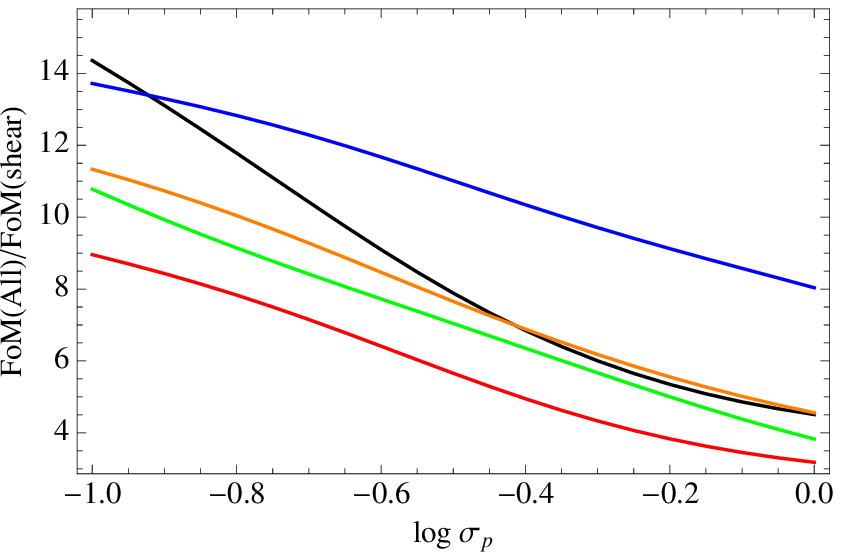} \\
\includegraphics[width=4.25cm]{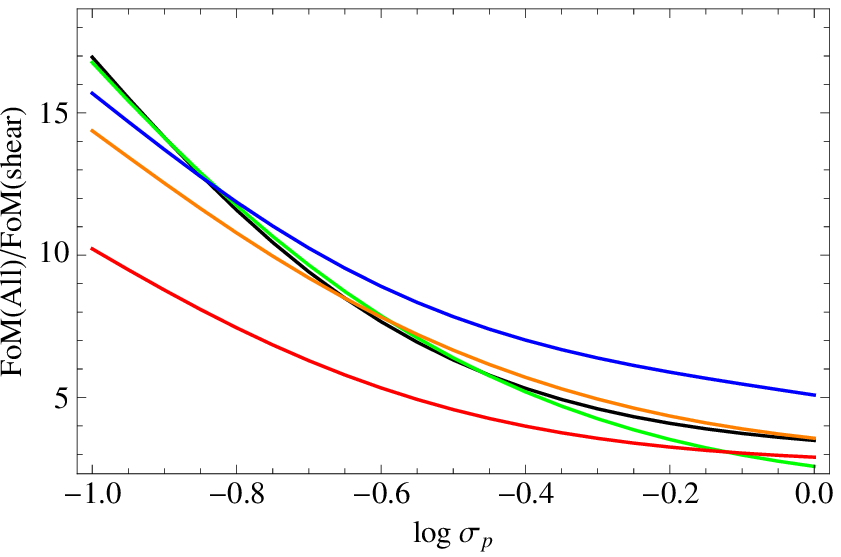}
\includegraphics[width=4.25cm]{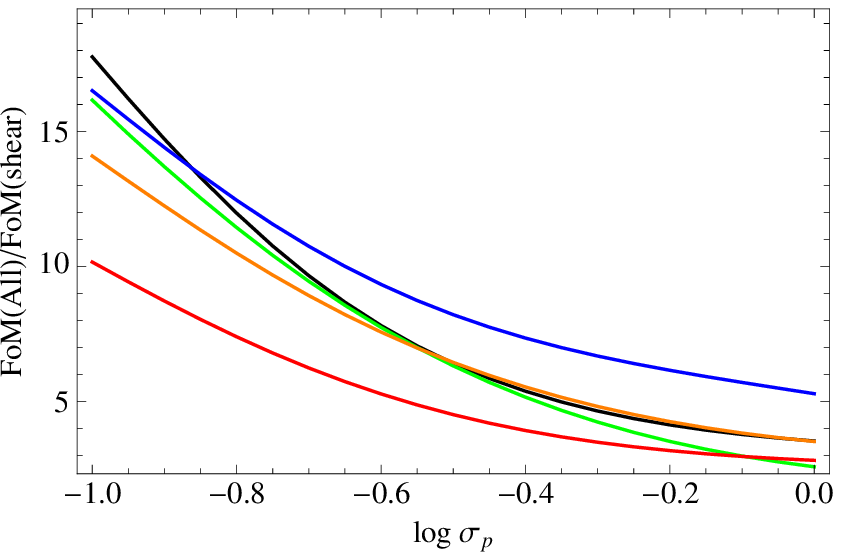}
\includegraphics[width=4.25cm]{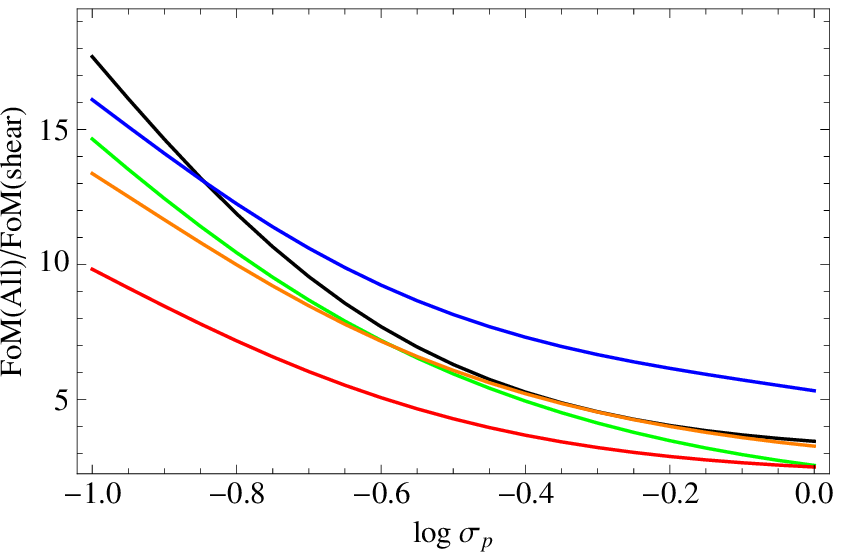}
\includegraphics[width=4.25cm]{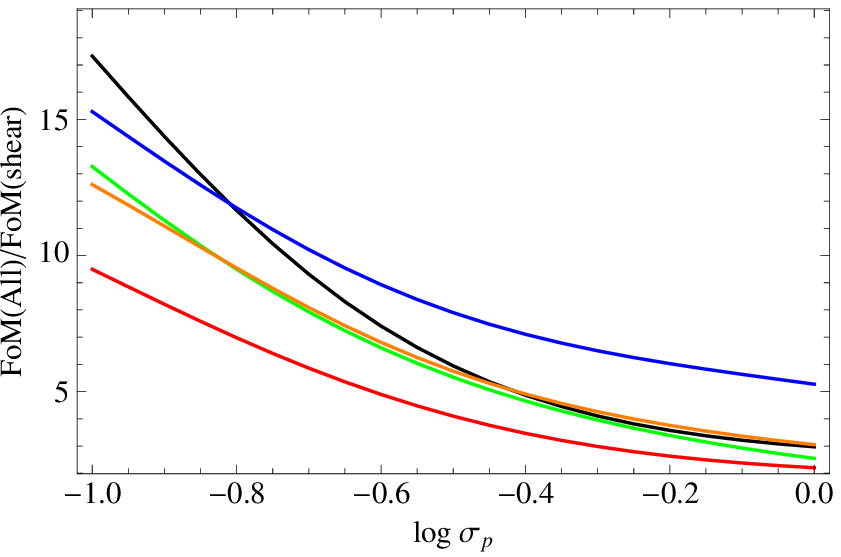} \\
\caption{Same as Fig.\,\ref{fig: fomratiovsmoments} but changing the smoothing angle range and using all HOM in combination with shear tomography (with 3 and 6 redshift bins results plotted in top and bottom panels). Black, green, blue, orange, red lines refer to the cases with $(\theta_{min}, \theta_{max}) = (2, 20), (2, 12), (4, 14), (6, 16); (8, 18) \ {\rm arcmin}$, respectively, with sampling $d\theta = 1, 2, 3, 4 \ {\rm arcmin}$ from left to right.}
\label{fig: fomratiovsrange}
\end{figure*}

Since we are interested in investigating how much the use of HOM improves the constraints with respect to the case when shear tomography only is used, we will always discuss ratios of quantities so that it is immediate to grasp any improvement. Although the details of an actual survey may be different from the idealized one we are considering, we are confident that taking the ratios partially washes out these differences making our results more general. However, we are well aware that this point should be addressed with care which is nevertheless outside the aim of this preliminary investigation. 

\begin{figure*}
\centering
\includegraphics[width=4.25cm]{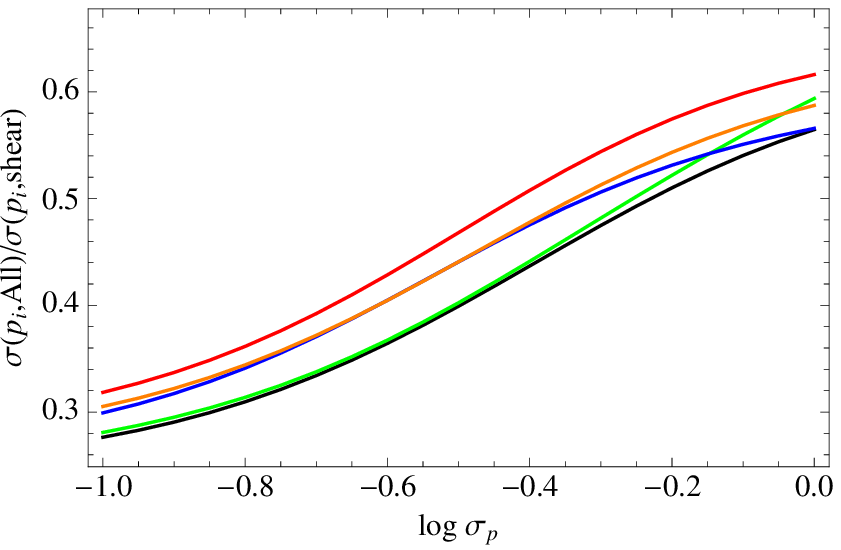}
\includegraphics[width=4.25cm]{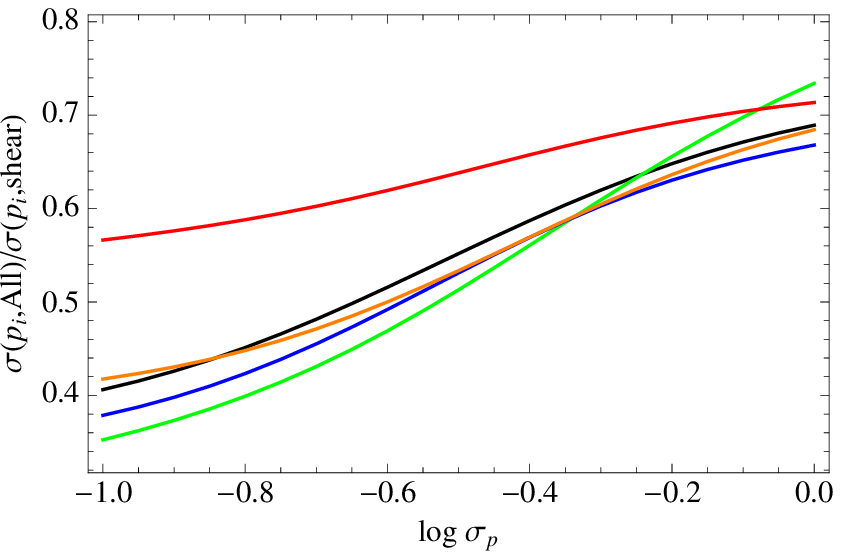}
\includegraphics[width=4.25cm]{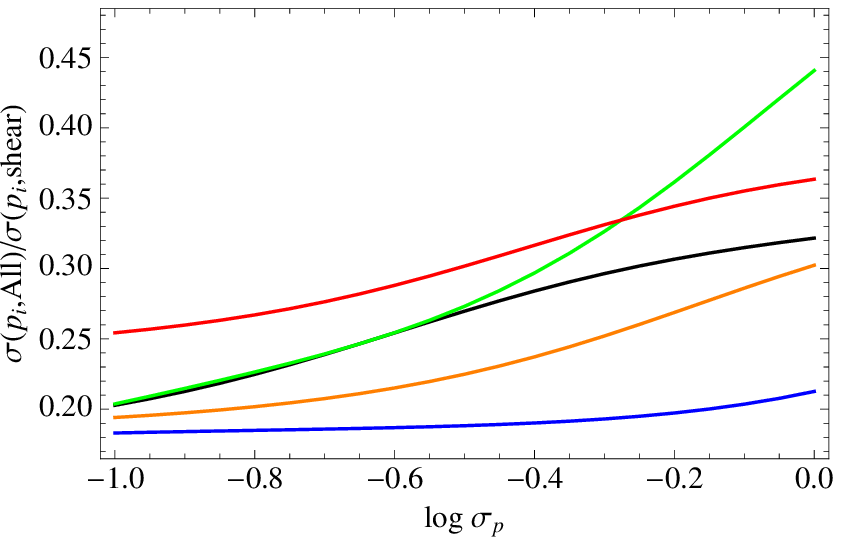}
\includegraphics[width=4.25cm]{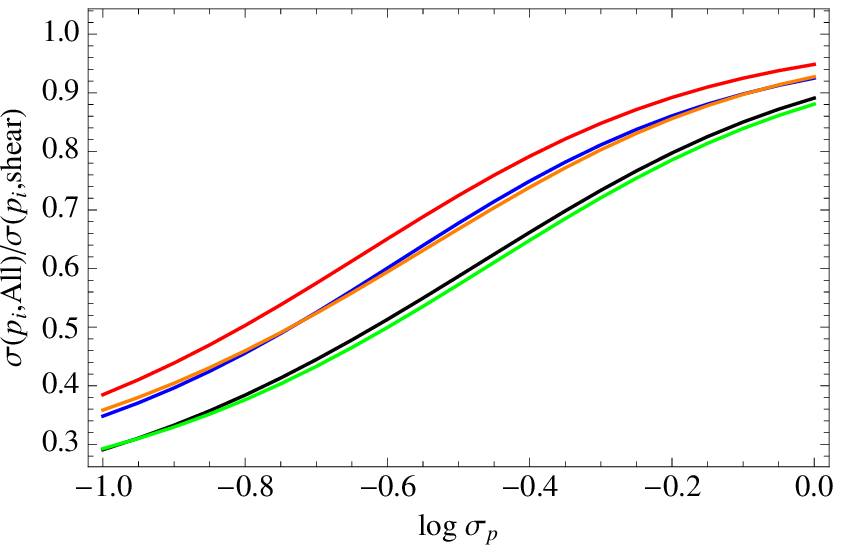} \\
\includegraphics[width=4.25cm]{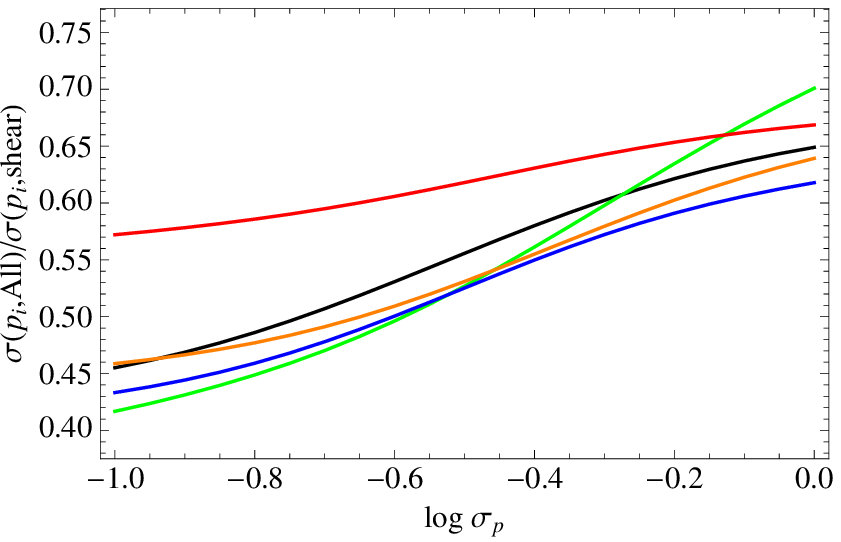}
\includegraphics[width=4.25cm]{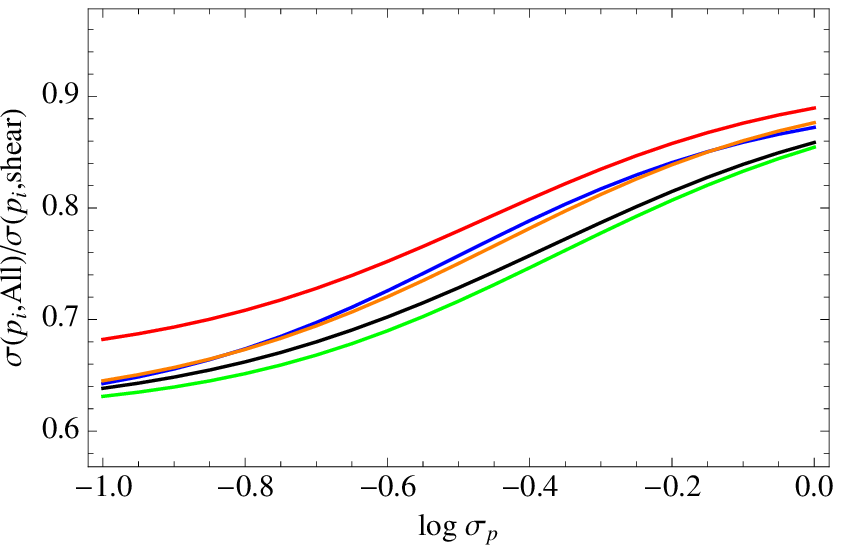}
\includegraphics[width=4.25cm]{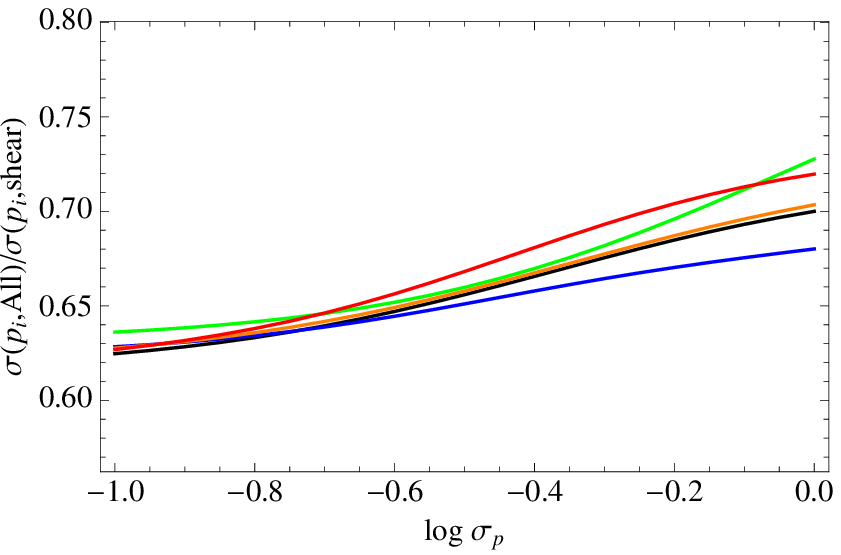}
\includegraphics[width=4.25cm]{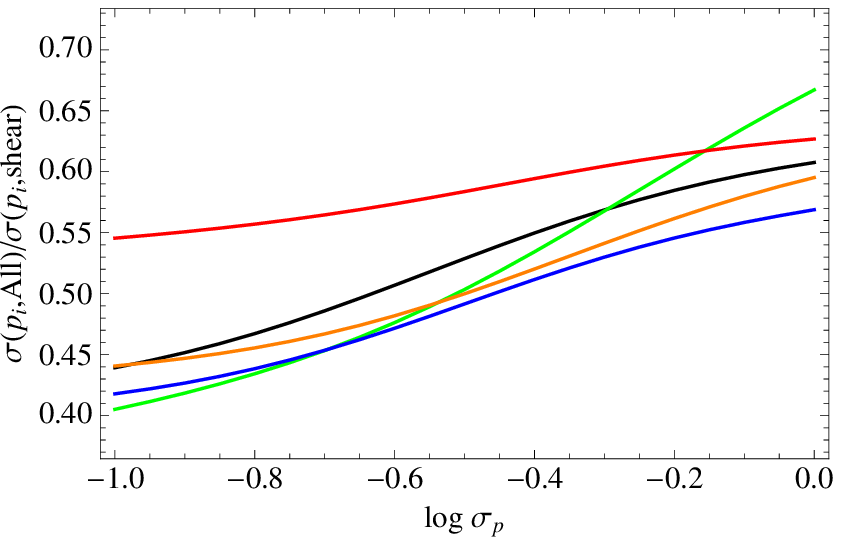} \\
\caption{Ratio between the marginalized errors on the cosmological parameters for the case with the full HOM dataset is used and shear tomography is computed splitting galaxies in six redshift bins. Same colour code as Fig.\,\ref{fig: fomratiovsrange} is used to denote the smoothing angle range used, while sampling step changes from left to right as in that same figure. Panels are for $(\Omega_M, w_0, w_a, w_p, \Omega_b, h, n_s, \sigma_8)$ from top left to bottom right, with $w_p$ the dark energy EoS at the pivot redshift.}
\label{fig: parratio}
\end{figure*}

With these caveats in mind, let us first look at Fig.\,\ref{fig: fomratiovsmoments} where we plot the ratio between the Figure of Merit  (hereafter, FoM) with and without using HOM, where we remind the reader that it is

\begin{displaymath}
{\rm FoM}= \sqrt{\frac{1}{{\rm det} {\bf C}(w_0, w_a)}}
\end{displaymath}
with ${\bf C}(w_0, w_a)$ the inverse of the Fisher matrix marginalized over all parameters but $(w_0, w_a)$. We hold fixed the smoothing angle range to the full one, and look at how the results change when the full HOM dataset or only a part of it is used. The most remarkable result we find is that the FoM increases by a factor $\sim 2$ even in the worst case scenario when only a very poor prior $(\sigma_p = 1)$ is set and 4th order moments only (blue line) are used. Results become roughly independent on the prior as far as $\sigma_p \le 0.03$ which is encouraging since it suggests that a realistic prior on the HOM nuisance parameters is enough to get a valuable FoM increase. Which combinations of HOM is best suited depends on the sampling step adopted. However, as a general rule, lines clearly separates in two groups depending on whether 2nd order moments are included or not in the combined shear + HOM dataset. When $\langle \kappa^2 \rangle(\theta)$ data are included, the FoM increase is maximized and is roughly the same no matter which sampling step is used or which other data are added. 

This result can be interpreted noting that both shear tomography and $\langle \kappa^2 \rangle(\theta)$ are second order quantities so that they are most sensible to the cosmological parameters than higher order ones. As such, the derivatives with respect to cosmological parameters entering the total Fisher matrix elements receive a further contribution thus ameliorating the overall result. Higher than 2nd order moments then help breaking the $\Omega_M$\,-\,$\sigma_8$ degeneracy which allows to better constrain these two quantities and, as a consequence, to also strengthen the constraints on the other parameters hence leading to a larger FoM. The breaking of the $\Omega_M$\,-\,$\sigma_8$ degeneracy takes place even when 2nd order moments are not used thus motivating the factor 2.5 boost of the FoM in these cases too. However, for all cases, the FoM ratio is maximized when the full HOM dataset is jointly fitted with shear tomography so that we take this as our final recommendation.

Let us now look at Fig.\,\ref{fig: fomratiovsrange} where we investigate the impact of changing the HOM data range assuming that all moments are used in combination with shear tomography. As soon as a prior with $\sigma_p \le 0.3$ is adopted, the preferred range is always the full one $(2, 20) \ {\rm arcmin}$ or its narrower version $(2, 12) \ {\rm arcmin}$. This is consistent with a picture where most of the cosmological information in encoded in the small smoothing angle regime which is expected given than the more the map is smoothed, the more difficult is to detect those deviations from non Gaussianity which bear the imprint of the underlying cosmological model. Which of the two ranges is most efficient in boosting the FoM then depends on the sampling step and the details (3 or 6 redshift bins) of the shear tomography, but we can advocate the use of the full one since it is the preferred case in all the panels but in the bottom left one. On the contrary, there is no significative dependence of the results on the sampling step $d\theta$. This is likely a consequence of the total Fisher matrix being dominated by the shear contribution which makes HOM important mainly to break degeneracies. Since this happens no matter which step is used, the differences become meaningless thus deleting any preference for any particular value.

\begin{figure*}
\centering
\includegraphics[width=5.5cm]{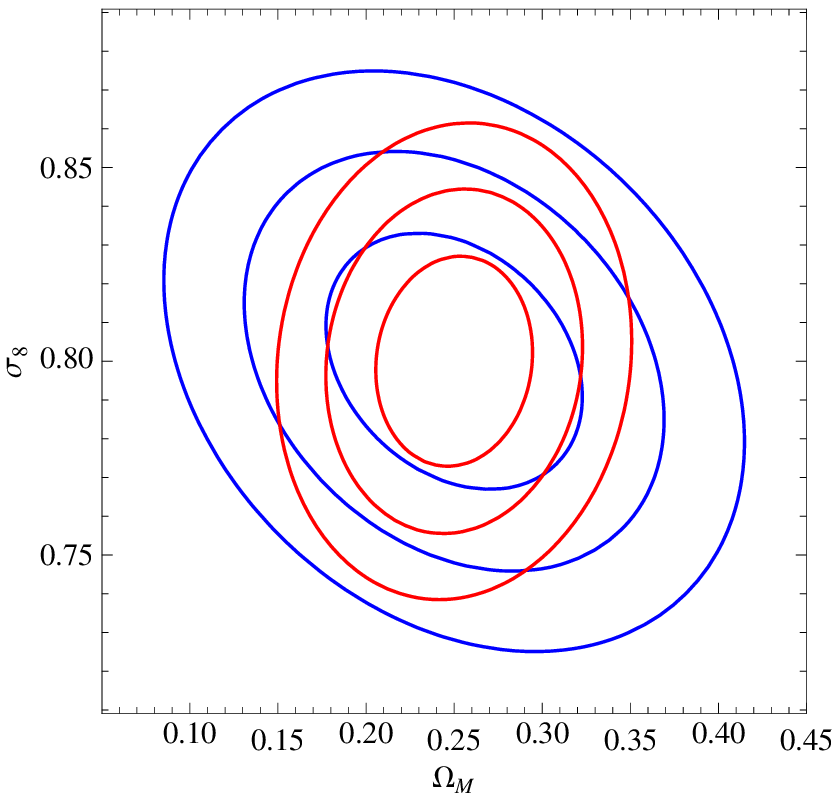}
\includegraphics[width=5.5cm]{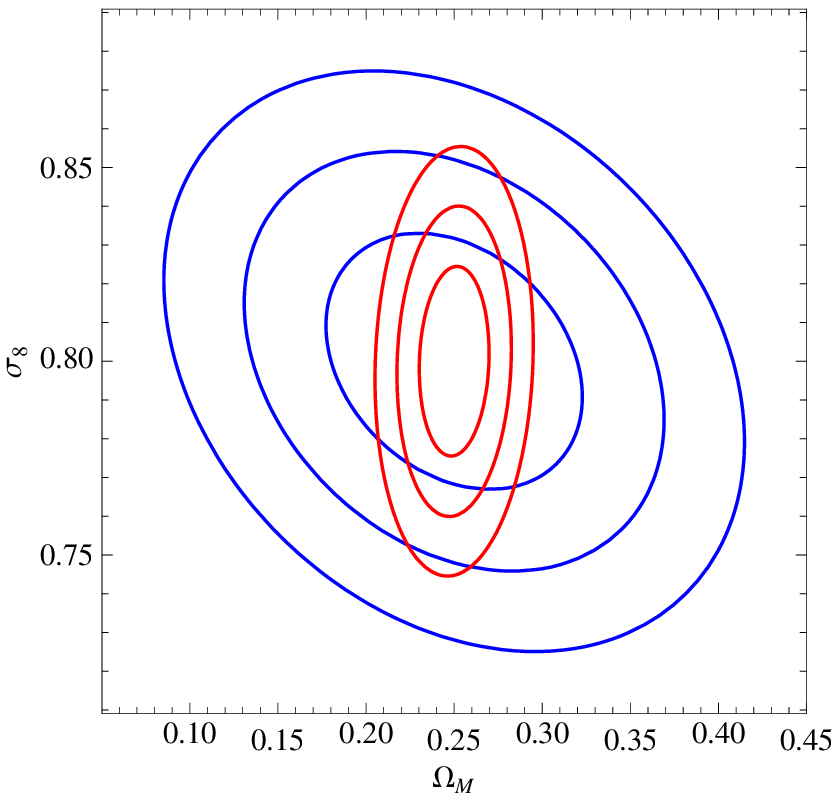}
\includegraphics[width=5.5cm]{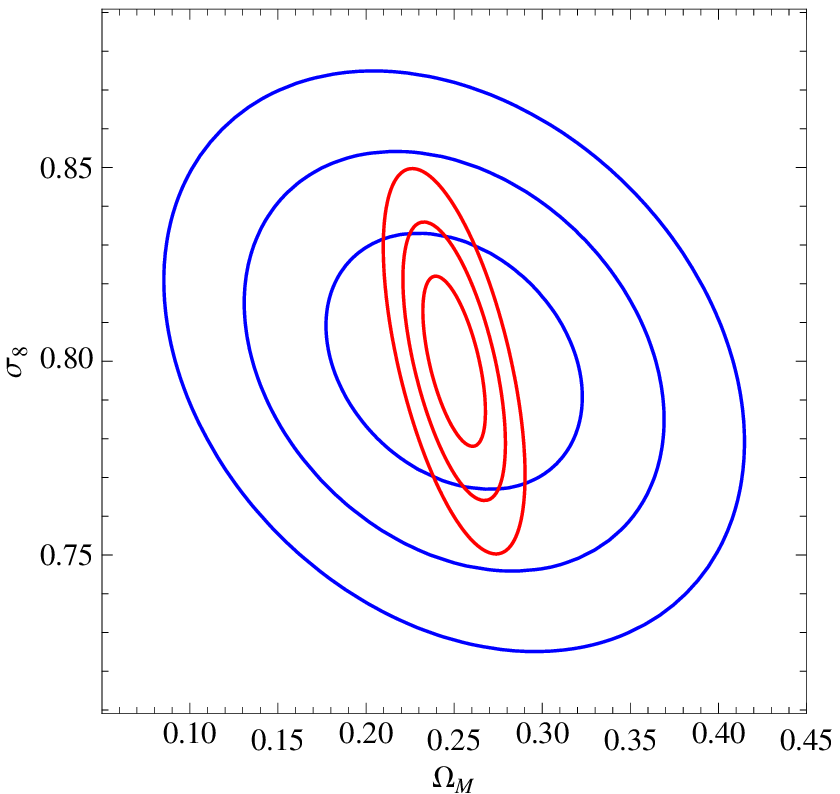} \\
\includegraphics[width=5.5cm]{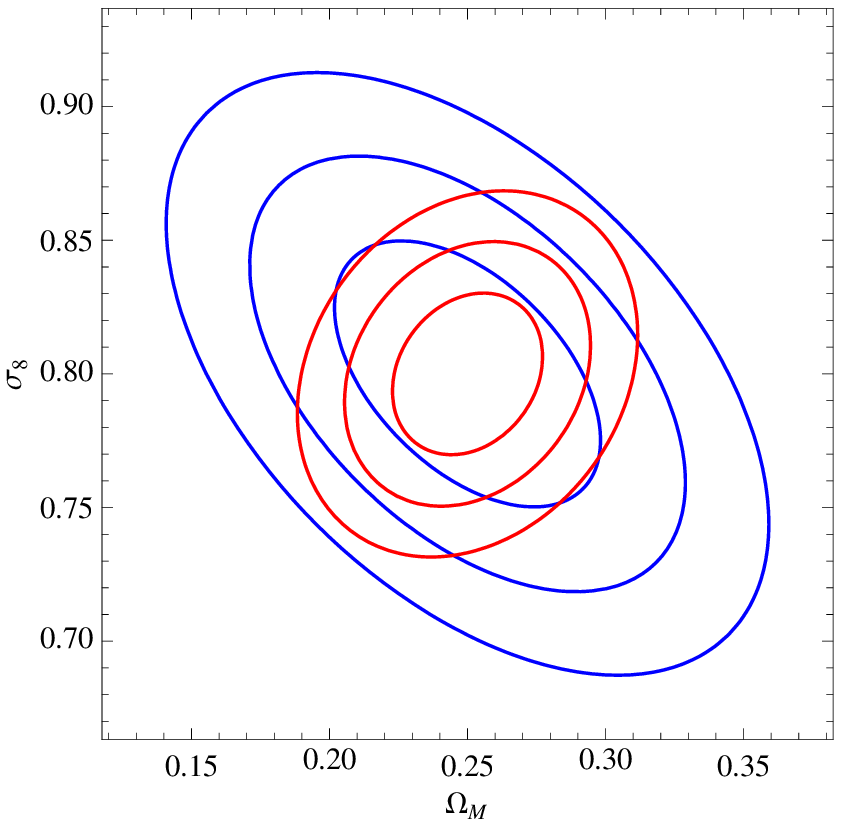}
\includegraphics[width=5.5cm]{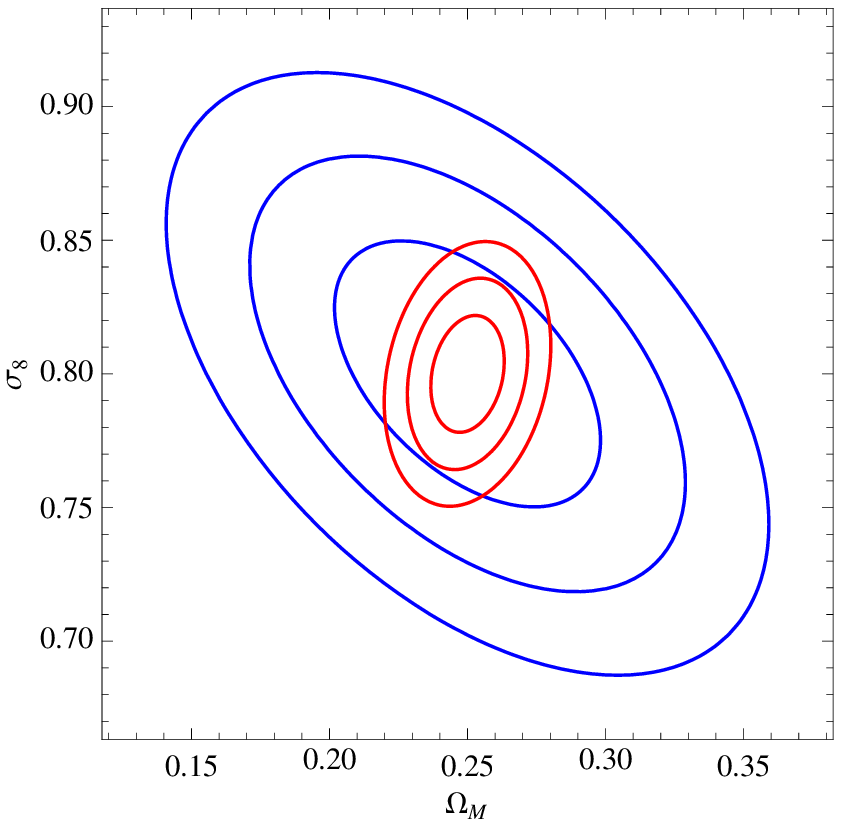}
\includegraphics[width=5.5cm]{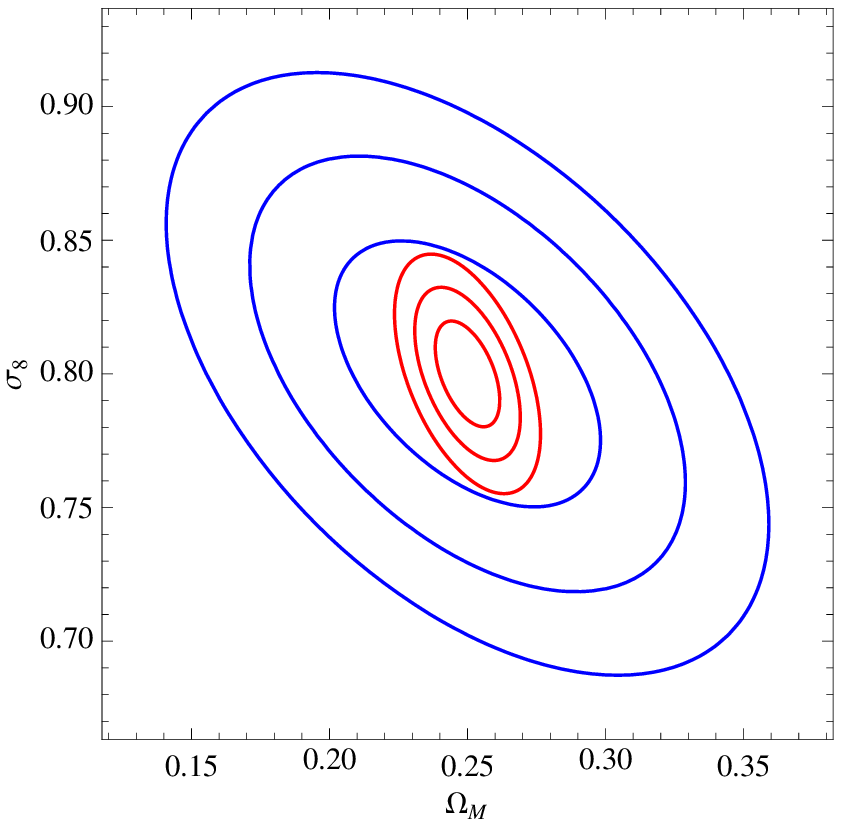} \\
\caption{$68, 95, 99\%$ confidence ranges in the $(\Omega_M, \sigma_8)$ plane as forecasted from the shear only (blue) and shear\,+\,HOM (red) Fisher matrix analysis. Top (bottom) panels refer to the case when 3 (6) redshift bins are used for tomography, while we use all HOM for the Gaussian case, the full smoothing angle range, a step $d\theta = 1 \ {\rm arcmin}$ and $\sigma_p = 1, 0.1, 0.01$ from left to right. }
\label{fig: contours}
\end{figure*}

The FoM is a global measure of the efficiency of a given set of cosmological probes to constrain the dark energy equation of state (EoS). Yet, it is interesting to look at the constraints on the single cosmological parameters and how they improve when HOM are added to shear tomography. Fig.\,\ref{fig: parratio} is the answer to this question when all HOM are combined with the six bins shear tomography considering the five different smoothing angle ranges sampled with a step $d\theta = 1 \ {\rm arcmin}$. As it is apparent, HOM are particularly efficient at reducing the errors on all parameters with a particular emphasis on EoS quantities $(w_0, w_a, w_p)$, the matter density $\Omega_M$, and the variance $\sigma_8$ of the linear matter power spectrum. It is worth noting that the range more convenient to reduce the error on a given parameter depend on the parameter itself, but the full $(2, 20)  \ {\rm arcmin}$ range is ranked first or second for all parameters. 

The reduction of the errors indeed comes from the HOM breaking the $\Omega_M$\,-\,$\sigma_8$ degeneracy. This can be efficiently seen from Fig.\,\ref{fig: contours} where blue (red) lines mark the $68, 95, 99\%$ confidence ranges in the $(\Omega_M, \sigma_8)$ plane for shear tomography with (without) the HOM contribution. It is clear that, no matter the details of the HOM configuration and whether 3 or 6 bins are used in shear tomography, the use of HOM definitely changes the orientation of the contours rotating them so that they become more parallel the coordinates axes. This is indeed what one expects when the $\Omega_M$\,-\,$\sigma_8$ degeneracy is broken. As a consequence, the constraints on $(\Omega_M, \sigma_8)$ improves which then start a sort of {\it chain reactions} leading to an improvement of the constraints on the other cosmological parameters too.

All the results discussed so far refer to the case when a Gaussian filter is used to smooth the map. Adopting a top hat filter does not change the main outcome of this analysis, i.e., HOM still increases the FoM by a factor 2.5 in the worst case scenario. However, two main differences occurs. First, the best strategy still claims for the use of 2nd, 3rd, and 4th order moments, but now giving away the fourth order ones leaves the FoM ratio almost unchanged. Second, the preferred smoothing angle range is now the $(8, 18) \ {\rm arcmin}$, and large $\theta_{min}$ values are typically preferred. Such an opposite result with respect to the Gaussian case can be traced back to the systematics errors coming from the calibration of fourth order moments. Indeed, $\rho_{rms}(n = 4)$ makes 4th order moments systematics comparable to statistical errors on small smoothing scales so that one has to move to the large ones to reduce the data covariance matrix elements and, hence, the overall FoM. Contrary to what has been suggested by the use of HOM to constrain $(\Omega_M, \sigma_8)$ only, we therefore conclude that a top hat filer can still provide a valid help to increase the cosmic shear tomography FoM.

\section{Convergence HOM systematics}

The results presented up to now are implicitly based on a number of realistic yet simplifiying assumptions. There are indeed some effects that we have neglected, but that can potentially impact our analysis introducing uncorrected for systematics. Some of them are discussed below trying to be as quantitative as possible. 

\subsection{Shear multiplicative and additive bias} 

The convergene map has been reconstructed from the noisy shear data implicitly assuming that these are a faithful reproduction of the true shear field. Actually, because of imperfections in the codes used to perform galaxy shape measurements and of other instrumental related systematics, the observed shear is a biased representation of the actual one. It is common practice \cite{shearbias} to model the relation between the observed and the actual shear as linear, i.e., $\gamma_{obs} = (1 + m) \gamma + c$, with $(m, c)$ the multiplicative and additive bias (to be not confused with the same quantities we have introduced in Sect.\,II). Considering that shear is a two components field, the $i$\,-\,th component will read

\begin{displaymath}
\gamma_{obs,i}(\vartheta_1, \vartheta_2) = [1 + m + m_i(\vartheta_1, \vartheta_2)] \gamma_1(\vartheta_1, \vartheta_2) + 
c + c_i(\vartheta_1, \vartheta_2)
\end{displaymath}
where $(\vartheta_1, \vartheta_2)$ are coordinates on the sky, and, without loss of generality, we have split the additive and multiplicative bias as the sum of a constant part common to both components and a varying part characteristic of each component. Moving to the Fourier space, we get

\begin{displaymath}
\hat{\gamma}_{obs,i}(\ell_1, \ell_2) = (1 + m) \hat{\gamma}_i(\ell_1, \ell_2) + c + \hat{{\cal{G}}}_i(\ell_1, \ell_2) + 
\hat{c}_i(\ell_1, \ell_2) \ ,
\end{displaymath}
with $(\ell_1, \ell_2)$ the component of the wavenumber vector, $\hat{f}(\ell_1, \ell_2) = {\cal{F}}[f(\vartheta_1, \vartheta_2)]$, and ${\cal{F}}$ the Fourier transform operator whose properties we have used to get the above relation. In particular, the function $\hat{{\cal{G}}}_i(\ell_1, \ell_2)$ is the convolution of the Fourier transfroms of the $i$\,-\,th components of the shear and multiplicative bias. According to the KS93 algorithm, Fourier transform of the convergence field will then be given by

\begin{displaymath}
\hat{\kappa}_{obs}(\ell_1, \ell_2) = \frac{\ell_1^2 - \ell_2^2 + 2 {\rm i} \ell_1 \ell_2}{\ell_1^2 + \ell_2^2} \ 
\hat{\gamma}_{obs} \ .
\end{displaymath}
Making some simple algebra, moving back to the real space and adding noise, the observed convergence field will finally read

\begin{eqnarray}
\kappa_{obs}(\vartheta_1, \vartheta_2) & = & (1 + m) \kappa_E(\vartheta_1, \vartheta_2) + c + {\cal{N}}(\vartheta_1, \vartheta_2) \nonumber \\
 & + & \kappa_{Emul}(\vartheta_1, \vartheta_2) + \kappa_{Eadd}(\vartheta_1, \vartheta_2) \nonumber \\
 & + &  {\rm i} \kappa_{B,tot}(\vartheta_1, \vartheta_2) 
\label{eq: kappaobssys}
\end{eqnarray}
where the real $E$ terms are given by 

\begin{displaymath}
\kappa_X(\vartheta_1, \vartheta_2) = {\cal{F}}^{-1}[\hat{\kappa}_X(\ell_1, \ell_2)]
\end{displaymath}
with $X = (E, Emul, Eadd)$ and

\begin{displaymath}
\hat{\kappa}_X(\ell_1, \ell_2) = 
\frac{\ell_1^2 - \ell_2^2}{\ell_1^2 + \ell_2^2} \hat{f}_{1}(\ell_1, \ell_2) 
- \frac{2 \ell_1 \ell_2}{\ell_1^2 + \ell_2^2} \hat{f}_{2}(\ell_1, \ell_2) 
\end{displaymath}
and $(\hat{f}_1, \hat{f}_2) = (\hat{\gamma}_{obs,1}, \hat{\gamma}_{obs,2}), (\hat{{\cal{G}}}_1, \hat{{\cal{G}}}_2), (\hat{c}_1, \hat{c}_2)$ for $X = E, Emul, Eadd$, respectively. A similar expression holds for the imaginary $B$ component which will still be the sum of three terms each one being the Fourier antitransform of a function having the following structure

\begin{displaymath}
\hat{\kappa}_Y(\ell_1, \ell_2) = 
\frac{\ell_1^2 - \ell_2^2}{\ell_1^2 + \ell_2^2} \hat{f}_{2}(\ell_1, \ell_2) 
+ \frac{2 \ell_1 \ell_2}{\ell_1^2 + \ell_2^2} \hat{f}_{1}(\ell_1, \ell_2)  \ .
\end{displaymath}
with $Y = B, Bmul, Badd$ and $(\hat{f}_1, \hat{f}_2)$ the same as before. 

Eq.(\ref{eq: kappaobssys}) reduces to our starting relation (\ref{eq: kappaobs}) when one assumes that the $B$ component vanishes, and the multiplicative and additive bias are not present so that both $\kappa_{Emul}$ and $\kappa_{Eadd}$ are identically null. Shear systematics therefore have a double effect. First, they introduce a further $B$ term so that their presence can be detected by the non vanishing of this part. Second, they add two further terms to the real $E$ part thus making the observed convergence field a biased reproduction of the actual one even if one assumes $m = c = 0$. Dropping the $E$ label for shortness, taking only the real part and subtracting the mean value, we then have 

\begin{equation}
\tilde{\kappa}_{obs} = (1 + m) \kappa + {\cal{N}} + \tilde{\kappa}_{mul} + \tilde{\kappa}_{add}
\label{eq: kappatildesys}
\end{equation}
with $\tilde{\kappa}_{mul} = \kappa - \langle \kappa_{mul} \rangle$ and $\tilde{\kappa}_{add} = \kappa - \langle \kappa_{add} \rangle$. Note that we can not assume a priori that the two systematics terms $(\kappa_{mul}, \kappa_{add})$ have null average so that they still enter $\tilde{\kappa}$ contrary to the constant offset term $c$. Eq.(\ref{eq: kappatildesys}) is the same as Eq.(\ref{eq: kappatilde}) provided one redefines the noise term as 

\begin{displaymath}
\tilde{{\cal{N}}} = {\cal{N}} + \tilde{\kappa}_{mul} + \tilde{\kappa}_{add} \ . 
\end{displaymath}
Under the reasonable assumption that the multiplicative and additive shear bias do not correlate with the signal and the noise, it is straightforward to show that one can repeat the full derivation of the calibration relations (\ref{eq: cal2nd})\,-\,(\ref{eq: cal4th}) carried out in Sect.\,II provided then noise moments are replaced with those of the effective noise ${\tilde{N}}$ which are a combination of the moments of the actual noise field ${\cal{N}}$ and the two systematics fields $(\tilde{\kappa}_{mul}, \tilde{\kappa}_{add})$. 

We therefore safely conclude that the presence of shear systematics has the only impact to change the nuisance parameters of the HOM calibration, but does not alter the main conclusion that HOM can be calibrated and hence used as cosmological tools. 

\subsection{Intrinsic alignment}

Even under the idealized conditions of no systematics from shape measurement, the observed shear can still be biased because of intrinsic alignment \cite{IA}. In the weak lensing limit, the observed ellipitcity $e$ of a galaxy is the sum of the intrinsic ellipticity $e_{int}$ and the shear $\gamma$, i.e., $e = e_{int} + \gamma$. Averaging over a large number of galaxies, one gets a shear estimator as $\gamma = \langle e \rangle$ under the assumption that galaxy are randomly oriented. Actually, beacause of intrinsic alignment (hereafter, IA), one has $\langle e_{int} \rangle \ne 0$ so that the observed shear is $\gamma_{obs} = \gamma + \gamma_{IA}$. It is straightforward to undestand that the $\gamma_{IA}$ terms originates an additive contribution in the convergence which now reads

\begin{displaymath}
\kappa_{obs} = (1 + m) \kappa  + {\cal{N}} + (\kappa_{IA} - \langle \kappa_{IA} \rangle)
\end{displaymath}
where we have already subtracted the mean value over the map so that it is $\langle \kappa_{obs} \rangle = 0$. Note that one can not assume a priori that $\langle \kappa_{IA} \rangle = 0$ so that we should carry this term along the derivation. For shortness of notation, we will however denote with $\kappa_{IA}$ the last term in the above sum. 

One can now follow the same steps carried on in Sect.\,II to derive the calibration relations (\ref{eq: cal2nd})\,-\,(\ref{eq: cal4th}), but there is now an important difference. While we can still assume that $\kappa$ and ${\cal{N}}$ are uncorrelated, this is not true for $\kappa$ and $\kappa_{IA}$ so that the expectation value of terms as $\kappa^n \kappa_{IA}^m$ has to be taken into account. Under the still realistic assumption that $\kappa_{IA}$ and ${\cal{N}}$ are uncorrelated, we finally end up with the following extended calibration relations\,:

\begin{eqnarray}
\langle \kappa_{obs}^2 \rangle & = &  (1 + m_2) \langle \kappa^2 \rangle + c_2 + \langle {\cal{N}}^2 \rangle \nonumber \\
 & + & 2(1 + m) \langle \kappa \kappa_{IA} \rangle \nonumber \\ 
 & + & \langle \kappa_{IA}^2 \rangle \ ,
\label{eq: cal2ndIA}
\end{eqnarray}

\begin{eqnarray}
\langle \kappa_{obs}^3 \rangle & = & (1 + m_3) \langle \kappa^3 \rangle + c_3 + \langle {\cal{N}}^3 \rangle \nonumber \\
 & + & 3(1 + m)^2 \langle \kappa^2 \kappa_{IA} \rangle + 3 (1 + m) \langle \kappa \kappa_{IA}^2 \rangle \nonumber \\
 & + & \langle \kappa_{IA}^3 \rangle \ ,
\label{eq: cal3rdIA}
\end{eqnarray}

\begin{eqnarray}
\langle \kappa_{obs}^4 \rangle & = & (1 + m_4) \langle \kappa^4 \rangle +  c_4 + \langle {\cal{N}}^4 \rangle  \nonumber \\
 & + & 6 [(1 + m_2) \langle \kappa^2 \rangle + c_2] \langle {\cal{N}}^2 \rangle \nonumber \\ 
 & + &  4(1 + m)^3 \langle \kappa^3 \kappa_{IA} \rangle + 4(1 + m) \langle \kappa \kappa_{IA}^3 \rangle \nonumber \\
 & + &  6(1 + m)^2 \langle \kappa^2 \kappa_{IA}^2 \rangle + 6 \langle \kappa_{IA}^2 \rangle \langle {\cal{N}}^2 \rangle \nonumber \\
 & + & \langle \kappa_{IA}^4  \rangle \ .  
\label{eq: cal4thIA}
\end{eqnarray}
For $\kappa_{IA}  = 0$, Eqs.(\ref{eq: cal2ndIA})\,-\,(\ref{eq:  cal4thIA}) reduce to the calibration relations (\ref{eq: cal2nd})\,-\,(\ref{eq:  cal4th}) we have used throughout the paper. When IA is present, additional terms appear which have to be dealt with. Some possible strategies are outlined below. 

\begin{itemize}

\item[i.)]{As a brute force approach, one can consider the additional terms as further nuisance parameters to be determined from the fit itself. Although this is a legitimate procedure, it is likely quite hard to actually implement. Indeed, the dimension of the nuisance vector ${\bf p}_n$ becomes dramatically close to that of the data vector thus forcing to use a smaller sampling step $d\theta$. Such a choice, however, will likely push the correlation matrix elements closer to 1 thus reducing the effective number of quasi\,-\,independent constraints. Second, the more the number of fitting parameters, the more the possibility of degeneracies among them. Both effects will operate to increase the forecasted errors on cosmological parameters thus making HOM less and less useful even when combined with shear tomography. Moreover, in a Fisher matrix forecast analysis, one should also find out fiducial values for the additional nuisance parameters which are actually hard to set.}

\item[ii.)]{To the opposite extreme, one can note that IA is a small scale effect so that its impact will be reduced when smoothing with a large enough filter aperture. One can therefore suppose that the smoothing procedure completely washes out the effect so that the terms related to IA in Eqs.(\ref{eq: cal2ndIA})\,-\,(\ref{eq: cal4thIA}) are negligible or can be compensated by a change in the remaining nuisance parameters. Ideally, one could also look for a suitably defined filter which makes this assumption comes true. Whether this is the case or not is, however, hard to say given the present day lack of any study of the impact of IA on map reconstruction.}

\item[iii.)]{Rather than considering the IA related terms as quantities to be fitted for, one can reduce the dimensionality of the problem by estimating them theoretically. To this end, one can consider the effective IA convergence field $\kappa_{IA}$ as an actual one and rely on the formalism developed in \cite{MJ01} to compute both the moments of $\kappa_{IA}$ and the expectation value of the cross terms $\langle \kappa^n \kappa_{IA}^m \rangle$. One should then replace the matter power spectrum $P(k, z)$ in Eqs.(\ref{eq: defkappatheta0}) with the IA power spectrum taking one of the models available in the literature \cite{IAmodels}. These are assigned by some characteristic parameters which then adds to the rest of nuisance parameters. Such an approach allows not only to reduce the dimension of ${\bf p}_n$, but makes the IA terms dependent on the cosmological parameters too thus adding further information in the HOM. However, to the best of our knwoledge, a theoretical estimate of the IA moments with the formalism of \cite{MJ01} has never been attempted nor validated so that it is still premature to judge whether such a theory based treatment of the impact of IA on HOM is viable or not.}

\end{itemize}
Which of the above strategy is best suited to tackle down the impact of IA on HOM is not possible to say at the moment of writing. Indeed, lensing lightcone simulations including IA in the shear field are needed to perform map reconstruction and HOM measurement and calibration and to test a theoretical approach to the IA moments estimate. We are nevertheless confident that IA systematics can be consistently included in the HOM framework as soon as suitable simulated dataset will be available.

\subsection{Measurement noise and denoising}

Eq.(\ref{eq: kappaobs}) has been taken as our starting point, but we have actually not shown that it actually holds since its derivation is straightforward. We have then included noise term in our simulated data by adding a random ellipticity to the value of the shear in each position of the map to take care of the non vanishing dispersion of the intrinsic ellipiticity distribution. It is worth noting, however, that this is not the only source of noise. One should indeed add the component coming from measurement errors typically related to instrumental effects (e.g., shape measurement will be degraded close to the edge of the image). The details of how to take care of this further component depend on the survey at hand, but some general comment are nevertheless possible and will be sketched below. 

To this end, we can note that the noise field may be considered as an additional field which enters the convergence map reconstruction. But this is exactly the same situation we have already dealt with in the previous paragraph for IA. We can therefore repeat the same derivation as above, but with the further simplification that this noise field can be considered uncorrelated from the convergence field. Eqs.(\ref{eq: cal2ndIA})\,-\,(\ref{eq: cal4thIA}) the reduces to

\begin{equation}
\langle \kappa_{obs}^2 \rangle = (1 + m_2) \langle \kappa^2 \rangle + c_2 + \langle {\cal{N}}^2 \rangle
+ \langle {\cal{M}}^2 \rangle \ ,
\label{eq: cal2ndNoise}
\end{equation}

\begin{equation}
\langle \kappa_{obs}^3 \rangle = (1 + m_3) \langle \kappa^3 \rangle + c_3 + \langle {\cal{N}}^3 \rangle 
+ \langle {\cal{M}}^3 \rangle \ ,
\label{eq: cal3rdNoise}
\end{equation}

\begin{eqnarray}
\langle \kappa_{obs}^4 \rangle & = & (1 + m_4) \langle \kappa^4 \rangle +  c_4 + \langle {\cal{N}}^4 \rangle  \nonumber \\
 & + & 6 [(1 + m_2) \langle \kappa^2 \rangle + c_2] \langle {\cal{N}}^2 \rangle \nonumber \\ 
 & + &  6[(1 + m)^2 \langle \kappa^2 \rangle + c_2] \langle {\cal{M}}^2 \rangle \nonumber \\
 & + & 6 \langle {\cal{M}}^2 \rangle \langle {\cal{N}}^2 \rangle + \langle {\cal{M}}^4  \rangle \ ,  
\label{eq: cal4thNoise}
\end{eqnarray}
where we have denoted with ${\cal{M}}$ the additional noise field after subtracting its average value. Eqs.(\ref{eq: cal2ndNoise})\,-\,(\ref{eq: cal4thNoise}) are actually the same as our standard calibration relations (\ref{eq: cal2nd})\,-\,(\ref{eq: cal4th}) provided defines a total noise field ${\cal{N}}_{tot} = {\cal{N}} + {\cal{M}}$ so that all the results we have discussed in previous sections still hold. As only difference, we get a increase in the magnitude of the noise moments. A second somewhat obvious difference will be the S/N ratio of the measured moments that will be smaller by an amount which depends on the ratio ${\cal{M}}/{\cal{N}}$ hence on the survey details. 

One can also wonder whether it is possible to fully remove the contribution of the noise moments to improve the convergence HOM measurements. Such a procedure has been suggested in \cite{Ludo2013} who have argued that a better estimate of the HOM is provided by the corrected moments

\begin{equation}
\langle \kappa_{corr}^2 \rangle = \langle \kappa_{obs}^2 \rangle - \overline{\langle \kappa_{rnd}^2 \rangle}
\label{eq: mu2denoise}
\end{equation}

\begin{equation}
\langle \kappa_{corr}^3  \rangle = \langle \kappa_{obs}^3 \rangle \ ,
\label{eq: mu3denoise}
\end{equation}

\begin{equation}
\langle \kappa_{corr}^4 \rangle = \langle \kappa_{obs}^4 \rangle 
- 6 \overline{\langle \kappa_{obs}^2 \rangle \langle \kappa_{rnd}^2 \rangle} 
-\overline{\langle \kappa_{rnd}^4 \rangle} \ ,
\label{eq: mu4denoise}
\end{equation}
where quantities with the label $rnd$ are measured over noise only maps, and the overline denotes average over a large number of noise only maps realizations.

In order to test whether this denoising procedure works, we first build 100 noise only shear maps keeping fixed both position and shape of galaxies on every $25 \ {\rm deg}^2$ patch,  but randomize their orientation. For a given filter and smoothing length, we then use these maps to estimate the terms entering Eqs.(\ref{eq: mu2denoise}) and (\ref{eq: mu4denoise}). Note that we are here considering only the intrinsic ellipticity noise (i.e., we set ${\cal{N}}_{tot} = {\cal{N}}$), but the results of this test should qualitatively hold whatever kind of noise is actually affecting the input shear map.

If denoising is actually working, we should expect that the corrected moments match the those measured on the noiseless convergence map. We therefore also perform map reconstruction from an idealized shear map with no noise. We then compare the 2nd and 4th order corrected moments\footnote{We do not test 3rd order moments since the denoising procedure does not apply any correction to them.} with those measured on this map to finally get

\begin{displaymath}
\overline{\langle \kappa_{corr}^2 \rangle/\langle \kappa_{free}^2 \rangle} = 0.88 \pm 0.01 \ ,
\ \overline{\langle \kappa_{corr}^4 \rangle/\langle \kappa_{free}^4 \rangle} = 0.17 \pm 0.02 
\end{displaymath}
with quantities with the label $free$ measured on the noise free map. These numbers clearly point at a failure of the denoising procedure since one should actually find values of order unity. On the contrary, such low numbers are an evidence that denoising is removing both the noise and (part of) the signal. As a possible way out, one could try to empirically modify Eqs.(\ref{eq: mu2denoise})\,-\,(\ref{eq: mu4denoise}) adding some multiplicative factors to the correction terms and tailoring them so that the noiseless moments are exactly recovered. However, this would ask for a large number of both noise and signal simulations to be sure that these corrective factors do not depend on the noise properties and the cosmological parameters, but are rather universal. This is outside our aims here so that we conservatively argue against using denoising in moments estimate.

\begin{figure*}
\centering
\includegraphics[width=4.0cm]{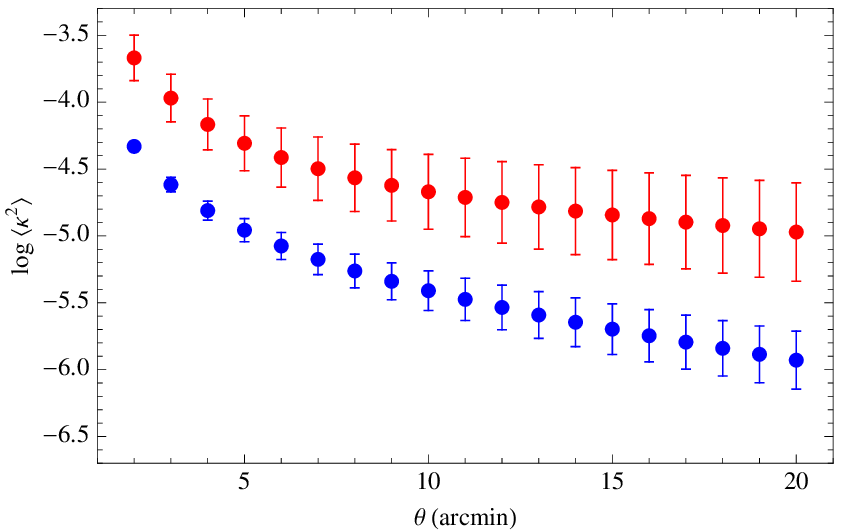}
\includegraphics[width=4.0cm]{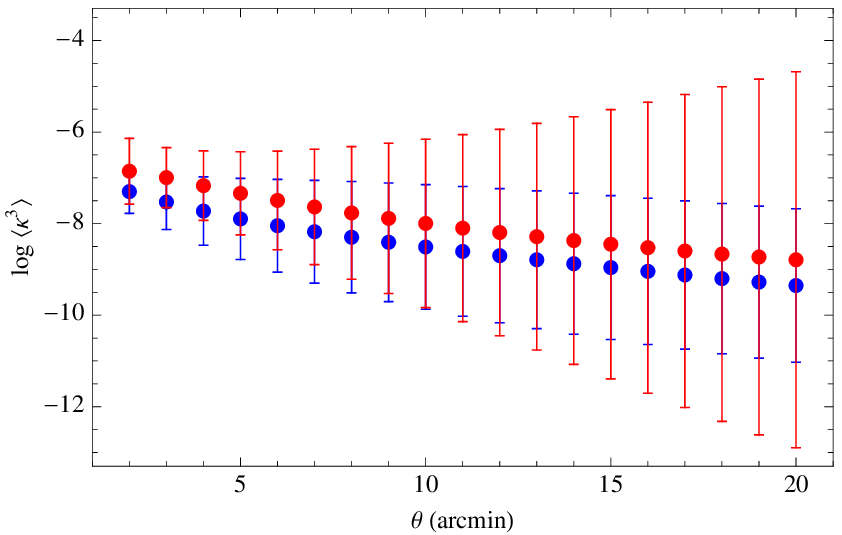}
\includegraphics[width=4.0cm]{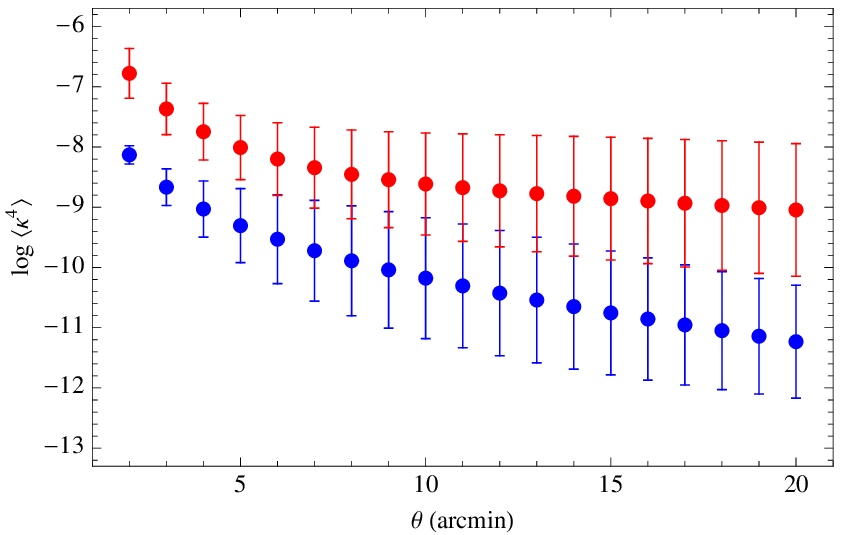}
\includegraphics[width=4.0cm]{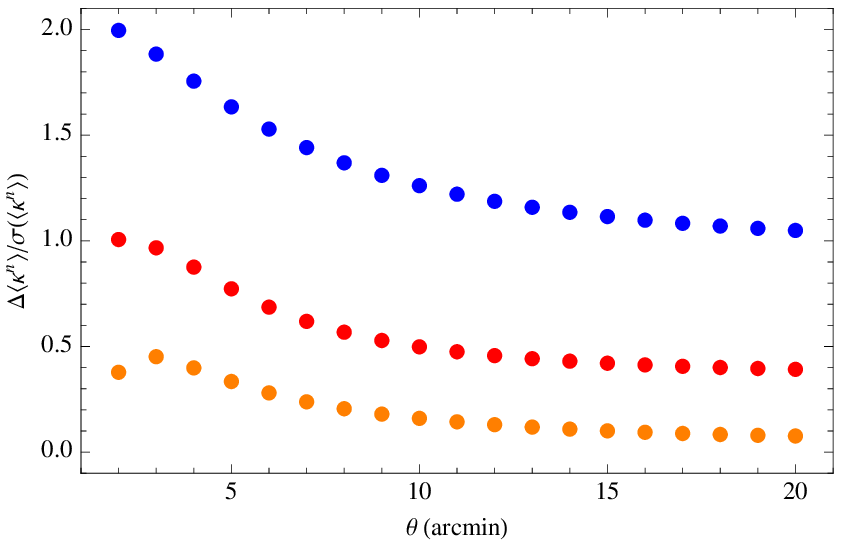} \\
\caption{Comparing convergence HOM measured on maps reconstructed from unmasked (blue) and masked (red) shear data. First three panels show the 2nd, 3rd, and 4th order moments, while the fourth one plots the difference between the two datasets normalized with respect to the statistical error (with blue, orange, red points referring to 2nd, 3rd, 4th order moments).}
\label{fig: maskmom}
\end{figure*}

\subsection{Masking}

Camera CCD defects, bright stars saturating the image, satellite trails, and various spurious artifacts must be masked out before any weak lensing analysis. On the contrary, we have here assumed that no masking is needed so that the full subfield area can be used to convert the shear field into the convergence map though the KS93 method. Actually, this is far from be true in realistic applications so that it is worth wondering whether and to which extent masking impacts HOM estimate. 

To this aim, one can first note that moments describe the shape of the convergence pdf so that they are global quantities. Unless masking preferentially selects areas with a given $\kappa$ value, it is therefore expected that the $\kappa$ distribution is overall unchanged so that HOM are unaffected by masking. Such a qualitative discussion would actually hold if masking were performed on the convergence map itself. On the contrary, masking is performed on the shear field introducing holes which must be filled somewhat in order to run the KS93 algorithm. Different techniques \cite{Ludo2013,Pires2009} have been developed to tackle this problem with the final aim of making the reconstructed convergence map as similar as possible to the actual one. As far as this goal is reached, we can still assume that the global properties of the convergence field are unaffected hence leading to no bias on the HOM.

In order to see whether this is indeed the case, we carry on a preliminary analysis using the filling form method already applied to reconstruct the convergence field from CFHTLensS data \cite{Ludo2013}. As a first step, we construct a simple mask to be applied on the catalog shear data by randomly positioning circles with radii between $0.18 \ {\rm arcsec}$ to $1 \ {\rm arcmin}$ in such a way that the total masked area is $20\%$. We generate this kind of mask for each of the 140 shear fields cutted from the MICECAT catalog and input them to the map reconstruction code. Moments are  then estimated from these reconstructed maps getting the results shown in Fig.\,\ref{fig: maskmom}. Note that this figure refers to moments estimated from a single subfield since this is the area used to generate the mask.

It is apparent that HOM from maps reconstructed from masked data are only shifted upwards with respect to the unmasked case with the shift being more or less statistically meaningful depending on the moment order. Such results are not fully unexpected and can be qualitatively explained as follows. First, we note that the second order moment is increased as a result of the smaller number of cells used to compute the convergence pdf distributions which reduces the statistics thus increasing the variance of the distribution becuause of the larger impact of the noise. This, however, does not significantly change the symmetry properties of the distribution so that the 3rd order moment is the least affected. On the contrary, the 4th order moment is increased as a consequence of the increase of the 2nd order one. Indeed, should the distribution be Gaussian, a factor 2 boost in the 2nd order moment translates into a factor 4 boost in the 4th order one. Although the convergence pdf is not Gaussian, the combination with the Gaussian noise makes this effect still partially at work thus motivating the increase we find. 

What is important to stress, however, is that masking does not invalidate our calibration approach. It is indeed still possible to rely on Eqs.(\ref{eq: cal2nd})\,-\,(\ref{eq: cal4th}) which is our fundamental assumption for the use of HOM as cosmological probes. Masking will change the values of the calibration parameters, but these quantities are actually of no interest for cosmology. Moreover, it is also worth noting that the use of inpainting techniques will likely reduce the difference among maps reconstructed from masked and unmasked data thus making the difference in calibration parameter likely negligible. While this is a point we will address in a separate paper, we can nevertheless safely conclude that masking introduces systematics in the analysis which do not bias the estimate of the cosmological parameters.

\section{Conclusions}

More and better data does not only imply more and better statistics. Improved observational techniques and wider area are two promises that ongoing and future lensing surveys will respect thus making available large dataset of unprecedented quality data. Rather than limiting ourselves to do better what has already been done up to now, one can think to take the next step forward. From the point of view of lensing studies, this means moving away from second order statistics (two points correlation function and cosmic shear tomography) and consider higher order estimators. Motivated by this idea, we have here explored the use of higher order moments of the convergence field to increase the FoM of lensing surveys. 

In order to make HOM viable cosmological tools, one has first to show that it is possible to match theory and observations, i.e., that reliable calibration relations exist so that the predictions for a given set of cosmological parameters are in agreement with the moments estimated from the corresponding convergence maps. Eqs.(\ref{eq: cal2nd})\,-\,(\ref{eq: cal4th}) are our answer to this question making it possible to connect the theoretical predictions from Eqs.(\ref{eq: kappa2nd})\,-\,(\ref{eq: kappa4th}) to the HOM estimated from the MICECAT simulated catalog data. The successful comparison with simulations have validated our approach so that we can safely rely on it to use HOM as cosmological probes.

A Fisher matrix analys has then shown that, while as standalone probe moments are unable to put meaningful constraints on $(\Omega_M, \sigma_8)$, the joint use of shear tomography and HOM can boost the FoM by up to a factor 14 and no less than a factor 2. Such an astonishing result is due to HOM breaking the $\Omega_M$\,-\,$\sigma_8$ degeneracy which makes it possible to strenghten the constraints on all the cosmological parameters. This degeneracy breaking takes place whatever smoothing angle range and sampling step is used, but we advocate the use of the full $(2, 20) \ {\rm arcmin}$ range with sampling $d\theta = 1 \ {\rm arcmin}$ since this choice guarantees a better statistics hence maximizing the FoM boost.

As encouraging as they are, these results have to be considered still preliminary and taken {\it cum grano salis}. First, we note that they refer to an idealized lensing survey having area, redshift distribution, and number density different from ongoing (DES, KIDS) and future (Euclid) surveys. In particular, the shear tomography only FoM is likely grossly underestimated because of the small number of redshift bins we have used. What is the impact of the number of bins can be appreciated noting that the shear FoM increases by a factor 2 when we move from 3 to 6 redshift bins. Since we have considered the FoM ratio, it is clear that, if the denominator is undersestimated, the full ratio is overestimated. It is, however, reassuring that the FoM ratio in Figs.\,\ref{fig: fomratiovsmoments} and \ref{fig: fomratiovsrange} is roughly the same between top and bottom panels suggesting that the FoM increase does not depend on the number of bins. However, changes in the redshift distribution and number density affect both shear tomography amd HOM and it is not possible to say in advance if the changes goes in the same direction thus leaving the FoM ratio unchanged. 

In order to test whether this is the case, one should repeat the analysis presented in this expoloratory paper based on simulations mimicking a given survey in terms of redshift distribution, area, and number density. A possible examples is the recently released Multidark weak lensing lightcone \cite{Gio15}. This will offer the possibility to not only extend the redshift range, but also adjust the source redshift distribution thus allowing to explore the impact of this quantity on the validity of our linear calibration relation. Moreover, being Multidark based on a different input cosmological model, we could also check whether the results in the present paper still hold when the background cosmology is changed. We stress that, in order for a such a test to be passed with green lights, it is not important that the calibration coefficients $(m, c )$ are the same since they will likely depend on both cosmology and the redshift range probed, but only that the relation between theoretical and observed moments can be approximated as a linear one with a small scatter. Although such a confirmation test is needed,  we are nevertheless confident that this is indeed the case since there is nothing in our procedure related to which cosmological model is the input for the simulation. What is the impact on the FoM ratio is, on the contrary, hard to reliably forecast.

Being a first step towards the use of HOM as supplementary cosmological probes, our analysis has been based on a number of assumptions which could be violated thus originating unaccounted for systematics. Some of them have been semi\,-\,quantitatively addressed in Sect.\,VII where we have shown that the main problems may come from IA only. How troublesome is IA in lensing analysis is, however, not an unexpected news being IA the main systematics also in standard shear tomography. As discussed in Sect.\,VIIB, taking care of IA asks for specifically designed simulations which implement it in the  shear field itself. Although unavailable at the moment of writing, work is in progress to make them available to the lensing community so that we are confident that time will come when this issue will be correctly taken into account. We note that a theoretical approach calibrated on simulations can turn IA from a problem to an opportunity since it should add further cosmology dependent terms in HOM thus increasing their constraining power and hence the joint FoM. 

As a conclusive remark, we want to stress that, as preliminary as they are, the results presented in this paper convincigly show that moving beyond second order statistics is an efficient way to deepen our understanding of the dark energy permeating the universe. It is therefore entirely possible that the wisest use of future data is not doing more of the same, but rather some of the higher.

\section*{Acknowledgments}

The authors acknowledge E. Jullo, T. Kitching, and S. Pires for comments on a preliminary version of the manuscript. We also thank F. Bernardeau for discussion on higher order perturbation theory, and C. Schimd for help with the effect of masking. We acknowledge the use of data from the MICE simulations, publicly available at {\tt http://www.ice.cat/mice}; in particular, we thank the MICE team for allowing us to use the MICECAT v2.0 catalog prior of publication. VFC and XE are funded by Italian Space Agency (ASI) through contract Euclid\,-\,IC (I/031/10/0) and acknowledge financial contribution from the agreement ASI/INAF/I/023/12/0. We acknowledge the support from the grant MIUR PRIN 2015 "Cosmology and Fundamental Physics: illuminating the Dark Universe with Euclid". MV acknowledges support from Funda\c{c}\~{a}o para a Ci\^{e}ncia e a Tecnologia (FCT) through the research grant IF/01518/2014.

\section*{Appendix A\,: calibration parameters}

We summarize here the results of the HOM calibration analysis. Tables \ref{tab: fithomnonoise} and \ref{tab: fithomnoise} report the median and the $68\%$ confidence range of the calibration parameters, and the rms of the percentage residuals for the idealized and noisy moments, respectively. Results for fits to limited smoothing angle ranges are also given. 

\begin{table*}
\begin{center}
\caption{Calibration parameters, $(m_n, c_n)$, and rms percentage residuals, $\rho_{rms} = rms[100 \times (1 - \langle \kappa^n \rangle(fit)/\langle \kappa^n \rangle(obs))]$, for different moments. Note that $c$ is in units of $(10^{-6}, 10^{-9}, 10^{-11})$ for $(\langle \kappa^2 \rangle, \langle \kappa^3 \rangle, \langle \kappa^4 \rangle)$, respectively.} 
\resizebox{18cm}{!}{
\begin{tabular}{cccccccccccccccc}
\hline \hline
\multicolumn{16}{c}{Gaussian filter} \\
\hline
$\theta_0$ range & \multicolumn{3}{c}{(2 - 20)} & \multicolumn{3}{c}{(2 - 12)} & \multicolumn{3}{c}{(4 - 14)} & \multicolumn{3}{c}{(6 - 16)} & \multicolumn{3}{c}{8 - 18} \\
\hline
$n$ & $m_n$ & $c_n$ & $\rho_{rms}$ & $m_n$ & $c_n$ & $\rho_{rms}$  & $m_n$ & $c_n$ & $\rho_{rms}$  & $m_n$ & $c_n$  & $\rho_{rms}$  & $m_n$ & $c_n$ & $\rho_{rms}$  \\
\hline
$2$ & $ 0.00 \pm 0.05$  & $-1.45 \pm 0.38$ & $4.88$ & $-0.08 \pm 0.06$  & $-1.45 \pm 0.66$ &  $3.30$ & $0.04 \pm 0.04$ & $-1.45 \pm 0.36$ & $1.68$ & $0.11 \pm 0.03$ & $-1.45 \pm 0.20$ & $0.93$ & $0.16 \pm 0.02$ & $-1.45 \pm 0.10$ & $0.50$ \\

$3$ & $0.46 \pm 0.05$ & $-2.82 \pm 0.29$ & $6.34$ & $0.48 \pm 0.08$ & $-2.82 \pm 0.12$ & $4.48$ & $0.59 \pm 0.03$ & $-2.82 \pm 0.25$ & $1.75$ & $0.54 \pm 0.05$ & $-2.82 \pm 0.36$ & $3.04$ & $0.43 \pm 0.07$ & $-2.82 \pm 0.34$ & $3.48$ \\

$4$ & $-0.19 \pm 0.13$ & $1.50 \pm 1.20$ & $21.5$ & $-0.31 \pm 0.12$ & $1.50 \pm 5.11$ & $15.9$ & $-0.04 \pm 0.10$ & $1.50 \pm 1.89$ & $ 8.07$ & $0.14 \pm 0.07$ & $1.50 \pm 0.80$ & $4.45$ & $0.25 \pm 0.05$ & $1.50 \pm 0.35$ & $2.49$ \\
\hline \hline
\multicolumn{16}{c}{Top Hat filter} \\
\hline
$\theta_0$ range & \multicolumn{3}{c}{(2 - 20)} & \multicolumn{3}{c}{(2 - 12)} & \multicolumn{3}{c}{(4 - 14)} & \multicolumn{3}{c}{(6 - 16)} & \multicolumn{3}{c}{8 - 18} \\
\hline
$n$ & $m_n$ & $c_n$ & $\rho_{rms}$ & $m_n$ & $c_n$ & $\rho_{rms}$  & $m_n$ & $c_n$ & $\rho_{rms}$  & $m_n$ & $c_n$  & $\rho_{rms}$  & $m_n$ & $c_n$ & $\rho_{rms}$  \\
\hline
$2$ & $-0.16 \pm 0.04$ & $0.74 \pm 0.65$ & $4.55$ & $-0.23 \pm 0.04$ & $0.74 \pm 0.93$ & $2.70$ & $-0.13 \pm 0.04$ & $ 0.74 \pm 0.57$ & $1.54$ & $-0.06 \pm 0.03$ & $0.74 \pm 0.41$ & $1.04$ & $0.00 \pm 0.03$ & $ 0.74 \pm 0.30$ & $0.75$ \\

$3$ & $0.30 \pm 0.09$ & $-1.46 \pm 1.94$ & $9.23$ & $0.21 \pm 0.11$ & $-1.46 \pm 5.36$ & $7.42$ & $0.42 \pm 0.06$ & $-1.46 \pm 1.93$ & $3.23$ & $0.53 \pm 0.03$ & $-1.46 \pm 0.77$ & $1.37$ & $0.57 \pm 0.01$ & $-1.46 \pm 0.19$ & $0.39$ \\

$4$ & $-0.48 \pm 0.09$ & $17.2 \pm 6.3$ & $21.2$ & $-0.55 \pm 0.08$ & $17.2 \pm 19.1$ & $14.5$ & $-0.39 \pm 0.07$ & $17.2 \pm 8.4$ & $8.44$ & $-0.26 \pm 0.06$ & $17.2 \pm 4.3$ & $5.47$  & $-0.15 \pm 0.05$ & $17.2 \pm 2.4$ & $3.70$ \\
\hline \hline
\end{tabular}}
\label{tab: fithomnonoise}
\end{center}
\end{table*}

\begin{table*}
\begin{center}
\caption{Same as Table\,\ref{tab: fithomnonoise} but for noisy moments. For each moment, there is now one further parameter, namely the order $n$ moment of the noise for a smoothing scale $\theta_0 = 1 \ {\rm arcmin}$. Note that $(c_2, c_3, c_4)$ are in units of $(10^{-6}, 10^{-9}, 10^{-11})$, while $(\nu_2, \nu_3, \nu_4)$ are in units of $(10^{-4}, 10^{-10}, 10^{-10})$, respectively.}
\resizebox{18cm}{!}{
\begin{tabular}{ccccccccccccccccccccc}
\hline \hline
\multicolumn{21}{c}{Gaussian filter} \\
\hline
$\theta_0$ range & \multicolumn{4}{c}{(2 - 20)} & \multicolumn{4}{c}{(2 - 12)} & \multicolumn{4}{c}{(4 - 14)} & \multicolumn{4}{c}{(6 - 16)} & \multicolumn{4}{c}{8 - 18} \\
\hline
$n$ & $m_n$ & $c_n$ & $\nu_n$ & $\rho_{rms}$ & $m_n$ & $c_n$ & $\nu_n$ & $\rho_{rms}$  & $m_n$ & $c_n$ & $\nu_n$ & $\rho_{rms}$  & $m_n$ & $c_n$ & $\nu_n$  & $\rho_{rms}$  & $m_n$ & $c_n$ & $\nu_n$ & $\rho_{rms}$  \\
\hline

$2$ & 
$-0.19_{-0.25}^{+0.28}$ & $-0.69_{-1.27}^{+1.35}$ & $0.21_{-0.21}^{+1.13}$ & $7.83$ & 
$-0.12_{-0.24}^{+0.27}$ & $-1.03_{-1.75}^{+2.14}$ & $0.14_{-0.14}^{+0.88}$ & $6.72$ &
$-0.25_{-0.22}^{+0.22}$ & $-0.72_{-1.11}^{+1.23}$ & $0.10_{-0.10}^{+0.97}$ & $4.44$ &
$-0.32_{-0.29}^{+0.34}$ & $-0.28_{-0.57}^{+0.78}$ & $0.15_{-0.15}^{+1.48}$ & $2.19$ &
$-0.42_{-0.25}^{+0.31}$ & $-0.10_{-0.55}^{+0.50}$ & $0.13_{-0.13}^{+1.5}$ & $3.03$ \\

$3$ &
$-0.53_{-0.10}^{+0.38}$ & $-0.61_{-0.52}^{+0.71}$ & $0.30_{-2.16}^{+1.97}$ & $6.10$ &
$-0.56_{-0.07}^{+0.27}$ & $-0.32_{-0.66}^{+0.72}$ & $0.12_{-4.97}^{+3.82}$ & $5.85$ &
$-0.53_{-0.09}^{+0.16}$ & $-0.78_{-0.50}^{+0.82}$ & $-0.35_{-1.16}^{+1.72}$ & $1.44$ &
$-0.56_{-0.11}^{+0.20}$ & $-0.48_{-0.56}^{+0.63}$ & $0.41_{-1.54}^{+1.07}$ &  $2.36$ &
$-0.59_{-0.12}^{+0.33}$ & $-0.41_{-0.42}^{+0.54}$ & $-0.07_{-0.67}^{+0.74}$ & $3.01$ \\

$4$ & 
$-0.11_{-0.33}^{+0.34}$ & $-0.10_{-3.15}^{+3.06}$ & $2.24_{-2.24}^{126}$ & $53.3$ &
$-0.14_{-0.30}^{+0.35}$ & $-0.04_{-6.14}^{+6.00}$ & $0.54_{-0.54}^{+6.85}$ & $24.6$ &
$-0.21_{-0.30}^{+0.37}$ & $0.33_{-2.33}^{+2.11}$ & $2.86_{-2.86}^{+205}$ & $11.6$ &
$-0.18_{-0.34}^{+0.36}$ & $-0.24_{-0.90}^{+1.07}$ & $1.07_{-1.07}^{+282}$ & $12.1$ &
$-0.24_{-0.33}^{+0.33}$ & $-0.07_{-0.66}^{+0.66}$ & $5.46_{-5.46}^{+431}$ & $4.18$ \\

\hline \hline
\multicolumn{21}{c}{Top Hat filter} \\
\hline
$\theta_0$ range & \multicolumn{4}{c}{(2 - 20)} & \multicolumn{4}{c}{(2 - 12)} & \multicolumn{4}{c}{(4 - 14)} & \multicolumn{4}{c}{(6 - 16)} & \multicolumn{4}{c}{8 - 18} \\
\hline
$n$ & $m_n$ & $c_n$ & $\nu_n$ & $\rho_{rms}$ & $m_n$ & $c_n$ & $\nu_n$ & $\rho_{rms}$  & $m_n$ & $c_n$ & $\nu_n$ & $\rho_{rms}$  & $m_n$ & $c_n$ & $\nu_n$  & $\rho_{rms}$  & $m_n$ & $c_n$ & $\nu_n$ & $\rho_{rms}$  \\
\hline

$2$ & 
$-0.03_{-0.24}^{+0.17}$ & $-1.67_{-2.71}^{+2.98}$ & $1.17_{-1.15}^{+3.25}$ & $ 8.56$ &
$-0.06_{-0.23}^{+0.28}$ & $-1.87_{-4.27}^{+5.50}$ & $1.42_{-1.40}^{+3.12}$ & $4.92$ &
$0.03_{-0.22}^{+0.13}$ & $-1.79_{-2.41}^{+2.75}$ & $0.51_{-0.50}^{2.64}$ & $2.66$ &
$-0.09_{-0.21}^{+0.15}$ & $-1.44_{-1.79}^{+2.29}$ & $0.32_{-0.32}^{+2.42}$ & $2.21$ &
$-0.16_{-0.17}^{+0.15}$ & $-1.12_{-1.52}^{+1.68}$ & $0.23_{-0.23}^{+2.44}$ & $1.22$ \\

$3$ &
$-0.58_{-0.15}^{+0.39}$ & $-0.02_{-1.62}^{+1.65}$ & $-2.67_{-4.89}^{+6.68}$ & $21.5$ &
$-0.58_{-0.16}^{+0.41}$ & $1.61_{-4.22}^{+3.43}$ & $-2.04_{-7.26}^{8.14}$ & $20.8$ &
$-0.59_{-0.04}^{+0.09}$ & $-0.12_{-0.74}^{+0.96}$ & $-1.18_{-4.18}^{+3.82}$ & $3.86$ &
$-0.55_{-0.05}^{+0.11}$ & $-1.02_{-0.57}^{+1.29}$ & $-0.81_{-3.28}^{+4.13}$ & $0.78$ &
$-0.54_{-0.05}^{+0.02}$ & $-1.25_{-0.26}^{+0.96}$ & $-0.73_{-2.06}^{+1.82}$ & $0.19$ \\

$4$ & 
$-0.13_{-0.27}^{+0.30}$ & $-4.63_{-33.7}^{+24.3}$ & $11.14_{-11.14}^{+611}$ & $42.2$ &
$-0.07_{-0.20}^{+0.22}$ & $-10.9_{-66.2}^{+80.5}$ & $6.51_{-6.51}^{+441}$ & $35.4$ &
$-0.19_{-0.30}^{+0.41}$ & $-5.62_{-13.5}^{+21.8}$ & $9.54_{-9.54}^{+786}$ & $13.0$ &
$-0.21_{-0.30}^{+0.31}$ & $-0.41_{-6.16}^{+ 5.91}$ & $18.6_{-18.6}^{+114}$ & $5.90$ &
$-0.15_{-0.35}^{+0.35}$ & $-0.32_{-4.10}^{+4.02}$ & $11.5_{-11.5}^{+1145}$ & $4.79$ \\

\hline \hline
\end{tabular}}
\label{tab: fithomnoise}
\end{center}
\end{table*}

Concerning the results for noisy moments, we remind the reader that 2nd and 4th order moments are jointly fit since the same parameters enter both calibration relations. As a side effect, the rms on 3rd order moments will typically be smaller than the one for 2nd and 4th order ones since in this case the fit has to adjust a smaller set of parameters and only minimize the residuals with respect to a single set of moments. A further remark concerns the $\nu_4$ parameter which turns out to be practically unconstrained at the upper end (with the upper limit of the $68\%$ confidence ranges being up to two orders of magnitude larger than the median value). Actually, we checked that this is a consequence of the very small contribution the noise gives to the overall fourth order moments so that even values of $\nu_4$ as large as its upper limit are negligible with respect to the signal. A similar discussion also holds for $\nu_3$. Although such large errors could be reduced by shrinking the statistical errors on moments with wider area surveys, we stress that we do not actually worry about the uncertainties on the calibration parameters since these quantities are marginalized over in the present Fisher matrix analysis and in future application to real data.

\section*{Appendix B\,: shear Fisher matrix}

We briefly sketch below how we computed the cosmic shear tomography Fisher matrix. Its elements are

\begin{eqnarray}
F_{\alpha \beta} & = & \sum_{\ell = \ell_{min}}^{\ell_{max}}{\frac{(2 \ell + 1) f_{sky} \Delta \ell}{2}} \nonumber \\
 & \times & 
{\rm Tr}{\left [ \frac{\partial {\bf C}(\ell)}{\partial p_\alpha} {\bf Cov}^{-1}(\ell)
\frac{\partial {\bf C}(\ell)}{\partial p_\beta} {\bf Cov}^{-1}(\ell) \right ]}
\label{eq: fmwl}
\end{eqnarray}
where $f_{sky}$ is the fraction of the sky covered by the survey, the sum is over the $\ell$ mode within the range $(\ell_{min}, \ell_{max})$, ${\bf C}(\ell)$ is the ${\cal{N}}_{bin} \times {\cal{N}}_{bin}$ lensing matrix whose elements ${\cal{C}}_{ij}(\ell)$ read

\begin{equation}
{\cal{C}}_{ij}(\ell) = \frac{c}{H_0} \int_{0}^{z_{h}}{\frac{{\cal{W}}_{i}^{\gamma}(z) {\cal{W}}_{j}^{\gamma}(z)}{E(z) \chi^2(z)} 
P\left [ \frac{\ell + 1/2}{\chi(z)}, z \right ] dz} 
\label{eq: defcij}
\end{equation}
having assumed no contribution from IA. These also enters the covariance matrix ${\bf Cov}(\ell)$ which is assumed to be Gaussian so that its elements are simply given by

\begin{equation}
{\bf Cov}_{ij}(\ell) = {\cal{C}}_{ij}(\ell) + \frac{\gamma_{int}^2 \delta_{ij}^{K}}{f_i n_g \times 3600 (180/\pi)^2}
\label{eq: defcov}
\end{equation}
with $\delta_{ij}^{K}$ the Kronecker symbol, and $f_i n_g$ the source number density for the $i$\,-\,th redshift bin. We will use equipopulated bins, it is $f_{i} = 1/{\cal{N}}_{bin}$, while we set $\gamma_{int} = 0.22$ for the variance of the intrinsic ellipticity distribution.

In Eq.(\ref{eq: defcij}), $P(k, z)$ is the same matter power spectrum we have used to compute the theoretical HOM, while ${\cal{W}}_i^{\gamma}(z)$ is the same as $W(z)$ in Eq.(\ref{eq: defkern}) provided one replaces $n(z)$ with $n_i(z)$, this latter quantity being the redshift distribution for the $i$\,-\,th bin. This is obtained from the full redshift distribution $n(z)$ convolving with the selection function of the $i$\,-\,th bin which also takes into account photometric redshift uncertainties. Finally, we set $(\ell_{min}, \ell_{max}) = (10, 5000)$ and use 50 equispaced logarithm bins in $\ell$ adjusting $\Delta \ell$ accordingly.

\end{document}